\documentclass[11pt,floatfix,prd,preprintnumbers,amssymb]{revtex4}
\pdfoutput=1

\usepackage{amsmath}
\usepackage{graphicx}
\usepackage{epsfig}
\usepackage{graphics}
\usepackage{epsfig}
\usepackage{psfrag}
\usepackage{rotating}
\usepackage{dcolumn}
\usepackage{bm}

\newcommand{\beq}{\begin{eqnarray}} 
\newcommand{\eeq}{\end{eqnarray}}

\newcommand{\ba}{\begin{array}}     
\newcommand{\ea}{\end{array}}     
\newcommand{\bd}{\begin{displaymath}}     
\newcommand{\ed}{\end{displaymath}}     
\newcommand{\be}{\begin{equation}}      
\newcommand{\ee}{\end{equation}}     
\newcommand{\bea}{\begin{eqnarray}}     
\newcommand{\eea}{\end{eqnarray}}    

\begin{document}
\baselineskip=16.5pt 
 
\rightline{LPT-ORSAY-11-02}

\vspace{0.7cm} 
 
\begin{center}

{\Large {\bf Discriminating four-dimensional supersymmetry 
\\ from its five-dimensional warped version}}
\vspace{0.7cm} 
 
{\large Charles Bouchart $^1$, Alexander Knochel $^2$, Gr\'egory Moreau $^1$}
 
\vspace{0.7cm}

1: {\it Laboratoire de Physique Th\'eorique, CNRS and Universit\'e Paris--Sud 11, \\ 
B\^at. 210, F--91405 Orsay Cedex, France.} 
 \\ 
2: {\it Institut f\"ur Theoretische Physik, Universit\"at Heidelberg, \\
Philosophenweg 19, D--69120 Heidelberg, Germany.}

\end{center} 
 
\vspace{0.5cm}  
\begin{abstract}   
\footnotesize{
In the scenario where only superpartners were produced at the Large Hadron Collider, how one could determine whether the supersymmetric 
model pointed out is four-dimensional or higher-dimensional ? We propose and develop a series of tests for discriminating between a pure supersymmetry (SUSY) 
and a SUSY realized within the well-motivated warped geometry \`a la Randall-Sundrum (RS). Two of these tests make use of some different patterns arising in the 
squark/slepton mass spectrum. The other distinctive RS SUSY feature is the possibly larger (even dominant) Higgs boson decay branching ratios into sleptons, 
compared to pure SUSY. Techniques for pinning up the presence of soft SUSY breaking terms on the TeV-brane are also suggested, based on the analysis of stop
pair production at the International Linear Collider.  
For all these phenomenological studies, we had first to derive the four-dimensional (4D) effective couplings and mass matrices of the sfermions and Higgs bosons in RS SUSY.
The localization of Higgs bosons, characteristic of RS, leads to singularities in their couplings which are regularized by the exchange contribution of infinite towers  
of Kaluza-Klein (KK) scalar modes with Dirichlet-Dirichlet boundary conditions. A general method is provided for this regularization, based on the 
completeness relation. The sfermion masses are obtained either from integrating out those specific KK towers or by treating their mixing effects. 
Finally, we show at the one-loop level how all quadratic divergences in the Higgs mass cancel out for any cut-off, due to 5D SUSY and to 5D anomaly cancellation;  
the analytical way followed here also allows a justification of the infinite KK summation required for the so-called KK regularization in 5D SUSY, which has   
motivated a rich literature.} 
\end{abstract} 
 
\maketitle

\newpage

\section{Introduction}
\label{intro}

Among the possible theories underlying the Standard Model (SM) of particle physics, the supersymmetric scenarios and the higher-dimensional 
models have several deep motivations. In particular, supersymmetry stabilizes the ElectroWeak Symmetry Breaking (EWSB) scale with respect to radiative 
corrections and when it is promoted to a local symmetry it provides a framework for the appearance of the graviton Lagrangian. 
On the other side, a warp extra-dimensional model has been proposed by Randall and Sundrum (RS) \cite{RS} 
\footnote{The RS scenarios with a fundamental Higgs boson on the so-called TeV-brane, the alternative models of gauge-Higgs unification \cite{GHunif} and the
Higgsless models \cite{Higgsless} can be thought of as warped extra-dimension models constituting dual descriptions -- through the $AdS/CFT$ correspondence
\cite{AdSCFT} -- of four-dimensional (4D) strongly coupled gauge theories (in the limit of a large number of colors) predicting the effective Higgs scalar field 
as a composite state (see e.g. Ref.~\cite{MHCM}).}
to explain the huge hierarchy between
the EWSB energy scale and the gravity scale while some RS extensions \cite{PomGher} give rise to purely geometrical interpretations of the fermion structure
in flavor space. 
Both supersymmetric and higher-dimensional scenarios can provide viable dark matter candidates (LSP, LKP) and allow for gauge unification.  
\\ Since the physics beyond the SM is still unknown, one should consider the possibility of an effective 
low-energy hybrid theory with both SUperSYmmetry (SUSY) and extra dimension(s).
More specifically, SUSY and extra dimensions could be crucial ingredients in a quantum description of gravity, 
as both indeed are in string theory. 
Moreover, D-branes of superstring theory provide a natural mechanism for the confinement of the SM fields in brane-models and constitute
thus another motivation for introducing SUSY in higher-dimensional scenarios.  
SUSY and extra dimensions could also be simultaneously involved
in the origin of the electroweak symmetry breaking \cite{EWSBI,EWSBII}. The higher-dimensional framework even provides
new and attractive ways of breaking supersymmetry, due to a more structured geometry of space-time; SUSY may be broken
either in the bulk (i.e. whole space) or on a brane, playing the role of the necessary hidden sector, and then the breaking mediated to the brane 
where are living the SM particles \cite{hidden}. A different approach towards the SUSY 
breaking \footnote{It has been shown recently that in the limit of an infinite Majorana mass term on the infra-red brane \cite{Steve}, the 
gaugino mediation mechanism through the bulk is comparable to a SUSY breaking with twisted boundary conditions.}
is to use specific Boundary Conditions (BC) for different fields (in the same supermultiplet) propagating along compactified 
extra dimensions \cite{BCmech}.
A famous example is the Scherk-Schwarz mechanism \cite{SSmech} which can be applied to the minimal supersymmetric SM (see e.g. 
\cite{SSmssm}) as well as to the Horava-Witten theory (see e.g. \cite{SShorava}) \footnote{In this context of extra dimensions, other new scenari
of SUSY breaking have arisen \cite{EWSBI} and in particular in the case of warped dimensions \cite{newSUSYbRS}.}.  
Finally, the supersymmetrization of extra-dimensional theories allows to have a realistic tension model \cite{tension}, 
to provide a new source for a tiny cosmological constant \cite{Schmidhuber},  
to alternatively achieve gauge unification \cite{unification} and to address the so-called $\mu$-problem in SUSY (see discussion below).   
\\ Hence, there exist several serious motivations for hybrid scenarios with both SUSY and (warped) extra dimensions. It was shown in \cite{AdSUSY}
that an Anti-de Sitter ($AdS$) space is compatible with SUSY. Later \cite{RSUSYspec}, the supermultiplet mass-spectrum was derived in $AdS_5$, 
and then the analysis extended to the case of a fifth dimension compactified on an orbifold \cite{PomGher}. The RS SUSY scenario was initially studied
with only gravity in the bulk \cite{RSUSYgrav} and finally with matter propagating in the bulk \cite{PomGher}. There exist also analyses
of GUT theories within RS SUSY frameworks \cite{RSUSYunif} and attempts of superstring realizations of the RS model \cite{RString}.

In this paper, we will study these supersymmetrized RS scenarios with matter/gauge fields in the bulk. 
Our first contributions are theoretical; we will write explicitly the complete 5D Lagrangian
for a Higgs scalar field confined on the TeV-brane -- as required to explain the discrepancy between EWSB and gravity scales 
and as motivated by the ${\rm SU(2)_R}$ breaking from bulk Yukawa interactions (forbidden trilinear chiral superfield couplings in $N=2$ 4D SUSY  
\cite{Strathdee}) --  
and, when deriving the 4D effective action, we will take into account for the first time the mixing among the several Higgs bosons occuring in the minimal SUSY extensions.   
\\ In particular, the 4D effective Higgs couplings to two scalar fields (squarks/sleptons and their Kaluza-Klein excitations), originating from the so-called D-terms, 
are not trivial to derive due to the localized aspect of the Higgs boson and due to contributions to these effective 
couplings from the tree level exchange of Kaluza-Klein (KK) scalar modes of $(--)$ chiral superfields
\footnote{We use a $N=1$ 4D superfield formalism as in Ref.~\cite{4DSuperFC}.}  
(i.e. additional superfields with Dirichlet-Dirichlet boundary conditions which are characteristic of 5D theories). 
We derive these 4D couplings which do not appear in literature and explain why this specific KK $(--)$ tower must be taken into account 
without any cut-off (a non-natural task within a non-renormalizable 5D theory).
\\ The derivation of Yukawa couplings between Higgs bosons and two scalars in the 4D effective Lagrangian is also subtle: singularities [Dirac peak functions
taken at the origin, $\delta(0)$'s, due to the Higgs local aspect] appear in these couplings -- after integrating out the auxiliary fields -- 
but those are not the sign of an incomplete theory. Indeed, those singularities 
are cancelled out by the infinite sum of KK $(--)$ scalar contributions. A related kind of cancellation was pointed out for some loop calculations 
in the string theory framework \cite{Sharpe} 
using either the expression of $\delta(0)$ in terms of a sum over the fifth component of the momentum \cite{hidden} (see Ref.~\cite{Garrie} 
for the warped background case) or a Gaussian brane distribution 
\cite{Gero}; in contrast, here we demonstrate the cancellation generally for the tree level couplings, 
using a simple method based on the completeness relation,   
by computing the Higgs couplings (including the Higgs-sfermion couplings and quartic Higgs interactions) 
in the flavor-motivated case of a warped extra dimension where only Higgs bosons are brane-localized \cite{PomGher}. 
\\ In addition, in the present work we will derive the 4D effective scalar mass matrices induced, after EWSB, by the localized Higgs Vacuum Expectation 
Values (VEV). In contrast with the above 4D couplings, these 4D masses computed at tree level do not depend on the energy scale and must be derived 
consistently either 
through the integration out of heavy KK $(--)$ scalar modes or through the scalar mixing with KK $(--)$ states, as we will show here. 
We will finally combine at first order these KK $(--)$ scalar effects on the 4D zero-mode scalar masses 
with the scalar mixing effects of KK $(++)$ states (having Neumann-Neumann boundary conditions).
Such a combination should have been done in Ref.~\cite{Gautam} [where effects of KK scalar modes with even 
$\mathbb{Z}_{2}$ parity (i.e. $(++)$ BC) are studied in a 5D SUSY context with brane-localized Yukawa interactions, 
but the above effects of KK scalars with an odd parity ($(--)$ BC) are considered neither in squark/slepton masses 
nor in Higgs couplings].

We will then use the new 4D effective Lagrangian, that we will have derived in the RS SUSY framework, for several phenomenological applications.
\\ The first application is an explicit and complete diagramatic computation of quantum corrections to the Higgs boson mass ($\delta m_{Higgs}$) 
at the one-loop level, including all kinds of KK contributions and each sector (Yukawa plus gauge couplings); 
we will show how the quadratically divergent parts can cancel each other due to 5D SUSY. This complete cancellation has not been shown  
before in warped SUSY models and it constitutes here an additional check of the obtained 4D effective Higgs couplings. 
More generically, we will clarify the connections between the cancellations of these 5D quadratic divergences and of the 5D triangular anomalies.   
Our way of finding the quadratic divergence cancellation brings some new light on the old debate about the validity of this
cancellation in $\delta m_{Higgs}$ within higher-dimensional SUSY models. In particular, the preliminary and complete 
calculation of 4D
effective Higgs couplings allows a more clear overview of the subtle points and in turn allows to address the `KK regularization' question and cut-off problems.   
\\ Our general results on the 
absence of quadratic divergences in $\delta m_{Higgs}$ for 5D SUSY models [with a soft breaking] is important in the following sense: 
it seems to mean that hybrid scenarios -- both higher-dimensional and supersymmetric -- must not necessarily rely on a geometrical background reducing the  
gravity scale down to the TeV scale in order to protect the Higgs mass against its quantum corrections (i.e. to not reintroduce the gauge hierarchy problem). 
One can thus imagine a 5D SUSY scenario where the Higgs mass is only protected by SUSY 
(which allows to avoid the remaining little hierarchy problem of pure 5D
models solving the gauge hierarchy) or a 5D SUSY scenario where in addition to SUSY protection the discrepancy between the fundamental
gravity and EWSB energy scales is explained 
by some geometrical feature like the warp factor: those two possible classes of higher-dimensional SUSY theories avoid to have two redundant solutions to the 
same Higgs mass instability problem [= fine-tuning problem].

The second application concerns the direct search at present and future high-energy colliders for the (necessary) physics standing beyond the SM;  
indeed, in the expected case where some signal for new physics would be discovered at the Large Hadron Collider (LHC) and then analyzed more precisely
at the International Linear Collider (ILC), the primary phenomenological work would be to identify exactly the nature of the new physics detected. Based on the fact that
today the two main types of new physics accessible at colliders are thought to be SUSY and the more recent paradigm of (warped) extra dimensions, three
interesting possibilities might arise. The more optimistic is that both superpartners (squarks, gauginos\dots) and KK excitations would be produced on-shell and observed,
proving then the existence of an higher-dimensional SUSY scenario. Another possibility is that only real KK excitations would be produced, however such a
situation would represent a good indication for the existence of a higher-dimensional non-SUSY theory 
as (in)direct constraints and gauge hierarchy considerations favor the mass regions around $10^2$ GeV for superpartners and
higher mass regions above ${\cal O}(1)$ TeV for first KK states (of the warped models). The last possibility -- among the cases of signals for new physics as it is predicted today -- 
is that only superpartners would be produced on-shell (and maybe some 
additional Higgs bosons characteristic of SUSY). Then an important and non-trivial question (see e.g. the related Ref.~\cite{HewSa}) 
would be: do the observed superpartners [either scalar or spinorial] belong to a pure SUSY theory or a (warped) higher-dimensional SUSY model ?    
\\ In the present paper, we will propose and develop some tests allowing to answer this question which reads in a more compact form as: 
{\it how one can distinguish between pure SUSY and warped SUSY at colliders ?} 
A simple test would be to look at the precise measurement of cross sections of stop quark pair production at ILC: 
the mixing of the stop quarks with KK excitations could lead to modifications of the production amplitude -- predicted in pure SUSY. 
However we will show that such RS corrections are too small to be observable.
Nevertheless, we will show how the ILC could allow to discriminate between different types of SUSY breaking in the RS context.
\\ Another example of proposed test relies on the measurement (at ILC or even at LHC) of the smuon masses: there are higher-dimensional minimal 
\footnote{`minimal' here means that the SUSY breaking masses and couplings are universal at the GUT scale.} SUperGRAvity (mSUGRA) scenarios where  
the mass splitting $m_{\tilde \mu_2}-m_{\tilde \mu_1}$ ($\tilde \mu_{1,2}$ are the two smuon eigenstates \footnote{Strictly speaking, $\tilde \mu_{1,2}$ 
denote the two lightest mass eigenstates which are generally mainly composed of the left and right (w.r.t. gauge group ${\rm SU(2)_L}$) 
smuon zero-modes while the heavier eigenstates are mainly made of their KK modes.}) can be larger than in the 4D mSUGRA scenarios \cite{NillesReport}, 
allowing then a discrimination between those two types of scenarios. The reason 
is that in mSUGRA the off-diagonal elements of the $2 \times 2$ smuon mass matrix -- partially responsible for the mixing and splitting -- are proportional to the muon 
Yukawa coupling constant which is suppressed compared to the top quark one, while in some RS frameworks \cite{PomGher}, the muon 5D Yukawa coupling is not
suppressed relatively to the top one (the lightness of the muon originates from its wave function overlap with the Higgs boson).
We will also show that RS SUSY models can lead to differences in the stop mass correlations ($m_{\tilde t_2}$ versus $m_{\tilde t_1}$) with respect to the pure SUSY case 
-- differences which are here not restricted to the mSUGRA model.
\\ Finally, there exist well-motivated warped geometrical setups of SUSY breaking where the Higgs boson couplings to sleptons can be  
significantly increased with respect to the conventional 4D SUSY frameworks. Hence, in such warped scenarios, the decay channels of the
heaviest neutral Higgs scalar (characteristic of SUSY and usually noted $H$)
\footnote{No deep modifications are expected for the branching ratios of the lightest neutral Higgs boson $h$, since its decays into two squarks/sleptons 
($\tilde q/\tilde \ell$) are kinematically closed due to the upper theoretical mass bound on $m_h$ GeV combined with 
the lower direct experimental bounds $m_{\tilde q/\tilde \ell} \gtrsim 10^2$ GeV \cite{PDG}.} 
into sleptons can have comparable or even larger 
branching ratios than the channels into (s)quarks and gauginos, leading to final states at LHC radically different from 
the 4D SUSY case. The produced $H$ bosons could initiate slepton cascade decays more often than in 4D SUSY and in turn give rise to an increase 
number of events with leptonic final states. 
We will illustrate this study by giving numerically all the $H$ boson branching ratios as a function of the mass, for characteristic parameter sets, 
and the cases of pseudo-scalar and charged Higgs fields will be briefly discussed.

At this level, one should mention related works. In fact, within the hybrid higher-dimensional SUSY context,
while the literature on various SUSY breaking mechanisms is rich, there exist few papers on phenomenological aspects.
Among those, we mention Ref.~\cite{Gautam} where radiative corrections to scalar masses (excluding quadratically
divergent parts) and to gauge/Yukawa couplings are calculated in order to {\it (i)} estimate the 5D MSSM upper limit on 
the lightest Higgs mass and {\it (ii)} constrain experimentally the universal scalar ($m_0$) and gaugino ($m_{1/2}$) masses.
See Ref.~\cite{DelgQuir} for another type of constraint on the Higgs mass in higher-dimensional SUSY models. 
In Ref.~\cite{Aldo}, based on 4D superfield actions, the evolution of neutrino masses/mixings
and confrontation with oscillation data were studied within a flat 5D MSSM.
See also Ref.~\cite{GautamBIS} for considerations on the proton decay suppression in SUSY grand unified theories 
with extra compact dimensions (and localized Yukawa couplings or not) 
as well as Ref.~\cite{Fichet} for phenomenology in SUSY gauge-Higgs unification scenarios,
Ref.~\cite{Ben} for SUSY composite Higgs models
and finally Ref.~\cite{AKnochel} for SUSY warped Higgsless models.

The organization of the paper is as follows. In Section \ref{theorypart}, After a description of the general higher-dimensional SUSY framework considered,
we derive the 4D effective couplings and mass matrices of sfermions and Higgs bosons, focusing on illustrative examples. In the first part of Section \ref{phenomenologypart}
(Subsection \ref{quantumpart}), the quadratic divergent contributions to the quantum corrections of the Higgs mass are calculated and the methods of calculation discussed. 
Up to this point, the results obtained are valid for any RS SUSY model -- and can be easily generalized to any higher-dimensional SUSY scenario --
with brane-Higgs bosons. In 
the three following subsections on collider phenomenology, a certain SUSY breaking setup must be chosen for the computations; 
in Subsection \ref{sfermionpart}, various distinctive effects of RS SUSY models are pointed out and quantified. Effects 
in the Higgs sector are studied in Subsection \ref{Higgspart}. 
Finally, in Subsection \ref{stopart}, we propose some tests, based on stop pair production at ILC, 
for distinguishing different RS SUSY realizations. We conclude in Section \ref{concludepart}.

\section{Theory}
\label{theorypart}                               

\subsection{The model}
\label{model}

\noindent {\bf Field content:}
In Appendix \ref{lagrangianpart}, we give the full 5D field Lagrangian for a toy model with a ${\rm U(1)}$ gauge symmetry applying on two chiral superfields $\Phi_L,\Phi^c_L$ 
(together with the associated $(--)$ superfields) and two 4D Higgs superfields $H^0_u,H^0_d$ localized on the TeV-brane within a warp background.   
With this appendix as a starting point, we will give throughout the paper the 4D version of this Lagrangian related to the Higgs boson or more precisely the parts which 
are not trivial/direct to derive from the 4D point of view. For these studied parts, we will extend the 4D Lagrangian to the phenomenological 
Minimal Supersymmetric Standard Model (pMSSM) \cite{pMSSM} field content and gauge symmetry.
\\ \\ 
\noindent {\bf Energy scales:}
The RS framework is constituted by a 5D theory where the extra dimension is warped and compactified over a $S^{1}/\mathbb{Z}_{2}$ orbifold. 
The non-factorizable metric is of type AdS
and the space-time, which is thus a slice of $AdS_5$, has two 4D boundaries: the Ultra-Violet (UV) boundary at the Planck scale
and the Infra-Red (IR) brane with an exponentially suppressed scale in the vicinity of the TeV scale. 
The Higgs boson has to be localized at this so-called TeV-brane if the EW scale is to be stabilized by such a geometrical structure.  
We consider the attractive RS version with all other fields propagating in the bulk \cite{PomGher}: this allows to suppress higher dimensional operators, potentially 
troublesome with respect to Flavor-Changing Neutral Current effects, by energy scales larger than the TeV scale. This feature has also the
advantage to possibly generate the fermion mass hierarchy and flavor structure by a simple geometrical mechanism \cite{PomGher,RSmass,RSmassBIS}. 
\\ 
More precisely, the gravity scale on the Planck-brane is $M_{\rm Planck}= 2.44\times
10^{18}$ GeV, whereas the effective scale on the TeV-brane $M_{\star}=e^{-\sigma(\pi R_c)} M_{\rm Planck}$ (with $\sigma(y)=k\vert y \vert$)
is suppressed by the warp factor which depends on the curvature
radius $1/k$ of $AdS_5$ and the compactification radius $R_c$.
For a product $k R_{c} \simeq 11$, $M_{\star}\!=\!{\cal O}(1)$ TeV allowing to
address the gauge hierarchy problem. We will take $k R_{c} \simeq 10.11$ so that
the maximum value of $M_{KK} \simeq 2.45 k e^{-\pi kR_{c}}$ [$M_{KK}$ is the first KK photon mass], fixed by the theoretical 
consistency bound $k<0.105 M_{\rm Planck}$, is $\sim 10$ TeV in agreement with 
the typical indirect limits from EW precision tests (see below). The beauty of the RS model is
to possess a unique fundamental energy scale $k \sim R_{c}^{-1} \sim M_{\rm Planck}$.
Besides, the parameters noted $c^\psi$ fix the 5D masses $m_\psi=c^\psi \partial_y \sigma$, affected to each fermion $\psi$, 
and thus control the field localizations in the bulk (and in turn the effective 4D masses). Those satisfy 
$\vert c^\psi \vert \! = \! {\cal O}(1)$ to avoid the introduction of new fundamental scales. The 5D masses for the scalar fields  
will be discussed throughout the paper.  
\\ \\
\noindent {\bf $\mu$-problem:} 
A usual problem of the supersymmetric theories is to explain why  
the $\mu$ parameter is of order of the EWSB scale (around the TeV), as imposed by the orders
of magnitudes of the masses involved in the minimization conditions for the Higgs potential. 
In RS SUSY, there is a simple way for generating a $\mu$-term at the EWSB scale as we discuss now
\footnote{We will not discuss the type of possibility suggested in Ref.~\cite{DelgQuiSS} 
which is characteristic of the Scherk-Schwarz mechanism not considered here. A related approach based on twisted BC for the Higgs fields
was proposed in Ref.~\cite{PartSusyRS}.}. If the Higgs superfields are confined on the TeV-brane (as motivated by the gauge hierarchy problem), 
the gauge invariant $\mu$-term in the 5D superpotential $W$ reads in terms of the 4D Higgs superfields as, 
\begin{eqnarray}
[ \mu H_u . H_d  ] \delta(y-\pi R_{c}) \in W \ , \ {\rm with} 
\ H_u = \left(\begin{array}{c} H_u^+ \\ H_u^0 \end{array}\right) , \ H_d = \left(\begin{array}{c} H_d^0 \\ H_d^- \end{array}\right) \  
{\rm and} \ \ H_u . H_d = \epsilon_{ab} H_u^a H_d^b \ ,    
\label{mu2scen} 
\end{eqnarray}
$H_{u,d}$ being ${\rm SU(2)_L}$ doublets, $a,b$ ${\rm SU(2)_L}$ indices and $\epsilon_{ab}$ the antisymmetric tensor defined by $\epsilon_{12}=1$. 
In analogy with Eq.(\ref{muAction}) for the ${\rm U(1)}$ model [equivalent to restrict the ${\rm SU(2)_L}$ model to the neutral Higgs couplings] which 
leads to the expression of the $\mu$ terms written with the fields in Eq.(\ref{ScaFieLag}), the term (\ref{mu2scen}) gives rise to the 4D mass terms  
for the neutral Higgs scalar fields in the potential (after field redefinition and inclusion of the warp metric factor),
$$
\vert \mu_{eff}\vert^2 \vert \phi_{H_u^0} \vert^2 + \vert \mu_{eff}\vert^2 \vert \phi_{H_d^0} \vert^2 \in V_{4D}, \ {\rm with} \ 
\vert \mu_{eff}\vert^2 = \int_{-\pi R_{c}}^{\pi R_{c}} dy \ \vert \mu \vert^2 \ e^{-2 \sigma(y)} \ \delta(y-\pi R_{c}) = \vert \mu e^{-\sigma(\pi R_{c})} \vert^2,
$$ 
so that the effective $\mu$ parameter reads as $\mu_{eff} = \mu e^{-k \pi R_{c}} \sim k e^{-k \pi R_{c}} \sim$ TeV.
Note the absence of $\delta(0)$ factors in $V_{4D}$ (or equivalently of $[\delta(y-\pi R_c)]^2$ terms in the 5D Lagrangian) 
after integration over $y$. 
In the paper, we will consider this setup with brane Higgs fields. 
\\ \\
\noindent {\bf Custodial symmetry:} 
In pure SUSY with $\tan \beta = 1$ \footnote{We use the conventional notations for the pMSSM Higgs scalar fields:
$\phi_{H^0_{u/d}}=(v_{u/d}+\phi_{h^0_{u/d}}+i \phi_{P^0_{u/d}})/\sqrt{2}$ ($\phi_{P^0_{u/d}}$ 
denoting the pseudo-scalars) with $\tan\beta =v_u/v_d$ and the squared VEV given by $v^2 = v_u^2 + v_d^2 \simeq (246 \ GeV)^2$.}, 
if the custodial symmetry ${\rm O(3) \approx SU(2)_V}$ -- resulting after EWSB from the global symmetry ${\rm O(4) \approx SU(2)_L \times SU(2)_R}$ -- 
was exactly respected it would protect the well known relation on gauge boson masses $\rho=m_W^2/(m_Z^2\cos^2\theta_W)=1$ 
against quantum corrections, as in the SM. Indeed, the two Higgs ${\rm SU(2)_L}$ doublets of the pMSSM can
form a $({\bf 2},{\bf 2})$ representation under the ${\rm SU(2)_L \times SU(2)_R}$ symmetry \cite{DreesHag}:
\begin{eqnarray}
{\cal H} \equiv (H_d \ H_u) = \left(\begin{array}{cc}  H_d^0 & H_u^+ \\ H_d^- & H_u^0 \end{array}\right).
\end{eqnarray} 
However, the custodial symmetry is broken in the gauge and Yukawa coupling sectors. Moreover, the 
deviations from $\tan \beta = 1$ and the presence of soft SUSY breaking terms for squarks/sleptons represent new sources of custodial symmetry (spontaneous)
breaking if those differ for the up-type and down-type (w.r.t. ${\rm SU(2)_L}$) scalars. Nevertheless, the loop level SUSY corrections to the precision EW observables
(contributing to $\delta \rho$) can pass the constraints from precision measurements, even easily in the case of equal soft terms for up and down scalars
\cite{DjouadiReviewII,PDG}.
\\ 
Concerning the tree level corrections to precision EW observables in warped models, the global fits of experimental data are satisfactory for large KK masses
(reducing the KK mixing effects): $M_{KK} \gtrsim 10$ TeV \cite{RSewpt}.
\\ 
In the present paper where we study hybrid scenarios with both superpatners and KK excitations, we will take identical soft parameters for up/down scalars and
KK masses just above $\sim 10$ TeV. We will not explore the whole parameter space and work out the precise global EW fits, but this realistic choice guarantees that 
the theoretical respective, and in turn total, 
corrections to SM observables have acceptable orders of magnitude given the experimental accuracy on these observables. There are (a priori) higher-order 
corrections involving both superpatners and KK modes, that we do not treat here. 
\\ 
Now one may wish to decrease the possible $M_{KK}$ values in order to improve the situation with respect to considerations on the `little hierarchy' problem
related to the Higgs mass.
In order to reduce the acceptable minimal $M_{KK}$ value from $\sim 10$ TeV down to $\sim 3$ TeV, an attractive \cite{CBGMI} possibility is to gauge the 
${\rm SU(2)_L \times SU(2)_R \times U(1)_X}$ symmetry in the bulk \cite{ADMS}.  
The SM gauge group is recovered after the breaking of the ${\rm SU(2)_R}$ group into another ${\rm U(1)_R}$, by boundary conditions.
\\
The question arising here is whether it is possible to gauge the ${\rm SU(2)_L \times SU(2)_R}$ symmetry within a SUSY context. 
First about the $\mu$-term: can it be written in the case of a bulk gauge custodial symmetry ? 
The question arises as ${\rm SU(2)_L \times SU(2)_R}$ must remain unbroken on the Higgs brane in order to protect precision EW observables. 
A simple possibility is to write down the following ${\rm SU(2)_R}$ invariant term of the superpotential giving rise to the usual $\mu$-term,
$$
-\frac{1}{2} \mu {\cal H}.{\cal H} \ \delta(y-\pi R_{c})= -\frac{1}{2}\mu (H_d.H_u - H_u.H_d ) \ \delta(y-\pi R_{c})= \mu H_u.H_d \ \delta(y-\pi R_{c})\ ,
$$ 
accordingly to Eq.(\ref{mu2scen}). Higgs couplings to matter invariant under
${\rm SU(2)_L \times SU(2)_R}$ can also be written with the ${\cal H}$ bidoublet giving rise in particular to the usual Yukawa couplings in the superpotential, as usually
 in warped models by promoting matter (superfield) multiplets to ${\rm SU(2)_R}$ representations (see e.g. \cite{ADMS,CBGMI}). 
 Finally, the gauge interaction sector of a SUSY theory as well as the soft breaking terms can also respect an additional 
${\rm SU(2)_R}$ gauge symmetry.  
\\
Hence, in this paper we will also consider the case of a bulk gauge ${\rm SU(2)_L \times SU(2)_R}$ symmetry with $M_{KK} \sim 3$ TeV 
and equal soft terms for up/down scalars, another setup which leads to realistic EW fits to data. 
\\ \\
\noindent {\bf SUSY breaking:} We adopt the classification of the three types of SUSY breaking framework where {\it (i)} the squark/slepton SUSY breaking masses
are induced at loop level by SUSY breaking gaugino masses as in \cite{PomGher}
 {\it (ii)} there exist tree level squark/slepton masses [in the bulk and/or brane-localized] resulting directly from a 5D SUSY breaking mechanism
 {\it (iii)} the squark/slepton masses are of KK type like in the scenario \`a la Scherk-Schwarz with SUSY breaking BC \cite{SSmech}.
This classification is motivated in particular by the fact that the framework with scalar SUSY breaking masses in the bulk leads to modifications of 
the wave functions and thus of the 4D effective couplings. Moreover this framework gives rise to an hard breaking of SUSY (reintroducing quadratic divergences
in the Higgs mass corrections)
which is acceptable in the present warped background where the Higgs mass is still protected by the reduced gravity scale.
Finally, this framework with bulk scalar SUSY breaking masses has the attractive feature to generate
scalar superpartner masses mainly through the Yukawa couplings like for SM fermions -- as will be described in details.
\\ We will consider the second class of SUSY breaking framework mentioned above, 
including the mSUGRA case where the SUSY breaking terms are universal and the soft trilinear scalar couplings (the $A$ couplings) 
are proportional to the Yukawa couplings. The first class is typically restricted to the kind of model described in \cite{PomGher} (with SUSY breaking
on the TeV-brane) while the third
class has a SUSY breaking mechanism deeply related to BC. We will show that the second type of SUSY breaking allows to develop
tests for distinguishing between pure SUSY and warped SUSY at colliders.

\subsection{4D scalar couplings}
\label{couplingpart}

In this section, we derive the 4D effective couplings of the brane-Higgs boson to two scalar superpartners (of 
type squark/slepton) in the RS SUSY framework, as well as the 4D Higgs self couplings. 

\subsubsection*{Scalar Yukawa couplings to two Higgs bosons}

In this first subsection, we calculate the 4D Yukawa-type coupling of a scalar field (e.g. $\phi_R$) to an   
Higgs boson (of $\phi_{H^0_u}$ type, as an illustrative example) within our simple ${\rm U(1)}$ model 
defined in Appendix \ref{SuperContent} corresponding the superfield action given in Appendix \ref{SuperAction}.

The obtained scalar Yukawa couplings are included in the following term of Eq.(\ref{ScaFieLagPrim}) 
in terms of the 5D fields, 
\begin{eqnarray}
\mathcal{L}_{5D} = 
- \sqrt{G} \left|{\cal Y}\delta(y-\pi R_c) \phi_{H_u^0}\phi_R 
- [\partial_y-(c_L+\frac{3}{2})(\partial_y\sigma)]\phi_L^c\right|^2.
\label{Yuk4Da}
\end{eqnarray}

\begin{figure}[!hc]
	\centering
	\vspace*{.5cm}
			\includegraphics[width=0.19\textwidth]{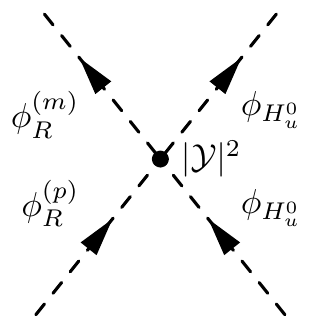}
			\hspace*{1.cm}
			\includegraphics[width=0.19\textwidth]{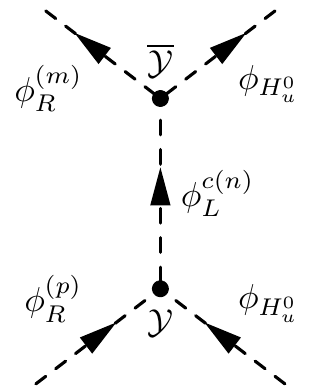}
\caption{\label{YukDrawn} \small{Feynman diagrams of the contributions to the 4D effective scalar Yukawa coupling
$\frac{\delta^4\mathcal{L}_{4D}}{\delta \phi_{H_u^0} \bar\phi_{H_u^0} \phi_R^{(p)} \bar\phi_R^{(m)}}$. 
The second indirect contribution is induced by the exchange of the KK tower of $(--)$ scalar superpartners $\phi_{L}^{c(n)}$.
The relevant coupling constants are described in details in text.}}
\end{figure}

Developing over the KK decomposition (\ref{PhiDecomp}) and integrating over the fifth dimension, 
the first squared term gives rise to the 4D scalar Yukawa coupling (after field redefinition), 
\begin{eqnarray}
\frac{\delta^4i\mathcal{L}_{4D}}{\delta \phi_{H_u^0} \bar\phi_{H_u^0} \phi_R^{(n)} \bar\phi_R^{(m)}} &=& 
- i |\mathcal{Y}|^2 \int_{-\pi R_{c}}^{\pi R_{c}} dy \delta^2(y-\pi R_c) \bar f^{++}_m(c_R;y) f^{++}_n(c_R;y) \nonumber \\
&=& - i |\mathcal{Y}|^2 \delta(0) \bar f^{++}_m(c_R;\pi R_c) f^{++}_n(c_R;\pi R_c),
\label{CompNotUsed}
\end{eqnarray}
whose corresponding diagram is drawn in Fig.(\ref{YukDrawn}).

It turns out that the crossed products in Eq.(\ref{Yuk4Da}) also bring a contribution to the 4D effective 
scalar Yukawa coupling, as we explain now; the crossed terms read as
\begin{eqnarray} 
\sqrt{G} \delta(y-\pi R_c) \bigg [ {\cal Y} \phi_{H_u^0} \phi_R[\partial_y-(c_L+\frac{3}{2})(\partial_y\sigma)]\bar\phi_L^c
+\bar{\cal Y} \bar\phi_{H_u^0} \bar\phi_R[\partial_y-(c_L+\frac{3}{2})(\partial_y\sigma)]\phi_L^c \bigg ]
\in \mathcal{L}_{5D}.
\label{5Dcross}
\end{eqnarray}
Combining the KK decomposition (\ref{PhiDecompBis}) of the 5D field $\phi_L^c$ together with the relation 
(\ref{UsefulRelatII}), which originates from the equation of motion, leads to
\begin{eqnarray}
[\partial_y-(c_L+\frac{3}{2})(\partial_y\sigma)]\phi_L^c(x^{\mu},y) 
&=& \sum_{n=1}^{\infty} \phi_L^{c(n)}(x^{\mu}) [\partial_y-(c_L+\frac{3}{2})(\partial_y\sigma)] f^{--}_n(c_L;y) \nonumber \\
&=& - e^{\sigma} \sum_{n=1}^{\infty} m_L^{(n)} \phi_L^{c(n)}(x^{\mu}) f^{++}_n(c_L;y), 
\end{eqnarray}
$m_L^{(n)}$ being the n{\it th} KK scalar mass. Using this relation in Eq.(\ref{5Dcross}) and developing
$\phi_R$ over its KK tower gives the 4D couplings (taking into account field redefinitions):
\begin{eqnarray} 
- \bigg [ &&
{\cal Y} \phi_{H_u^0} \sum_{m=0,n=1}^{\infty} m^{(n)}_L f^{++}_m(c_R;\pi R_c) \bar f^{++}_n(c_L;\pi R_c)
\phi_R^{(m)} \bar \phi_L^{c(n)}  \quad + \nonumber \\
&&\bar{\cal Y} \bar\phi_{H_u^0} \sum_{m=0,n=1}^{\infty} m^{(n)}_L \bar f^{++}_m(c_R;\pi R_c) f^{++}_n(c_L;\pi R_c)
\bar\phi_R^{(m)} \phi_L^{c(n)} 
\quad \bigg ]
\in \mathcal{L}_{4D}.
\label{4Dcross}
\end{eqnarray}
These two types of 4D coupling to KK modes give rise to new contributions to scalar Yukawa couplings through the 
exchange of the KK tower states $\phi_L^{c(n)}$, as exhibits the second Feynman diagram of Fig.(\ref{YukDrawn})
\footnote{Note that the $\phi_L^{c(n)}$ fields, whose presence reflects the vectorial aspect of the 5D theory
(like for the $\Sigma^{(n)}$ fields), enter indirectly -- via some exchanges -- the couplings in the
4D chiral Lagrangian.}.
The resulting new contributions to 4D scalar Yukawa couplings are given in the following, taking real wave functions,
\begin{eqnarray}
\frac{\delta^4i\mathcal{L}_{4D}}{\delta \phi_{H_u^0} \bar\phi_{H_u^0} \phi_R^{(p)} \bar\phi_R^{(m)}} \bigg \vert_{indirect} =
- |\mathcal{Y}|^2 \sum_{n\geq1} \frac{i \ m^{(n)2}_L}{k^2 - m^{(n)2}_L} 
f^{++}_p(c_R;\pi R_c) f^{++}_m(c_R;\pi R_c) \bigg ( f^{++}_n(c_L;\pi R_c) \bigg )^2 \nonumber\\ 
\end{eqnarray}
$k^\mu$ being the $\phi_L^{c(n)}$ four-momentum.
These couplings can be rewritten as follows, according to the completeness relation of Eq.(\ref{completeness})
together with the 5D propagator definition [$G_5^f(k^2;y,y')$] from Appendix \ref{propapator}, 
\begin{eqnarray}
&& \frac{\delta^4i\mathcal{L}_{4D}}{\delta \phi_{H_u^0} \bar\phi_{H_u^0} \phi_R^{(p)} \bar\phi_R^{(m)}} \bigg \vert_{indirect}  \nonumber\\
&=& - i |\mathcal{Y}|^2 \sum_{n\geq0} \left\{ \frac{k^2}{k^2 - m^{(n)2}_L} -1 \right\} 
f^{++}_p(c_R;\pi R_c) f^{++}_m(c_R;\pi R_c) \bigg ( f^{++}_n(c_L;\pi R_c) \bigg )^2 \nonumber\\
&=& - i |\mathcal{Y}|^2 f^{++}_p(c_R;\pi R_c) f^{++}_m(c_R;\pi R_c) k^2 G_5^{f^{++}(c_L)}(k^2;\pi R_c,\pi R_c)
+ i |\mathcal{Y}|^2 \delta(0) f^{++}_p(c_R;\pi R_c) f^{++}_m(c_R;\pi R_c) \nonumber\\
&=&  i |\mathcal{Y}|^2 f^{++}_p(c_R;\pi R_c) f^{++}_m(c_R;\pi R_c) 
[\delta(0) - k^2 G_5^{f^{++}(c_L)}(k^2;\pi R_c,\pi R_c)].  
\label{CompleteUsed}
\end{eqnarray}

Summing the two contributions, from Eq.(\ref{CompNotUsed}) and Eq.(\ref{CompleteUsed}), the divergent $\delta(0)$ terms 
cancel each other, and we obtain the 4D effective scalar Yukawa couplings:
\begin{eqnarray}
\frac{\delta^4i\mathcal{L}_{4D}}{\delta \phi_{H_u^0} \bar\phi_{H_u^0} \phi_R^{(p)} \bar\phi_R^{(m)}} \bigg \vert_{total} = 
- i |\mathcal{Y}|^2 f^{++}_p(c_R;\pi R_c) f^{++}_m(c_R;\pi R_c) k^2 G_5^{f^{++}(c_L)}(k^2;\pi R_c,\pi R_c).
\label{HYukCoupl}
\end{eqnarray}
Hence, starting from couplings in a 5D SUSY theory, we have derived consistent 4D effective couplings.
More precisely, at this level the Lagrangian of Eq.(\ref{HYukCoupl}) given for any energy $k^2$ 
still describes a real 5D SUSY theory (the KK sum in
$G_5^{f^{++}(c_L)}(k^2;\pi R_c,\pi R_c)$ is infinite) but its form is given from a 4D point of view (4D fields
are used).

An interesting check of Eq.(\ref{HYukCoupl}) is the following one.
In the low--energy limit, $k^2 \ll m^{(n)2}_L$ ($n\geq1$), 
only the zero-mode in the 5D propagator survives at zero{\it th} order, 
so that the superpartner coupling above simplifies to
\footnote{The 5D Yukawa coupling constant $\mathcal{Y}$ has dimension -1 in energy for the simplified scheme with $\hbar=c=1$, 
while $\mathcal{Y}_{4D}$ is dimensionless.}:
\begin{eqnarray}
\frac{\delta^4i\mathcal{L}_{4D}}{\delta \phi_{H_u^0} \bar\phi_{H_u^0} \phi_R^{(0)} \bar\phi_R^{(0)}} \bigg \vert_{total} \to
-i|\mathcal{Y}|^2 \left( f^{++}_0(c_R;\pi R_c) f^{++}_0(c_L;\pi R_c) \right)^2 = -i\ |\mathcal{Y}_{4D}|^2
\label{equiv4DyukF}
\end{eqnarray}
i.e. exactly the squared $\mathcal{Y}_{4D}$ Yukawa coupling of associated 
fermion zero-modes (with a brane-Higgs) \cite{CBGMI}, 
as one expects in a pure 4D SUSY theory (where all KK states decouple).

\subsubsection*{D-term couplings to two Higgs bosons}

Now, we derive the 4D couplings, issued from D-terms, of scalar fields (continuing on the $\phi_R$ example) to the two 
$\phi_{H_u^0}$ bosons in the ${\rm U(1)}$ model of Appendices \ref{SuperContent} and \ref{SuperAction}.

The obtained D-term couplings of $\phi_R$, in terms of 5D fields, are included in Eq.(\ref{ScaFieLagPrim}):  
\begin{eqnarray}
\mathcal{L}_{5D} = -  {\sqrt{G}\over2} \left| [\partial_y - 2(\partial_y\sigma)]\Sigma 
- g \Big( q_R \Delta_{\phi_R} + q_{H_u^0} \phi_{H_u^0} \bar\phi_{H_u^0} \delta(y-\pi R_c) \Big) \right|^2 .
\label{DtermI}
\end{eqnarray}

The 4D effective D-term couplings are deduced from the above 5D Lagrangian (taking real gauge coupling constants): 
\begin{eqnarray}
\frac{\delta^4i\mathcal{L}_{4D}}{\delta \phi_{H_u^0} \bar\phi_{H_u^0} \phi_R^{(n)} \bar\phi_R^{(m)}} &=& 
- i  q_{H_u^0} q_R \vert g\vert ^2
\int_{-\pi R_{c}}^{\pi R_{c}} dy \delta(y-\pi R_c) \bar f^{++}_m(c_R;y) f^{++}_n(c_R;y) \nonumber \\
&=& - i  q_{H_u^0} q_R \vert g\vert ^2 \bar f^{++}_m(c_R;\pi R_c) f^{++}_n(c_R;\pi R_c).
\label{CompNotUsedBis}
\end{eqnarray}

\begin{figure}[!hc]
	\centering
	\vspace*{.5cm}
			\includegraphics[width=0.19\textwidth]{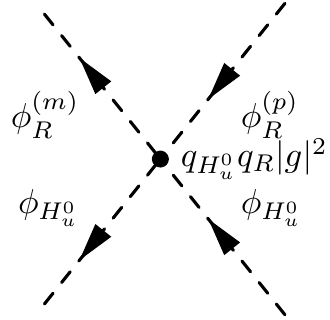}
			\hspace*{1.cm}
			\includegraphics[width=0.19\textwidth]{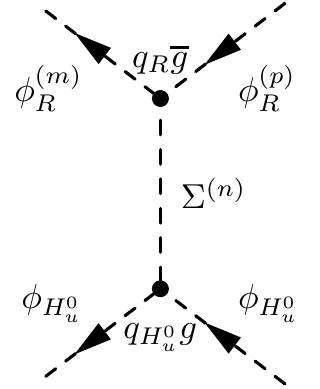}
\caption{\label{DtDrawn} \small{Feynman diagrams of the contributions to the 4D effective scalar gauge coupling
$\frac{\delta^4\mathcal{L}_{4D}}{\delta \phi_{H_u^0} \bar\phi_{H_u^0} \phi_R^{(p)} \bar\phi_R^{(m)}}$. 
The second indirect contribution is induced by the exchange of the KK tower of $(--)$ scalar modes $\Sigma^{(n)}$.
The relevant coupling constants are described in details in text.}}
\end{figure}

As for the Yukawa couplings, there are additional contributions to the 4D effective D-term couplings. Indeed,    
other couplings arising from Eq.(\ref{DtermI}) are
\begin{eqnarray} 
\sqrt{G} g (q_R \bar \phi_R \phi_R + q_{H_u^0} \bar \phi_{H_u^0} \phi_{H_u^0}\delta(y-\pi R_c)) [\partial_y - 2(\partial_y\sigma)]\Sigma
+ h.c. \in \mathcal{L}_{5D}.
\label{5Dterms}
\end{eqnarray}
The KK decomposition of the 5D field $\Sigma$ in Eq.(\ref{SigmaDecomp}) together with the relation (\ref{UsefulRelat}) 
allow to write
\bea
[\partial_y - 2(\partial_y\sigma)] \Sigma(x^{\mu},y) = \sum_{n=1}^{\infty} \Sigma^{(n)}(x^{\mu}) [\partial_y - 2(\partial_y\sigma)] g^{--}_n(y)
= - e^{2 \sigma} \sum_{n=1}^{\infty} M^{(n)} \Sigma^{(n)}(x^{\mu}) g^{++}_n(y)
\label{TwoTim}
\eea
$M^{(n)}$ being the n{\it th} KK gauge boson mass. Inserting this relation in Eq.(\ref{5Dterms}) and developing
$\phi_R$ over its KK tower gives the 4D couplings for the redefined fields:
\begin{eqnarray} 
- q_R g\sum_{m,p=0;n=1}^{\infty} M^{(n)} \bar\phi_R^{(m)} \phi_R^{(p)} 
\Sigma^{(n)} {\cal F}_R^{mnp} \nonumber\\
- q_{H_u^0} g \bar \phi_{H_u^0} \phi_{H_u^0} \sum_{n=1}^{\infty} M^{(n)} \Sigma^{(n)} g^{++}_n(\pi R_c) + h.c. 
\in \mathcal{L}_{4D}
\label{4Dterms}
\end{eqnarray}
with ${\cal F}_R^{mnp} = \int_{-\pi R_c}^{\pi R_c} dy \bar f^{++}_m(c_R;y) g^{++}_n(y) f^{++}_p(c_R;y)$.
These 4D couplings induce new contributions to the D-term couplings via the exchange of the KK modes $\Sigma^{(n)}$, as shown in Fig.(\ref{DtDrawn}).
These contributions read as, 
\begin{eqnarray}
\frac{\delta^4i\mathcal{L}_{4D}}{\delta \phi_{H_u^0} \bar\phi_{H_u^0} \phi_R^{(p)} \bar\phi_R^{(m)}} \bigg \vert_{indirect} =
- q_R q_{H_u^0} \vert g\vert ^2 \sum_{n\geq1} \frac{i}{k^2 - M^{(n)2}} M^{(n)2} {\cal F}_R^{mnp} g^{++}_n(\pi R_c) 
\end{eqnarray}
$k^\mu$ being the $\Sigma^{(n)}$ four-momentum.
Let us rewrite these couplings with the help of the completeness relation (\ref{completeness}) and 5D propagator, 
\begin{eqnarray}
\frac{\delta^4i\mathcal{L}_{4D}}{\delta \phi_{H_u^0} \bar\phi_{H_u^0} \phi_R^{(p)} \bar\phi_R^{(m)}} \bigg \vert_{indirect} 
&=& - i q_R q_{H_u^0} \vert g\vert ^2 \sum_{n\geq0} \left\{ \frac{k^2}{k^2 - M^{(n)2}} -1 \right\} 
{\cal F}_R^{mnp} g^{++}_n(\pi R_c) \nonumber\\
&=& - i q_R q_{H_u^0} \vert g\vert ^2 \int_{-\pi R_c}^{\pi R_c} dy \bar f^{++}_m(c_R;y) f^{++}_p(c_R;y) k^2 G_5^{g^{++}}(k^2;y,\pi R_c) \nonumber\\
&& + i q_R q_{H_u^0} \vert g\vert ^2 f^{++}_m(c_R;\pi R_c) f^{++}_p(c_R;\pi R_c).  
\label{CompleteUsedBis}
\end{eqnarray}

Then, the complete 4D couplings are of course obtained by summing the two contributions (\ref{CompNotUsedBis}) and (\ref{CompleteUsedBis}),
\begin{eqnarray}
\frac{\delta^4i\mathcal{L}_{4D}}{\delta \phi_{H_u^0} \bar\phi_{H_u^0} \phi_R^{(p)} \bar\phi_R^{(m)}} \bigg \vert_{total} = 
- i q_R q_{H_u^0} \vert g\vert ^2 \int_{-\pi R_c}^{\pi R_c} dy \bar f^{++}_m(c_R;y) f^{++}_p(c_R;y) k^2 G_5^{g^{++}}(k^2;y,\pi R_c)
\label{HDtCoupl}
\end{eqnarray}
having taken into account the canceling terms. Once more, starting from couplings in a 5D SUSY theory, we have derived consistent 4D effective couplings; 
the Lagrangian (\ref{HDtCoupl}) corresponds to a real 5D SUSY theory but written from the 4D point of view.

We finish this part by making the same check as in previous subsection. In the low--energy limit, $k^2 \ll M^{(n)2}$ ($n\geq1$), 
only the zero-mode in the 5D propagator survives at zero{\it th} order, so that the superpartner couplings (\ref{HDtCoupl}) simplify to 
\begin{eqnarray}
\frac{\delta^4i\mathcal{L}_{4D}}{\delta \phi_{H_u^0} \bar\phi_{H_u^0} \phi_R^{(0)} \bar\phi_R^{(0)}} \bigg \vert_{total} \to
- i \ q_R q_{H_u^0} \frac{\vert g\vert ^2}{2 \pi R_c}          
= - i \ q_R q_{H_u^0} \vert g_{4D}\vert ^2  
\end{eqnarray}
since the gauge boson zero-mode profile encoded in $g^{++}_0(y)=1/\sqrt{2 \pi R_c}$ is flat along the fifth dimension. 
Thus the couplings in this limiting case correspond rigorously to the dimensionless $- iq_R q_{H_u^0} \vert g_{4D}\vert ^2$ gauge coupling product 
of associated fermions, as expected in a pure 4D SUSY theory.

\subsubsection*{Comments on the obtained 4D couplings to Higgs bosons}

It is interesting to note that 
in order to obtain the 4D effective couplings to two Higgs bosons [Yukawa couplings and D-terms] which are consistent 
(i.e. without the $\delta(0)$ divergences and recovering the 4D SUSY  
couplings in the limit $k^2 \ll m^{(n)2}, M^{(n)2}$), we had to use the completeness relation which relies 
on an infinite sum over the KK levels. The reason is that these couplings belong
to the 4D effective Lagrangian of a fundamental 5D SUSY theory -- the infinite KK tower reflects this 5D aspect.    

Besides, no truncating cut-off must be applied when implementing the completeness relation 
(for the sum not involving KK masses in Eq.(\ref{CompleteUsed}) and
Eq.(\ref{CompleteUsedBis})), otherwise the consistent 4D effective couplings cannot 
be obtained while there is no reason why a non-renormalizable theory -- as we are considering 
here -- could not possess a 4D description. The reason for not applying a cut-off 
(due to the non-renormalizable aspect of the 5D SUSY theory) is that in 
Eq.(\ref{CompNotUsed})-(\ref{CompleteUsed}) and
Eq.(\ref{CompNotUsedBis})-(\ref{CompleteUsedBis}) one is integrating the fifth dimension and summing {\it all} 
the exchanged heavy KK states to get an effective 4D vision of the {\it 5D theory}. Only once the 4D
effective couplings are obtained so that the 4D description is completed, the cut-off 
must be put, e.g. in Eq.(\ref{HYukCoupl}) [on $G_5^{f^{++}(c_L)}(k^2;\pi R_c,\pi R_c)$], to take into account the 
non-renormalizable aspect of the 5D SUSY theory.

\subsubsection*{Final single couplings to the Higgs boson $H$}

In this part, we deduce from the last three subsections the total 4D effective scalar couplings to the single Higgs boson $H$ in the more realistic framework where
the gauge symmetry group is as in the SM (EWSB has occurred) and the superfield content is extended to the pMSSM one. In particular, this framework is based on the 
coexistence of two complex Higgs ${\rm SU(2)_L}$ doublets $H_u$, $H_d$ of superfields
with opposite hypercharges ($Y_{H_u}=-Y_{H_d}=+1$) which guarantees the absence 
of chiral anomalies originating from triangular fermionic loops. The five scalar degrees of freedom, not absorbed in the longitudinal polarizations of the massive 
gauge bosons, constitute the five physical Higgs bosons: two CP-even neutral Higgs fields $h$ (the lightest one) and $H$, one pseudo-scalar $A$ boson and one pair 
of charged scalar particles $H^{\pm}$. We will give explicitly the $H$ couplings in the RS SUSY framework, and similar couplings hold for the other Higgs fields. All
Higgs bosons are assumed to be stuck on the TeV-brane. 
\\ 
We will focus on the top quark ($t$) superpartner couplings to $H$ as an illustrative example, since the couplings of other scalar superpartners are analog.
The n{\it th} KK level modes of the stop fields are denoted $\tilde{t}^{(n)}_L$ and $\tilde{t}^{(n)}_R$ respectively for the superpartners of the Left-handed ($t_L$)
and Right-handed ($t_R$) top quarks. $\tilde{t}^{(n)}_L$ and $\tilde{t}^{(n)}_R$ are similar to the 4D $(++)$ scalar fields $\phi^{(n)}_L$ and $\phi^{(n)}_R$ 
defined in Eq.(\ref{eq:phiL}), Eq.(\ref{eq:phiR}) and Eq.(\ref{PhiDecomp}) for the ${\rm U(1)}$ model, except of course with respect to the gauge quantum numbers. 
In the interaction basis $\{ \tilde{t}^{(0)}_L , \tilde{t}^{(0)}_R , \tilde{t}^{(1)}_L, \tilde{t}^{(1)}_R \}$ (generalization to higher KK states is straightforward), 
the stop-stop couplings to a single $H$ boson appearing after EWSB are encoded in the matrix [in our notations $H$ denotes the scalar field]:
{\small \begin{eqnarray}
i \ {\cal C}_{H\tilde{t}\tilde{t}}  \equiv   \ \left|{\cal Y}\right|^2 \sin \alpha \ v_u \ \times
\nonumber
\end{eqnarray}
\begin{eqnarray}
\left(\begin{array}{cccc}
(f_L^{0})^2 \, k^2 \, G_5^{f^{++}(c_{\tilde t_{R}})}(k^2;\pi R_c,\pi R_c) & 0 & f^0_L f^1_L (f^0_R)^2 & 0 \\
0 & (f_R^{0})^2 \, k^2 \, G_5^{f^{++}(c_{\tilde t_{L}})}(k^2;\pi R_c,\pi R_c)  & 0 &  f^0_R f^1_R(f^0_L)^2 \\
f^0_L f^1_L(f^0_R)^2  & 0 &  (f^1_L f^0_R)^2 & 0 \\
0 &  f^0_R f^1_R(f^0_L)^2  & 0 &  (f^0_L f^1_R)^2 
\end{array}\right) 
\nonumber
\end{eqnarray}
\begin{eqnarray}
- \ \frac{{\cal Y} \cos \alpha \ \mu_{eff}}{\sqrt{2}} 
\left(\begin{array}{cccc}
0 & f^0_L f^0_R & 0 & f^0_L f^1_R \\
f^0_L f^0_R & 0 & f^1_L f^0_R & 0 \\
0 & f^1_L f^0_R & 0 & f^1_L f^1_R \\
f^0_L f^1_R & 0 & f^1_L f^1_R & 0
\end{array}\right) \ + \ g_Z^2 \ \pi R_c \ \cos(\alpha+\beta) \ v \ \times
\nonumber
\end{eqnarray}
\begin{eqnarray}
\left(\begin{array}{cccc}
Q_Z^{t_L} \int dy f_L^{0}(y)^2 \, k^2 \, G_5^{g^{++}}(k^2;y,\pi R_c) & 0 & 0 & 0 \\
0 & -Q_Z^{t_R} \int dy f_R^{0}(y)^2 \, k^2 \, G_5^{g^{++}}(k^2;y,\pi R_c) & 0 & 0 \\
0 & 0 & Q_Z^{t_L}/2\pi R_c & 0 \\
0 & 0 & 0 & -Q_Z^{t_R}/2\pi R_c 
\end{array}\right) 
\nonumber
\end{eqnarray}
\begin{eqnarray}
+ \ \frac{A e^{-k\pi R_c} \sin \alpha}{\sqrt{2}}
\left(\begin{array}{cccc}
0 & f^0_L f^0_R & 0 & f^0_L f^1_R \\
f^0_L f^0_R & 0 & f^1_L f^0_R & 0 \\
0 & f^1_L f^0_R & 0 & f^1_L f^1_R \\
f^0_L f^1_R & 0 & f^1_L f^1_R & 0
\end{array}\right)  ,  
\nonumber \\ 
\label{HCouplMat}
\end{eqnarray}}
where we have used the compact notation e.g. $f_{L/R}^{n}=f^{++}_n(c_{\tilde t_{L/R}};\pi R_c)$. Besides, based on the 5D superpotential of Eq.(\ref{muAction}) 
and on the deduced field Lagrangian in Eq.(\ref{ScaFieLag}), the 4D effective $\mu$ parameter appearing above reads as (after field redefinition and inclusion of the metric 
warp factor):
\begin{eqnarray}
\mu_{eff} =  \mu e^{-k \pi R_c} \sim k e^{-k \pi R_c} \sim  \mbox{TeV}. 
\label{mueff}
\end{eqnarray}
No $\delta(0)$ factors appear after integration over $y$. 
The soft trilinear scalar coupling constant in Eq.(\ref{HCouplMat}) is taken at $A \sim 1$ to avoid the introduction of a new scale in the bulk.
This 4D effective coupling matrix is deduced from Eq.(\ref{HYukCoupl}) for the Yukawa couplings and from Eq.(\ref{HDtCoupl}) for the D-term couplings
(where the effects of the $(--)$ KK towers of $\phi^c_{L/R}$ and $\Sigma$ type fields, as defined in Eq.(\ref{eq:phicL})-(\ref{eq:phicR}) 
and Eq.(\ref{SuperVectorialbis}), have been taken into account).
This coupling matrix, that will be taken at the energy $k^2=m_H^2$ (for the $Z$ coupling constant $g_Z=g_{4D}/\cos\theta_W$ 
and top charges to the $Z$ gauge boson: $Q_Z^{t_{L/R}}=I_{3L}^{t_{L/R}}-Q^{t_{L/R}}_{e.m.}\sin^2\theta_W$), can be easily generalized e.g. to sbottoms.

Let us comment more precisely on the 5D effects. 
The Higgs mixing between $\phi_{H^0_d}$ and $\phi_{H^0_u}$ into the mass eigenstates $h$ and $H$ is parametrized by the mixing angle
noted here, as usually, $\alpha$ \cite{DjouadiReviewII}. $\sin\alpha$, which enters the above coupling matrix, 
receives some corrections in the present 5D framework, as the 
$\vert \phi_{H_u^0} \vert^2 \vert \phi_{H_d^0} \vert^2$ and $\vert \phi_{H_{u,d}^0} \vert^4$ couplings do so (see next subsection).
\\ Moreover, the stop-stop-Higgs couplings are affected by the mixing between the stops and their KK $(++)$ excitations.
The $H\tilde t_i\tilde t_j$ couplings, where $\tilde t_i$ [$i=1,\dots,4$] are the stop mass eigenstates, are obtained after transformation from the
basis $\{ \tilde{t}^{(0)}_L , \tilde{t}^{(0)}_R , \tilde{t}^{(1)}_L, \tilde{t}^{(1)}_R \}$ to the stop mass basis (rotation matrices being obtained from 
diagonalizing the stop mass matrix given later).
\\ The third effect is the exchange of KK $(--)$ modes, encoded in the 5D propagators $G_5$ appearing in the matrix (\ref{HCouplMat}).
Such KK $(--)$ contributions to couplings between $H$ and KK stops would represent higher-order corrections to the $H\bar{\tilde{t}}^{(0)}_{L/R}\tilde{t}^{(0)}_{L/R}$ 
couplings and are thus not written in matrix (\ref{HCouplMat}).
\\ All these heavy KK mixing and KK exchange effects will not be computed numerically as they are sub-leading compared to other direct 5D effects in 
the structure of zero-mode $H\bar{\tilde{f}}^{(0)}_{L/R}\tilde{f}^{(0)}_{L/R}$ couplings ($\tilde{f} \equiv$ sfermion), that will be studied in details in Section \ref{Higgspart}.

\subsubsection*{Self couplings of the Higgs boson}

Finally, we derive the non-trivial 4D quartic couplings e.g. of the Higgs boson $\phi_{H_u^0}$,  
still within the context of the toy model defined in Appendix \ref{lagrangianpart}. A totally similar
study could be made for the $\phi_{H_u^0} \bar\phi_{H_u^0} \phi_{H_d^0} \bar\phi_{H_d^0}$ couplings
(without combinatorial factor neither the additional $\Sigma^{(n)}$ exchange in the t-channel discussed below).

As above we start from the 5D couplings included in Eq.(\ref{ScaFieLagPrim}):  
\begin{eqnarray}
\mathcal{L}_{5D} = - {\sqrt{G}\over2} \left | [\partial_y - 2(\partial_y\sigma)]\Sigma 
- g \Big( q_{H_u^0} \phi_{H_u^0} \bar\phi_{H_u^0} \delta(y-\pi R_c) \Big) \right|^2 .
\label{DtermII}
\end{eqnarray}

The 4D quartic couplings directly deduced from this Lagrangian are: 
\begin{eqnarray}
\frac{\delta^4i\mathcal{L}_{4D}}{\delta \phi_{H_u^0}^2 \bar\phi_{H_u^0}^2} &=& 
- {i\over2}  q_{H_u^0}^2 \vert g\vert^2 \delta(0) \times 4.
\label{CompNotUsedTer}
\end{eqnarray}
We have included the combinatorial factor $4$ as we consider a process rather than a coupling here since  
Eq.(\ref{CompNotUsedTer}) will be combined with other contributions.

\begin{figure}[!hc]
	\centering
	\vspace*{.5cm}
			\includegraphics[width=0.19\textwidth]{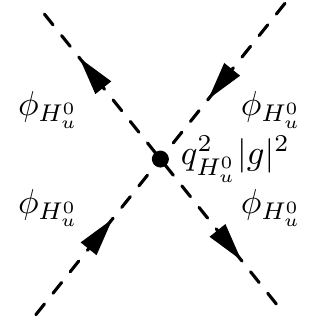}
			\hspace*{1.cm}
			\includegraphics[width=0.19\textwidth]{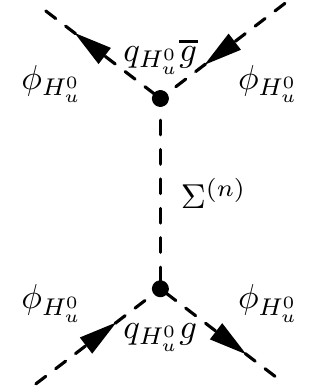}	
			\hspace*{1.cm}	
			\includegraphics[width=0.25\textwidth]{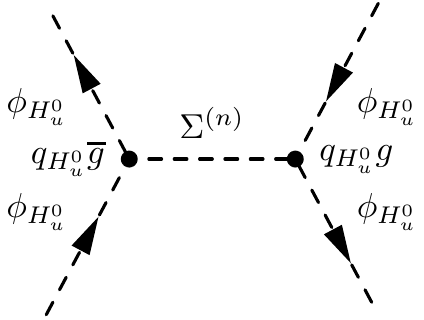}	
\caption{\label{SelfDrawn} \small{Feynman diagrams of the contributions to the 4D effective self scalar gauge coupling
$\frac{\delta^4\mathcal{L}_{4D}}{\delta \phi_{H_u^0}^2 \bar\phi_{H_u^0}^2}$. 
The second and third indirect contributions are induced by the exchanges of the KK tower of $(--)$ scalar modes $\Sigma^{(n)}$
in the s-channel and t-channel, respectively (the choice of calling `s-channel' or `t-channel' a given diagram depends on which final state is considered).
The relevant coupling constants are described in details in text.}}
\end{figure}

Indeed, here again, there exist additional contributions to the 4D couplings as Eq.(\ref{DtermII}) also induces the $\phi_{H_u^0}$ couplings of Eq.(\ref{5Dterms})
and in turn of Eq.(\ref{4Dterms}). These later 4D couplings induce the two following new contributions to the quartic terms, via two possible exchanges of the KK 
$\Sigma^{(n)}$ [see Fig.(\ref{SelfDrawn})],
\begin{eqnarray}
\frac{\delta^4i\mathcal{L}_{4D}}{\delta \phi_{H_u^0}^2 \bar\phi_{H_u^0}^2} \bigg \vert_{indirect} =
&-& q_{H_u^0}^2 \vert g\vert^2 \sum_{n\geq1} \frac{i}{k^2 - M^{(n)2}} M^{(n)2} \bigg ( g^{++}_n(\pi R_c) \bigg )^2 \nonumber\\ 
&-& q_{H_u^0}^2 \vert g\vert^2 \sum_{m\geq1} \frac{i}{q^2 - M^{(m)2}} M^{(m)2} \bigg ( g^{++}_m(\pi R_c) \bigg )^2 
\end{eqnarray}
$k^\mu$ being the $\Sigma^{(n)}$ four-momentum in the s-channel while $q^\mu$ represents the $\Sigma^{(n)}$ momentum in the t-channel. 
Rewriting these couplings with the completeness relation in mind gives,
\begin{eqnarray}
& & \frac{\delta^4i\mathcal{L}_{4D}}{\delta \phi_{H_u^0}^2 \bar\phi_{H_u^0}^2} \bigg \vert_{indirect}   \nonumber\\
&=& -  i q_{H_u^0}^2 \vert g\vert^2 \sum_{n\geq0} \left\{ \frac{k^2}{k^2 - M^{(n)2}} -1 \right\} \bigg ( g^{++}_n(\pi R_c) \bigg )^2 
-  i q_{H_u^0}^2 \vert g\vert^2 \sum_{m\geq0} \left\{ \frac{q^2}{q^2 - M^{(m)2}} -1 \right\} \bigg ( g^{++}_m(\pi R_c) \bigg )^2 
\nonumber\\
&=&  i q_{H_u^0}^2 \vert g\vert^2 [ 2 \delta(0) - k^2 G_5^{g^{++}}(k^2;\pi R_c,\pi R_c) - q^2 G_5^{g^{++}}(q^2;\pi R_c,\pi R_c) ].  
\label{CompleteUsedTer}
\end{eqnarray}

Adding the contributions (\ref{CompNotUsedTer}) and (\ref{CompleteUsedTer}) gives the complete 4D quartic terms:
\begin{eqnarray}
\frac{\delta^4i\mathcal{L}_{4D}}{\delta \phi_{H_u^0}^2 \bar\phi_{H_u^0}^2} \bigg \vert_{total} = 
-  i q_{H_u^0}^2 \vert g\vert^2 k^2 G_5^{g^{++}}(k^2;\pi R_c,\pi R_c)
-  i q_{H_u^0}^2 \vert g\vert^2 q^2 G_5^{g^{++}}(q^2;\pi R_c,\pi R_c).  
\label{HQuarCoupl}
\end{eqnarray}
Note the cancellation of $\delta(0)$ terms which leads to consistent 4D effective couplings.

In the check of the low--energy limit, $k^2,q^2 \ll M^{(n)2}$ [$n \neq 0$], the Higgs coupling (\ref{HQuarCoupl}) is reduced to the following form  
(the combinatorial factor $4$ is taken off to get the pure quartic Lagrangian coupling),
\begin{eqnarray}
\frac{\delta^4i\mathcal{L}_{4D}}{\delta \phi_{H_u^0}^2 \bar\phi_{H_u^0}^2} \bigg \vert_{total} \to
- \frac{i}{2} q_{H_u^0}^2 \frac{\vert g\vert^2}{2 \pi R_c} = - \frac{i}{2} \ q_{H_u^0}^2 \vert g_{4D}\vert^2
\end{eqnarray}
recalling that $g^{++}_0(y)=1/\sqrt{2 \pi R_c}$. 
The quartic Higgs coupling in this limit corresponds thus well to the exact squared gauge coupling expected in a pure 4D SUSY theory.

\begin{figure}[!hc]
	\centering
	\vspace*{.5cm}
			\includegraphics[width=0.19\textwidth]{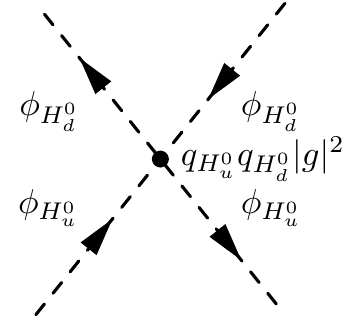}
			\hspace*{1.cm}
			\includegraphics[width=0.19\textwidth]{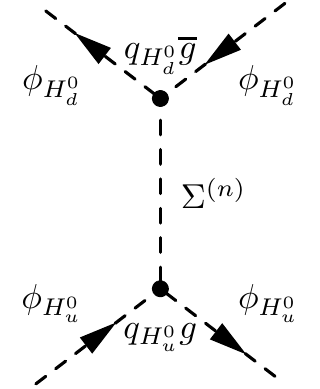}
\caption{\label{SelfDrawnBis} \small{Feynman diagrams of the contributions to the 4D effective scalar gauge coupling
$\frac{\delta^4i\mathcal{L}_{4D}}{\delta \phi_{H_u^0} \bar\phi_{H_u^0} \phi_{H_d^0} \bar\phi_{H_d^0}}$. 
The second indirect contribution is induced by the exchange of the KK tower of $(--)$ scalar modes $\Sigma^{(n)}$.}} 
\end{figure}

For completeness (and it will prove to be useful for the following), we give the result for the $\phi_{H_u^0} \bar\phi_{H_u^0} \phi_{H_d^0} \bar\phi_{H_d^0}$
coupling, obtained through the same method,
\begin{eqnarray}
\frac{\delta^4i\mathcal{L}_{4D}}{\delta \phi_{H_u^0} \bar\phi_{H_u^0} \phi_{H_d^0} \bar\phi_{H_d^0}} \bigg \vert_{total} = 
-  i q_{H_u^0}q_{H_d^0} \vert g\vert^2 k^2 G_5^{g^{++}}(k^2;\pi R_c,\pi R_c).  
\label{HBiCoupl}
\end{eqnarray}
This coupling is obtained from the two contributions drawn in Fig.(\ref{SelfDrawnBis}).

\subsection{Scalar mass matrix}
\label{matrixpart}

We now calculate the 4D effective mass matrix for sfermions induced by brane-Higgs bosons within RS SUSY. 
One must develop different methods than in the above approach of 4D Higgs couplings.

\subsubsection*{Effect of the $\phi^c$ KK tower on $\phi^{(0)}$ masses through mixing}
\label{DDmixing}

In this subsection, we derive the 4D masses coming from Yukawa interactions for the $\phi_R$ scalar field  
[the same analysis can be done for $\phi_L$] in the toy model of Appendices \ref{SuperContent} and \ref{SuperAction},
assuming that the $\phi_{H_u^0}$ boson acquires a VEV $v_u = \sqrt{2} \langle \phi_{H_u^0} \rangle \sim 10^2$ GeV.

In addition to the $\bar \phi_R^{(0)} \phi_R^{(0)}$ mass, proportional to $({\cal Y} v_u)^2$, 
which is obtained directly from the Lagrangian (\ref{ScaFieLagPrim}), the exchanges of the KK modes $\phi_L^{c(n)}$ also contribute to this mass
as illustrates Fig.(\ref{YukDrawn}) in the case where $\phi_{H_u^0}$ acquires a VEV.
In order to compute the whole zero-mode mass, one needs to estimate the mixing between $\phi_R^{(0)}$ and $\phi_L^{c(n)}$, a mixing induced by the VEV
at the origin of the additional contributions as shows Fig.(\ref{YukDrawn}) -- analogously to the mixing with the heavy Majorana neutrino in the type I See-saw model
\cite{Hambye}. For that purpose, we write down the complete 4D mass matrix in the infinite basis 
$\vec\phi = (\phi^{(0)}_R,\phi^{c(1)}_L,\phi^{c(2)}_L,\dots)$ and search for the smallest eigenvalue.
Indeed, the lightest eigenstate is typically mainly composed by the $\phi_R^{(0)}$ state, given usually the realistic KK $\phi_L^{c(n)}$ masses which are around a 
few TeV at least and hence much larger than the VEV-induced $\phi_R^{(0)}$ mass. 
From Lagrangian (\ref{ScaFieLagPrim}), this 4D mass matrix reads as $-\bar{\vec\phi} {\cal M}^2_{RR} \vec\phi^t \in {\cal L}_{4D}$ with 
\footnote{Note that taking purely real wave functions and Yukawa coupling constants 
simplifies the approach by avoiding a {\it bi}-diagonalization, as it renders the mass matrix fully real and symmetric.} 
{\small$$
{\cal M}^2_{RR} = 
$$
\bea        
\left( 
\begin{array}{cccc}
{\cal Y}^2 \hat v_u^2 \int dy \delta^2 f^0_R(y)^2 &  {\cal Y} \hat v_u m^{(1)}_L \int dy' \delta f^0_R(y') f^1_L(y') 
       & {\cal Y} \hat v_u m^{(2)}_L \int dy' \delta f^0_R(y') f^2_L(y') & \dots \\
{\cal Y} \hat v_u m^{(1)}_L \int dy' \delta f^0_R(y') f^1_L(y') & m^{(1)2}_L & 0 & \dots \\
{\cal Y} \hat v_u m^{(2)}_L \int dy' \delta f^0_R(y') f^2_L(y') & 0 & m^{(2)2}_L & \dots \\
\vdots & \vdots & \vdots & \ddots 
\end{array}
\right)
\label{MassMatRR} \\ \nonumber
\eea}
where we have used again the compact notation $f_{L/R}^{n}(y)=f^{++}_n(c_{L/R};y)$, $f_{L/R}^{cn}(y)=f^{--}_n(c_{L/R};y)$ and $\hat v_u = v_u / \sqrt{2}$, 
$\delta = \delta(y-\pi R_c)$. For writing the off-diagonal elements, we have made use of relation (\ref{UsefulRelatII}) which can be written in a compact form as 
$D'_5 f^{c \ n}_L(y) = - m^{(n)}_L e^{\sigma} f^n_L(y)$ and then have redefined the scalar wave functions with $e^{\sigma}$ factors as usual in RS.
\\
The generalized characteristic equation, of which the (squared mass) eigenvalues $m^2$ are solutions, reads as
\bea
\bigg ( 
{\cal Y}^2 \hat v_u^2 \int dy \delta^2 f^0_R(y)^2 - m^2 - \sum_{n=1}^\infty \frac{[{\cal Y} \hat v_u m^{(n)}_L \int dy' \delta f^0_R(y') f^{n}_L(y')]^2}{m_L^{(n)2}-m^2}
\bigg ) \ \Pi_{n=1}^\infty (m_L^{(n)2}-m^2)= 0.
\label{CharEq}
\eea
As $m^2 = m_L^{(n)2}$ leads to divergences in the above equality, those are not solutions. Hence Eq.(\ref{CharEq}) simplifies to
\bea
{\cal Y}^2 \hat v_u^2 \int dy \delta^2 f^0_R(y)^2 - m^2 + \sum_{n=1}^\infty \frac{[{\cal Y} \hat v_u m^{(n)}_L \int dy' \delta f^0_R(y') f^{n}_L(y')]^2}{m^2-m_L^{(n)2}} = 0.
\label{CharEqBis}
\eea
Using here also the equality $m_L^{(n)2}/(m^2-m_L^{(n)2})=-1+m^2/(m^2-m_L^{(n)2})$ together with the completeness 
relation applied as $\sum_{n=0}^\infty f^{n}_L(y) f^{n}_L(y') = \delta(y-y')$, one can rewrite 
\bea
{\cal Y}^2 \hat v_u^2 \delta(0) f^{0}_R(\pi R_c)^2 - m^2 - {\cal Y}^2 \hat v_u^2 f^{0}_R(\pi R_c)^2 \delta(0)  
+  {\cal Y}^2 \hat v_u^2 m^2 f^{0}_R(\pi R_c)^2 G_5^{f^{++}(c_L)}(m^2;\pi R_c,\pi R_c) = 0 \nonumber 
\eea
\bea 
m^2 \Big (1 -  {\cal Y}^2 \hat v_u^2 f^{0}_R(\pi R_c)^2 G_5^{f^{++}(c_L)}(m^2;\pi R_c,\pi R_c) \Big )= 0.
\label{DelCan}
\eea
Since $m^2 = 0$ is not a physically acceptable solution, 
\bea
m^2 = \frac{f^{0}_L(\pi R_c)^2}{\frac{1}{{\cal Y}^2 \hat v_u^2 f^{0}_R(\pi R_c)^2} - \sum_{n=1}^\infty \frac{f^{n}_L(\pi R_c)^2}{m^2-m_L^{(n)2}}} .
\label{lastequation}
\eea
The check at this level is that in the decoupling limit $m_L^{(n)} \to \infty$ for any $n\geq 1$, we recover the equality between
the Yukawa mass for (zero-mode) fermions and their scalar superpartner, as expected in a 4D SUSY theory:
\bea
m^2 \to {\cal Y}^2 \hat v_u^2 f^{0}_R(\pi R_c)^2 f^{0}_L(\pi R_c)^2 = {\cal Y}_{4D}^2 \hat v_u^2 = m_{fermion}^2.
\eea
The divergence cancellation in Eq.(\ref{DelCan}) and this 4D SUSY limiting case confirm that our solution for the smallest 
eigenvalue will be consistent. 
This consistency is due to the infinite aspect of the $\phi^{c(n)}_L$ basis considered here,
in analogy to the calculation of Yukawa couplings in Section \ref{couplingpart}.
Now in the case (of realistic scenarios) where the eigenvalue of the lightest eigenstate $m^2_{lightest}$ 
is much smaller than $m_L^{(1)2}$, one obtains at leading order from Eq.(\ref{lastequation}):
\bea
m^2_{lightest} \simeq \frac{f^0_L(\pi R_c)^2}{\frac{1}{{\cal Y}^2 \hat v_u^2 f^{0}_R(\pi R_c)^2} + \sum_{n=1}^\infty \frac{f^{n}_L(\pi R_c)^2}{m_L^{(n)2}}} .
\label{LightSol}
\eea
An eigenvalue $m^2$ much smaller than $m_L^{(1)2}$ can only be the smallest one since all the others are larger than $m_L^{(1)2}$. As a matter of fact, at leading order
in $f^{0}_L(\pi R_c)^2/f^{n}_L(\pi R_c)^2$ for $n\geq 1$, Eq.(\ref{lastequation}) can be rewritten as  
\bea
m^2 \Big ( - \frac{1}{{\cal Y}^2 \hat v_u^2 f^{0}_R(\pi R_c)^2} + \sum_{n=1}^\infty \frac{f^{n}_L(\pi R_c)^2}{m^2-m_L^{(n)2}} \Big ) \simeq 0.
\eea
Here the solution $m^2 \simeq 0$, at leading order in $f^{0}_L(\pi R_c)^2/f^{n}_L(\pi R_c)^2$, corresponds to the lightest solution (\ref{LightSol}). Concentrating on the
other solutions, those satisfy
\bea
1 = {\cal Y}^2 \hat v_u^2 f^{0}_R(\pi R_c)^2 \sum_{n=1}^\infty \frac{f^{n}_L(\pi R_c)^2}{m^2-m_L^{(n)2}} .
\label{HeavySol}
\eea  
For this sum to be equal to unity, at least one of the terms must be positive, that is to say that $m^2-m_L^{(n)2} > 0$ for at least one value of $n\geq 1$. 
Even if it occurs for $n=1$, one would obtain that $m^2 > m_L^{(1)2}$ which means that all the solutions $m^2$ of Eq.(\ref{HeavySol}) have to be larger than $m_L^{(1)2}$
at least. The above method was inspired from an higher-dimensional analysis performed in \cite{MultiBrane}.

\subsubsection*{Effect of the $\Sigma$ KK tower on $\phi^{(0)}$ masses through integration out}
\label{DDintegration}

Let us calculate the 4D zero-mode masses due to SUSY D-terms for $\phi_L$ (to vary our examples), still in the context of Appendix \ref{lagrangianpart}.
There exist contributions from the exchange of the KK modes $\Sigma^{(n)}$ if $\phi_{H_u^0}$ acquires a VEV, as is illustrated in Fig.(\ref{DtDrawn}) replacing $\phi_R$
by $\phi_L$. Such contributions to the $\bar \phi^{(0)}_L \phi^{(0)}_L$ mass are not arising from a mixing between $\phi^{(0)}_L$ and some KK excitations (as happens
in previous subsection for Yukawa terms). These low-energy contributions must instead be calculated by integrating out the $\Sigma^{(n)}$ fields, 
exactly like the heavy triplet scalar is integrated out within the type II See-saw scenario \cite{Hambye}.
Hence we derive here the complete $\vert \phi_{H_u^0} \vert^2 \bar \phi_L^{(i)} \phi_L^{(j)}$ couplings induced by integrating out all the $\Sigma^{(n)}$ modes,
which contribute to the $\bar \phi^{(0)}_L \phi^{(0)}_L$ mass after $\phi_{H_u^0}$ gets its VEV.

As we are going to concentrate on the explicit derivation of the various contributions to the $\vert \phi_{H_u^0} \vert^2 \bar \phi_L^{(i)} \phi_L^{(j)}$ couplings,
we will consider a Lagrangian part depending only on the fields $\phi_{H_u^0}$, $\phi_L$ and $\Sigma$. This Lagrangian is obtained from
Eq.(\ref{ScaFieLag}) and Eq.(\ref{AppZpar}) in terms of the 5D fields: 
\begin{eqnarray}
- e^{4\sigma} \mathcal{L}_{5D}^{D-terms} &=& - \frac{e^{2\sigma}}{2} \partial_\mu \Sigma \partial^\mu \Sigma  
+ {1\over2} \left\vert D^k_5 \Sigma \right\vert^2 
- g \Big( q_L \overline{\phi}_L \phi_L   
+ q_{H_u^0} \delta \overline{\phi}_{H_u^0} \phi_{H_u^0}  \Big) D^k_5 \Sigma  
+ g^2 q_L q_{H_u^0} |\phi_L\phi_{H_u^0}|^2 \delta.
\nonumber
\end{eqnarray}
Using once more Eq.(\ref{TwoTim}), replacing the 5D fields by their KK decomposition,
redefining them with warp factors (to recover the canonical 4D kinetic terms) 
and applying the orthonormalization condition (\ref{orthonormalization}), one gets the 4D Lagrangian 
\begin{eqnarray}
\mathcal{L}_{4D}^{D-terms} = {1\over2} \sum_{n=1}^{\infty} \left\vert \partial_\mu \Sigma^{(n)} \right\vert^2 
-{1\over2} \sum_{n=1}^{\infty} M^{(n)2} \Sigma^{(n)2}  
- q_L g \sum_{i,j=0}^{\infty} \sum_{n=1}^{\infty} 
{\cal T}_L^{ijn} M^{(n)} \Sigma^{(n)} \overline{\phi}_L^{(i)} \phi_L^{(j)} \nonumber \\
- q_{H_u^0} g \sum_{n=1}^{\infty} g^{++}_n(\pi R_c) M^{(n)} \Sigma^{(n)} \overline{\phi}_{H_u^0} \phi_{H_u^0}    
 - q_L q_{H_u^0} g^2 \sum_{i,j=0}^{\infty} \overline{\phi}_L^{(i)} \phi_L^{(j)} 
f^{++}_i(c_L;\pi R_c)f^{++}_j(c_L;\pi R_c) \overline{\phi}_{H_u^0} \phi_{H_u^0} \nonumber \\
\label{MainScaFieLag}
\end{eqnarray}
with ${\cal T}_L^{ijn}=\int_{-\pi R_c}^{\pi R_c} dy f^{++}_i(c_L;y)f^{++}_j(c_L;y) g^{++}_n(y)$.

The equation of motion for each field $\Sigma^{(n)}$ ($n\geq 1$) is then given by,
\begin{eqnarray}
-\Big( \partial_\mu \partial^\mu + M^{(n)2} \Big) \Sigma^{(n)}
= q_L g \sum_{i,j=0}^{\infty} {\cal T}_L^{ijn} M^{(n)} \overline{\phi}_L^{(i)} \phi_L^{(j)}
+ q_{H_u^0} g g^{++}_n(\pi R_c) M^{(n)} \overline{\phi}_{H_u^0} \phi_{H_u^0}, 
\end{eqnarray}
which can be rewritten at second order in $\partial_\mu \partial^\mu / M^{(n)2}$,
\begin{eqnarray}
\Sigma^{(n)} \simeq - \frac{g}{M^{(n)}}
\Big[ 1-\frac{\partial_\mu \partial^\mu}{M^{(n)2}}+\Big\{\frac{\partial_\mu \partial^\mu}{M^{(n)2}}\Big\}^2 \Big] 
\Big(q_L \sum_{i,j=0}^{\infty} {\cal T}_L^{ijn} \overline{\phi}_L^{(i)} \phi_L^{(j)}
+ q_{H_u^0} g^{++}_n(\pi R_c) \overline{\phi}_{H_u^0} \phi_{H_u^0} \Big).
\label{EOMSigma}
\end{eqnarray}

Now we integrate out all the $\Sigma^{(n)}$ fields
by inserting Eq.(\ref{EOMSigma}) into Eq.(\ref{MainScaFieLag}) which gives rise to the following terms in the Lagrangian, 
restricting ourselves to the first order in $\partial_\mu \partial^\mu / M^{(n)2}$,
\begin{eqnarray}
 &   &
q_{H_u^0} q_L g^2 \sum_{n=1}^{\infty} g^{++}_n(\pi R_c)  
\overline{\phi}_{H_u^0} \phi_{H_u^0} \sum_{i,j=0}^{\infty} {\cal T}_L^{ijn} \overline{\phi}_L^{(i)} \phi_L^{(j)}\cr
&- &  q_L q_{H_u^0} g^2 \sum_{i,j=0}^{\infty} \overline{\phi}_L^{(i)} \phi_L^{(j)} 
f^{++}_i(c_L;\pi R_c)f^{++}_j(c_L;\pi R_c) \overline{\phi}_{H_u^0} \phi_{H_u^0} \cr
&+&  q_{H_u^0} q_L g^2 \sum_{i,j=0;n=1}^{\infty} g^{++}_n(\pi R_c) {\cal T}_L^{ijn}
\frac{\partial_\mu (\overline{\phi}_{H_u^0} \phi_{H_u^0})
\partial^\mu (\overline{\phi}_L^{(i)} \phi_L^{(j)})}{M^{(n)2}}  \ \ \in \mathcal{L}_{4D}^{D-terms}.
\label{IntLag}
\end{eqnarray}
The first term in this Lagrangian part can be rewritten, according to the orthonormalization condition 
(\ref{orthonormalization}) and completeness relation (\ref{completeness}), as
\bea
&&
q_{H_u^0} q_L g^2 \sum_{n=1}^{\infty} g^{++}_n(\pi R_c)  
\overline{\phi}_{H_u^0} \phi_{H_u^0} \sum_{i,j=0}^{\infty} {\cal T}_L^{ijn} \overline{\phi}_L^{(i)} \phi_L^{(j)}\cr
&&
=
q_{H_u^0} q_L g^2 \int_{-\pi R_c}^{\pi R_c} dy [\delta(y-\pi R_c) -g^{++}_0(\pi R_c) g^{++}_0(y)] 
\overline{\phi}_{H_u^0} \phi_{H_u^0} \sum_{i,j=0}^{\infty} f^{++}_i(c_L;y)f^{++}_j(c_L;y) 
\overline{\phi}_L^{(i)} \phi_L^{(j)} 
\cr
&&
= q_{H_u^0} q_L g^2 \sum_{i,j=0}^{\infty} f^{++}_i(c_L;\pi R_c)f^{++}_j(c_L;\pi R_c)  
\overline{\phi}_{H_u^0} \phi_{H_u^0} \overline{\phi}_L^{(i)} \phi_L^{(j)}
- q_{H_u^0} q_L {g^2\over2\pi R_c}   
\overline{\phi}_{H_u^0} \phi_{H_u^0} \sum_{i=0}^{\infty} \overline{\phi}_L^{(i)} \phi_L^{(i)}.\nonumber \\
\label{InfSum}
\eea
Then plugging this expression into Eq.(\ref{IntLag}) brings,
\begin{eqnarray}
 &  - & q_{H_u^0} q_L {g^2\over2\pi R_c}   
\overline{\phi}_{H_u^0} \phi_{H_u^0} \sum_{i=0}^{\infty} \overline{\phi}_L^{(i)} \phi_L^{(i)} 
\cr
&+& q_{H_u^0} q_L g^2 \sum_{i,j=0;n=1}^{\infty} g^{++}_n(\pi R_c) {\cal T}_L^{ijn}
\frac{\partial_\mu (\overline{\phi}_{H_u^0} \phi_{H_u^0})
\partial^\mu (\overline{\phi}_L^{(i)} \phi_L^{(j)})}{M^{(n)2}}  \ \ \in \mathcal{L}_{4D}^{D-terms} .
\label{FinalIntLag}
\end{eqnarray}
At this stage, a useful check is to recover the 4D SUSY Lagrangian -- in our simple toy model
with a ${\rm U(1)}$ gauge symmetry and a minimal matter content -- by taking the limiting case 
$M^{(n)},m_{L/R}^{(n)}\to\infty$ ($n\geq 1$). Indeed, in this case Eq.(\ref{FinalIntLag}) simplifies to the
4D SUSY Lagrangian
\begin{eqnarray}
\mathcal{L}_{4D}^{D-terms} & \to  &
- q_{H_u^0} q_L {g^2\over2\pi R_c}   
\overline{\phi}_{H_u^0} \phi_{H_u^0} \overline{\phi}_L^{(0)} \phi_L^{(0)}  =
- q_{H_u^0} q_L g_{4D}^2   
\overline{\phi}_{H_u^0} \phi_{H_u^0} \overline{\phi}_L^{(0)} \phi_L^{(0)}
\label{FinalIntLag4D}
\end{eqnarray}
where $g_{4D}$ represents the 4D effective gauge coupling constant. This test confirms the consistency of the obtained
4D couplings in Eq.(\ref{FinalIntLag}).
This consistency relies on the full summation over the infinite $\Sigma$ KK tower in Eq.(\ref{InfSum}),
similarly to the derivation of D-term couplings in Section \ref{couplingpart}.

In conclusion, the KK corrections with respect to the 4D SUSY $\vert \phi_{H_u^0} \vert^2 \vert\phi_L^{(0)} \vert^2$ coupling
arise at the order $1/M^{(n)2}$.   
At this order $1/M^{(n)2}$ (and above) the corrections in Eq.(\ref{FinalIntLag}) affect the 
$\vert \phi_{H_u^0} \vert^2 \vert\phi_L^{(0)} \vert^2$ coupling but not the consequent masses as the `$\partial_\mu$'
acting on the constant $\phi_{H_u^0}$ VEV vanishes.  
This means that there are no KK tower-induced corrections to any D-term mass at order $1/M^{(n)2}$
with respect to 4D SUSY.

\subsubsection*{Complete scalar mass matrices}

We have done all the necessary preliminary calculations to obtain different 4D mass contributions so that we can 
now write the whole 4D effective scalar mass matrix within the pMSSM. 
We will give the stop mass matrix as an example, the other scalar superpartner masses being easily deducible from it.
This $\tilde{t}$ mass matrix in the interaction basis $ \vec{\tilde{t}} = \{ \tilde{t}^{(0)}_L , \tilde{t}^{(0)}_R , \tilde{t}^{(1)}_L , \tilde{t}^{(1)}_R \}$ reads as
$-\bar{\vec{\tilde{t}}} {\cal M}^2_{\tilde{t}\tilde{t}} \vec{\tilde{t}}^t \in {\cal L}_{4D}$ with, 
\small{\begin{eqnarray}
&&  \  \ Ê\  \  \ Ê\  \  \ Ê\  {\cal M}^2_{\tilde{t}\tilde{t}}  \  \ Ê\ \equiv  \  \ Ê\
\left(\begin{array}{cccc}
0 & 0 & 0 & 0 \\
0 & 0 & 0 & 0 \\
0 & 0 & m^{(1)2}_L & 0 \\
0 & 0 & 0 & m^{(1)2}_R
\end{array}\right) \ + \ {\cal Y}^2 \hat v_u^2 \nonumber
\end{eqnarray}}
\small{\begin{eqnarray}
&  \times &
\left(\begin{array}{cccc}
(f^0_L f^0_R)^2 [1-{\cal Y}^2 \hat v_u^2 (f^0_L)^2\sum_{n=1}^\infty \frac{(f^{n}_R)^2}{m_R^{(n)2}}] & 0 & f^0_L f^1_L (f^0_R)^2 & 0 \\
0 & (f^0_L f^0_R)^2 [1-{\cal Y}^2 \hat v_u^2 (f^0_R)^2\sum_{n=1}^\infty \frac{(f^{n}_L)^2}{m_L^{(n)2}}] & 0 & f^0_R f^1_R(f^0_L)^2 \\
f^0_L f^1_L(f^0_R)^2 & 0 & (f^1_L f^0_R)^2 & 0 \\
0 & f^0_R f^1_R(f^0_L)^2 & 0 & (f^0_L f^1_R)^2
\end{array}\right) \nonumber
\end{eqnarray}}
\small{\begin{eqnarray}
&-&
{{\mu_{eff} {\cal Y} \ \hat v_u}\over{\tan\beta}} 
\left(\begin{array}{cccc}
0 & f^0_Lf^0_R & 0 & f^0_Lf^1_R \\
f^0_Lf^0_R & 0 & f^1_Lf^0_R & 0 \\
0 & f^1_Lf^0_R & 0 & f^1_Lf^1_R \\
f^0_Lf^1_R & 0 & f^1_Lf^1_R & 0
\end{array}\right) 
\quad+\quad
\cos2\beta \ m_Z^2
\left(\begin{array}{cccc}
Q_Z^{t_L} & 0 & 0 & 0 \\
0 & -Q_Z^{t_R} & 0 & 0 \\
0 & 0 & Q_Z^{t_L} & 0 \\
0 & 0 & 0 & -Q_Z^{t_R}
\end{array}\right) \nonumber
\\
&+&
\tilde{m} e^{-2k\pi R_c}
\left(\begin{array}{cccc}
f^0_Lf^0_L & 0 & f^0_Lf^1_L & 0 \\
0 & f^0_Rf^0_R & 0 & f^0_Rf^1_R \\
f^0_Lf^1_L & 0 & f^1_Lf^1_L & 0 \\
0 & f^0_Rf^1_R & 0 & f^1_Rf^1_R
\end{array}\right)
\quad+\quad
A e^{-k\pi R_c} \ \hat v_u 
\left(\begin{array}{cccc}
0 & f^0_Lf^0_R & 0 & f^0_Lf^1_R \\
f^0_Lf^0_R & 0 & f^1_Lf^0_R & 0 \\
0 & f^1_Lf^0_R & 0 & f^1_Lf^1_R \\
f^0_Lf^1_R & 0 & f^1_Lf^1_R & 0
\end{array}\right),
\nonumber  \\ 
\label{WholeMassMat}
\end{eqnarray}}
with the notation $f_{L/R}^{n}=f^{++}_n(c_{\tilde t_{L/R}};\pi R_c)$. $m_Z$ is the $Z^0$ gauge boson mass, 
$\mu_{eff} =  \mu e^{-k \pi R_c} \sim  \mbox{TeV}$ [{\it c.f.} Eq.(\ref{mueff})], 
$A \sim 1$ as above
and to not introduce a new bulk scale, the soft mass localized on the TeV-brane for bulk stops (see later discussion for the alternative possibilities 
of SUSY breaking masses either on the Planck-brane or in the bulk) is taken at $\tilde{m} \sim k$. We have taken a unique soft mass $\tilde{m}$ for
having a more simple matrix here but the above mass matrix is easily extended to the case $\tilde{m}_L \neq \tilde{m}_R$ (respective soft
masses for $\tilde{t}^{(0)}_L$ and $\tilde{t}^{(0)}_R$).

The fourth mass matrix of Eq.(\ref{WholeMassMat}) originates from the D-terms while 
the Yukawa masses (second matrix) have been generalized from Eq.(\ref{LightSol}). 
We have noted $\tilde{t}^{(0)}_L$ and $\tilde{t}^{(0)}_R$ the first two states of the basis but one should clarify the point that $\tilde{t}^{(0)}_{L/R}$
possess in fact small admixtures of the $\tilde{t}^{c(n)}_{L/R}$ KK states (identical to the fields noted $\phi^{c(n)}_{L/R}$ defined in the KK decomposition
(\ref{PhiDecompBis})), a kind of mixing described in the study of mass matrix (\ref{MassMatRR}) and at the origin of the two corrections to the first elements in the
second matrix of Eq.(\ref{WholeMassMat}).
The KK sum in these corrections for the stop, typically localized near the TeV-brane, must be cut at $\sim 2 M_{KK}^{(1)}$ as is usual in this 5D framework 
(see the perturbativity considerations on the top Yukawa coupling e.g. in \cite{CBGMI}).

In conclusion, the effect of $(--)$ KK towers on the $\tilde{t}^{(0)}$ mass is taken into account at first order in $1/M^{(n)2}$ (or $1/m_{L/R}^{(n)2}$) 
via the corrective terms in the second matrix of Eq.(\ref{WholeMassMat}), 
whereas the $(--)$ KK effects in D-term masses have been shown above to vanish at this first order (see Eq.(\ref{FinalIntLag})). 
The mixing effect of $(++)$ KK towers on the $\tilde{t}^{(0)}$ mass 
is taken into account at first order by diagonalizing the mass matrix (\ref{WholeMassMat}) 
which includes the first KK modes.

\section{Phenomenology}
\label{phenomenologypart}

\subsection{Quantum corrections to the Higgs mass}
\label{quantumpart}

In this section, we compute explicitly the quantum corrections -- at the one-loop level -- to the $\phi_{H^0_u}$ Higgs mass 
(similar corrections hold for the complex $\phi_{H^0_d}$ boson) in the model defined in Appendix \ref{lagrangianpart}  
without $\mu$-term for simplification reason 
and without introducing any (soft) SUSY breaking mechanism. Indeed, our goal in this part is to 
study generically the possible re-introduction, 
due to the existence of warped extra dimensions, of the gauge hierarchy problem in a SUSY framework. 
For that purpose, we focus on the quadratic divergent contributions exclusively, 
using the 4D effective couplings derived in previous sections.

Before starting let us discuss the general aspect of this loop analysis, through an overview of the assumptions made.
First, our results on the cancellation of quadratic divergences and on the necessity for some cancellation 
conditions in certain cases, although obtained within a minimal SUSY model, 
also apply to the pMSSM gauge group, Lagrangian and field content. Indeed those results rely (partially) 
on the gauge symmetry structure whatever gauge group it is [in particular abelian or not].
\\ Secondly, there are no masses for the Higgs bosons in our framework (as neither $\mu$-terms nor soft scalar
masses) and in turn the Higgs fields do not acquire VEV. Nevertheless, all our conclusions on quadratic 
cancellations still hold, for example, in the pMSSM after spontaneous EWSB. 
The vanishing mass hypothesis also allows us to work in the Higgs rest frame 
where the external four-momentum are exactly equal to zero, a Lorentz transformation 
choice which does not affect our conclusions due to the invariant nature of the (Higgs) mass.
\\ Finally, we work generically without choosing a specific gauge; the gauge choice will be 
parametrized by the non-fixed $\lambda$ quantity entering the n$th$ KK gauge boson propagator: 
$$
\frac{-i}{k^2-M^{(n)2}} \left[ \eta_{\mu\nu} 
+ \frac{1-\lambda}{\lambda}\frac{k_{\mu}k_{\nu}}{k^2-{1\over\lambda}M^{(n)2}} \right].
$$
Recall that for instance $\lambda=0$ ($\lambda=1$) corresponds to the Landau (Feynman) gauge
which is characterized by $\partial^\mu A_\mu=0$ as considered in the other parts of this paper 
together with $A_5=0$ -- for a gauge boson field $A_M$.
We thus find that the $\lambda$-dependent terms vanish which means that the physical result on 
Higgs mass corrections is not gauge-dependent, as expected. This approach confers to the analysis, and 
a fortiori to the result, a general character.
Although we do not fix $\lambda$, or equivalently $Im(z)$ (which determines the $A_\mu$ field in 
Eq.(\ref{SuperVectorial})), we choose to work in the Wess-Zumino gauge where the extra fields 
that generically appear in the $V$ expression have been transformed to zero (see Eq.(\ref{SuperVectorial}))
due to the specific choices of $Re(z)$, $\eta$, $f$ -- if a gauge transformation
on generic vectorial (chiral) superfields $V$ ($\Omega$) is defined by $V \to V+{\cal Z}+\bar{\cal Z}$ ($\Omega \to \Omega+\sqrt 2 \partial_y {\cal Z}$), 
the chiral superfield ${\cal Z}$ involving the scalar field $z$, spinor $\eta$ and auxiliary $f$. We choose the Wess-Zumino gauge
for simplicity in the loop calculation but the physical result that we obtain would clearly be the same in 
another gauge.

\subsubsection*{Yukawa coupling sector}

Starting with the fermion contributions,
the first step is to get the 4D effective Yukawa couplings of the fermions to the Higgs boson $\phi_{H^0_u}$; from the Lagrangian obtained in 
Eq.(\ref{FermLag}), we know that those couplings after field redefinition (absorbing the 
$\sqrt{G}$ factor) are $\lambda_{nm} \equiv - \int dy {\cal Y} f_L^{n}(y)f_R^{m}(y)\delta(y-\pi R_c)$, assuming a real Yukawa coupling for simplicity.  
\\ The loop diagrams involving Yukawa couplings and resulting in quadratic divergences are those drawn in Fig.(\ref{Fig4f}) and Fig.(\ref{Fig5s}).
From the formal point of view, for a loop having fermions of different masses as in Fig.(\ref{Fig4f}), say $m_{f^1}$ and $m_{f^2}$, running in, 
one can write the loop integral as (calling $\lambda$ the Yukawa coupling constants):
\begin{figure}[!hc]
	\centering
	\vspace*{.5cm}
			\includegraphics[width=0.30\textwidth]{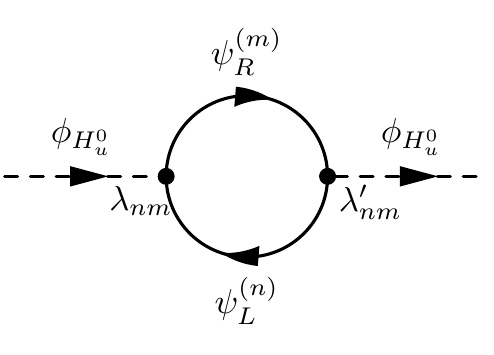}
\caption{\label{Fig4f} \small{Quadratically divergent quantum corrections at one-loop 
to the $\phi_{H^0_u}$ mass due to the exchange of KK Dirac fermions $\psi_{L/R}^{(n)}$.
Couplings are described in text.}}
\end{figure}
\begin{eqnarray}
&& - \int\frac{d^4k}{(2\pi)^4} \mbox{Tr}\left[ \left(-i\lambda\right)\left(\frac{1-\gamma_5}{2}\right) \left(\frac{i}{\displaystyle{\not}k-m_{f^1}}\right) \left(-i\lambda
\right)\left(\frac{1+\gamma_5}{2}\right) \left(\frac{i}{\displaystyle{\not}k-m_{f^2}}\right) \right] \nonumber\\ &&
= - 2 \lambda^2 \int\frac{d^4k}{(2\pi)^4} \left[ \frac{k^2}{(k^2-m_{f^1}^2)(k^2-m_{f^2}^2)} \right] 
\end{eqnarray}
The four-momentum of the fields exchanged in the loop is generically noted $k^\mu$.
Note the presence of the $\frac{1\pm\gamma_5}{2}$ chirality projectors, since such scalar couplings flip the fermion chirality, as well as the overall
minus sign due to the Fermi-Dirac statistics which is crucial for the SUSY cancellation of divergences. 
Now expanding over different KK fermion towers results in the following contribution to the $\phi_{H^0_u}$ Higgs mass, respecting the
correct order between the discrete KK sum and the four-momentum loop integral,
\begin{eqnarray}
\mathcal{I}_F = - 2 \sum_{\{n,m\}=0,0}^{N,M} \int\frac{d^4k}{(2\pi)^4} \left[ \frac{k^2}{(k^2-m_L^{(n)\,2})(k^2-m_R^{(m)\,2})} \right] \lambda_{nm} \lambda'_{nm}
\end{eqnarray}
where $\lambda'_{nm} \equiv - \int dy' {\cal Y} f_L^{n}(y')f_R^{m}(y')\delta(y'-\pi R_c)$ 
and $m_{L/R}^{(n)}$ [with $m_{L/R}^{(0)}=0$] are the KK fermion masses.
We have truncated the KK summation at the indices $N,M$ such that $m_L^{(N)}, m_R^{(M)} \sim \Lambda$ 
and consistently the momentum integration at the cut-off of the 5D theory $\Lambda$ (one has typically 
$\Lambda \sim M^{(2)}_{KK}$, the second KK gauge mass).  
The reason being that the non-renormalizable 5D SUSY theory is only valid below this cut-off.
The integral gives us :
\begin{eqnarray}
\mathcal{I}_F &=& - \ 2{\cal Y}^{\,2} \sum_{\{n,m\}=0,0}^{N,M} \int\frac{d^4k}{(2\pi)^4} k^2 \int dy \int dy' 
\left[ \frac{f_L^{n}(y)f_L^{n}(y')}{k^2-m_L^{(n)\,2}} \frac{f_R^{m}(y)f_R^{m}(y')}{k^2-m_R^{(m)\,2}} \right] \delta(y-\pi R_c) \ \delta(y'-\pi R_c) \nonumber\\
& = & - \ 2{\cal Y}^{\,2} \sum_{\{n,m\}=0,0}^{N,M} 
\int\frac{d^4k}{(2\pi)^4} k^2 \frac{f_L^{n}(\pi R_c)f_L^{n}(\pi R_c)}{k^2-m_L^{(n)\,2}} \frac{f_R^{m}(\pi R_c)f_R^{m}(\pi R_c)}{k^2 - m_R^{(m)\,2}} .
\label{ToAdd}
\end{eqnarray}

We now calculate the scalar superpartner contributions. 
First one has to derive the 4D effective Yukawa coupling of the matter scalars to the Higgs boson $\phi_{H^0_u}$; we have obtained it 
in Eq.(\ref{HYukCoupl}) 
renaming it now for commodity as $\tilde \lambda_{n}|_{L/R} = - i {\cal Y}^2 \left[f_{L/R}^{n}(\pi R_c)\right]^2 \, k^2 \, 
G_5^{f^{++}_{c_{R/L}}}(k^2;\pi R_c,\pi R_c)$;
the KK scalar contributions to the $\phi_{H^0_u}$ mass only come from these `diagonal' $nn$ Yukawa couplings as shows Fig.(\ref{Fig5s}).
As explained at the end of the part commenting the scalar 4D couplings to two Higgs bosons in Section \ref{couplingpart}, 
the KK summation in $G_5^{f^{++}_{c_{R/L}}}(k^2;\pi R_c,\pi R_c)$ has to be truncated and we truncate it consistently 
at the index $N$ such that $m_L^{(N)} \sim \Lambda$.
\begin{figure}[!hc]
	\centering
	\vspace*{.5cm}
			\includegraphics[width=0.30\textwidth]{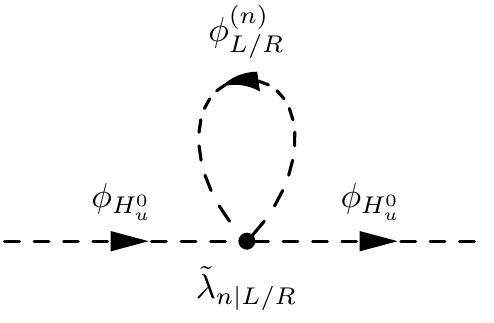}
\caption{\label{Fig5s} \small{Quadratically divergent quantum corrections at one-loop 
to the $\phi_{H^0_u}$ mass due to the exchange of KK scalar superpartners $\phi_{L/R}^{(n)}$.
Couplings are described in text.}}
\end{figure}
\\ From Fig.(\ref{Fig5s}), we see that expanding over different KK scalar towers results in the contribution:
\begin{eqnarray}
\mathcal{I}_S = \sum_{n=0}^{N} \int\frac{d^4k}{(2\pi)^4} \left\{
\frac{i \tilde \lambda_{n}|_L}{k^2-m_L^{(n)\,2}} +
\frac{i \tilde \lambda_{n}|_R}{k^2-m_R^{(n)\,2}}
\right\}
\label{WithSum}
\end{eqnarray}
where $m_{L/R}^{(n)}$ are the KK scalar masses. As there is no SUSY breaking here the fermions and their scalar superpartners have identical
KK masses and wave functions. 
We have cut the KK tower and momentum integration at $\Lambda$ ($\sim M^{(2)}_{KK}$) as for the fermion contribution. 
We get
\begin{eqnarray}
\mathcal{I}_S &=& {\cal Y}^{\,2} \sum_{n=0}^{N} \int\frac{d^4k}{(2\pi)^4} \left[ 
\frac{\left[f_L^{n}(\pi R_c)\right]^2}{k^2-m_L^{(n)\,2}} k^2 \, G_5^{f^{++}_{c_R}}(k^2;\pi R_c,\pi R_c) 
+ \frac{\left[f_R^{n}(\pi R_c)\right]^2}{k^2-m_R^{(n)\,2}} k^2 \, G_5^{f^{++}_{c_L}}(k^2;\pi R_c,\pi R_c)  
\right] \nonumber \\
&=& 2 {\cal Y}^{\,2}  \sum_{\{n,m\}=0,0}^{N,M} \int\frac{d^4k}{(2\pi)^4} k^2 \frac{\left[f_R^{n}(\pi R_c)\right]^2}{k^2-m_R^{(n)\,2}} \, 
  \frac{\left[f_L^{m}(\pi R_c)\right]^2}{k^2-m_L^{(m)\,2}} 
\label{ToInvert}
\end{eqnarray}
after inverting a {\it finite} summation (in 5D propagators) with the {\it cut} integration over $k$.

One thus finds $\mathcal{I}_F + \mathcal{I}_S = 0$ due to SUSY. Strictly speaking,
we have in fact obtained this quadratic divergence cancellation analytically in the generic case of any cut-off 
(i.e. for any $\Lambda$ value). Was this result predictable~? In the limiting case $\Lambda < M_{KK}$ 
-- the first KK gauge mass $M_{KK}$ is the smallest KK mass among bosons and fermions -- where $n=m=0$, 
one recovers the 4D SUSY model (with a cut-off) so that the above result of quadratic cancellation constitutes only a good check.    
On the opposite side, for $\Lambda \to \infty$ assuming a known UV completion for the 5D theory (or more realistically $\Lambda \simeq M_{\star}$ in our RS context) 
so that the theory is totally 5D (or only up to the effective gravity scale), it is not surprising to find that the pure 5D SUSY guarantees the quadratic cancellation.  
Finally, for any intermediate cut-off truncating KK sums and loop integrations, 
our generic result is that the cancellation systematically occurs [the same cut-off $\Lambda$ must 
be applied on Eq.(\ref{ToAdd}) and Eq.(\ref{ToInvert}) so that these expressions remain exactly opposite] in what could be called a `truncated 5D theory'; 
this constitutes the proof of a supersymmetric cancellation KK level by KK level (i.e. the n$th$ KK scalar contribution compensates the n$th$ KK fermion 
contribution), a non-obvious result. A similar result holds for the quadratic divergence cancellation in the gauge coupling sector treated below.

\subsubsection*{Comments about the cancellation of quadratic divergences}

Based on the above example of cancellation of quadratic divergences in the Yukawa coupling sector (compensation between fermions and their scalar
superpartner), we discuss here why our approach brings some new light
on the old debate about this cancellation in Higgs mass quantum corrections in higher-dimensional SUSY theories (the SUSY breaking aspect
is not considered here) with localized Higgs interactions.

First there were questions \cite{Ghilencea,Kim} on the sense of the ``KK regularization'' \cite{KKreg} in higher-dimensional SUSY theories. 
The KK regularization is the divergence cancellation which relies on performing first the infinite summation of loop-exchanged KK states and secondly the infinite four-momentum 
loop-integral; this order corresponds to a non-justified inversion in the analytical computation of an {\it infinite} number of KK contributions at the one-loop level to the
Higgs mass.
\\ We have shown, in the part commenting the scalar 4D couplings to two Higgs bosons in Section \ref{couplingpart}, that writing the scalar effective 4D 
couplings requires to perform an infinite summation on KK tower without applying any cut-off. Only once one has derived these scalar effective 4D couplings,
the analytical loop computation -- in a 4D framework -- of KK scalar contributions to the Higgs mass can be started: that is the correct order. Hence, the loop four-momentum 
integration must be performed {\it after} the aforementioned infinite KK summation, as exhibits Eq.(\ref{WithSum}) (where an infinite summation has already been calculated to
obtain $\tilde \lambda_{n}|_{L/R}$), even if this is in contrast with a naive thinking. 
In Eq.(\ref{ToInvert}), the summations in $G_5$ -- which originate from the remaining summation in $\tilde \lambda_{n}|_{L/R}$ -- are finite so  
it makes sense to invert those with the cut integration, in order to obtain the quadratic divergence cancellation with Eq.(\ref{ToAdd})  
\footnote{In fact, within our approach we don't have to make {\it abnormally} an infinite KK summation before a loop integration.}.

Related doubts pointed out in Ref.~\cite{Ghilencea} concerned the effect of the necessary cut-off, due to the non-renormalizable aspect of 5D SUSY models,  
which prevents from making any infinite KK summation and in turn to find the quadratic cancellation: indeed, the authors of \cite{Ghilencea} have demonstrated that
the cancellation results partially from SUSY and partially from a compensation between the quadratic terms of a 
finite number of KK modes with masses below the cut-off and the logarithmic terms of an infinite number of KK states above it. 
It was believed that the procedure in Ref.~\cite{Ghilencea} based on a sharp cut of the KK tower, spoiling the supersymmetry of the underlying theory, 
could prevent from quadratic cancellations. However, the proper time cut-off (made separately from the KK level truncation) 
not spoiling four-dimensional symmetries can be applied \cite{Koba} and similar problems arise: quantum corrections
to the Higgs mass become insensitive to details of physics at the UltraViolet (UV) scale only under certain conditions.
Another alternative to the sharp KK tower truncation is the suppression by a Gaussian brane distribution \cite{Gero} (the couplings of high KK modes are suppressed by a 
finite width of the brane) which indeed allows to recover a finite Higgs mass -- including the cut-off effect but keeping an infinite sum so evading the drawbacks outlined in 
\cite{Ghilencea}. 
Nevertheless, it appears also in Ref.~\cite{Gero} that other distributions leading to a linear sensitivity on the momentum cut-off exist, a remark forbidding a general conclusion
\footnote{The approach developed in Ref.~\cite{Gero} also shows that the Higgs mass corrections at higher order (two loops) 
potentially give rise to linear divergences which are in fact cancelled by linear threshold effects of Yukawa and gauge couplings \cite{Gero2loops}.}. 
These two works \cite{Koba,Gero} have thus not really solved the problems raised in \cite{Ghilencea} on a general justification of the finiteness of the Higgs mass.  
\\ The way to describe the cut-off problem of Ref.~\cite{Ghilencea} within our framework is as follows. An infinite KK summation has to be 
computed for applying the completeness relation and hence for finding the $\tilde \lambda_{n}|_{L/R}$ form obtained in 
Eq.(\ref{HYukCoupl}) 
and used in Eq.(\ref{WithSum}) to give rise to the quadratic cancellation (between $\mathcal{I}_S$ and $\mathcal{I}_F$),  
however, the infinite character seems meaningless as the KK states with masses above $\Lambda$ make no sense in a theory valid only up to the cut-off scale.
In fact it turns out, as we have discussed in the part commenting the scalar 4D couplings (Section \ref{couplingpart}), 
that an infinite KK summation must really be calculated in order to write down a consistent 4D
Lagrangian for the fundamental 5D SUSY theory; the cut-off is indeed applied 
but only after the 4D couplings have been derived (which requires an infinite summation computation) and it is applied 
on the remaining not-computed sums in the coherent 4D framework
\footnote{Hence, our approach to the cut-off problem (w.r.t. divergence cancellations) 
is different from Ref.~\cite{Koba,Gero}: we do not deal with the kind of cut-off that is applied but instead we justify why 
this cut-off must not be applied systematically on KK sums.}.

We have not brought arguments against the claims of Ref.~\cite{Ghilencea} but we have proposed a different approach avoiding their problems (for the KK regularization)
and we have justified the required infinite KK summation (including the cut-off aspect).  
Therefore the quadratic cancellation appears to be well treated in our context and thus to be meaningful. Similar arguments hold for the independent quadratic cancellation 
in the gauge coupling sector (involving also Higgs boson, higgsino, gauge boson and gaugino contributions to the Higgs mass) that will be treated in the following subsection.
\\ There exist other approaches like the Pauli-Villars regularization \cite{PV} or the elegant 5D (mixed position-momentum space) framework \cite{5D},  
based on formally correct treatments of the divergences avoiding subtleties on KK excitations and 4D Lagrangians, 
which also conclude positively on the validity of quadratic cancellations.

\subsubsection*{Gauge coupling sector}

\begin{figure}[!hc]
	\centering
	\vspace*{.5cm}
		\includegraphics[width=0.30\textwidth]{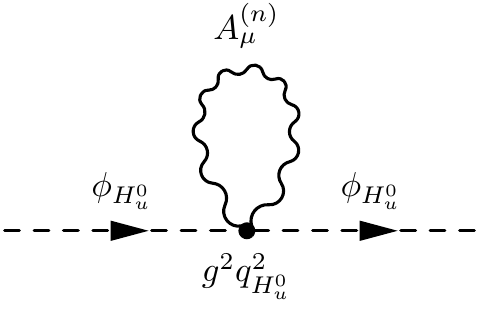}
\caption{\label{Fig6b} \small{Quadratically divergent quantum corrections at one-loop 
to the $\phi_{H^0_u}$ mass due to the exchange of KK gauge bosons $A_{\mu}^{(n)}$.
Couplings are described in text.}}
\end{figure}

\begin{figure}[!hc]
	\centering
	\vspace*{.5cm}
		\includegraphics[width=0.30\textwidth]{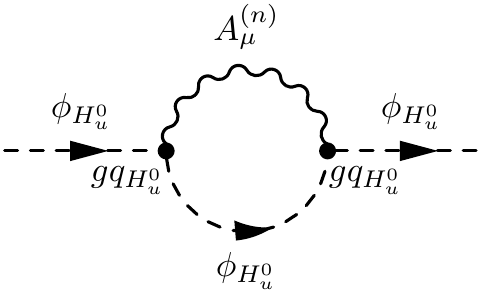}
	\caption{\label{Fig7b} \small{Quadratically divergent quantum corrections at one-loop 
to the $\phi_{H^0_u}$ mass due to the exchange of KK gauge bosons $A_{\mu}^{(n)}$ together with the Higgs boson itself.
Couplings are described in text.}}
\end{figure}

\begin{figure}[!hc]
	\centering
	\vspace*{.5cm}
		\includegraphics[width=0.3\textwidth]{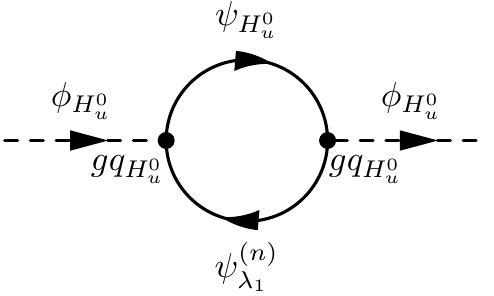}
	\caption{\label{Fig8g} \small{Quadratically divergent quantum corrections at one-loop 
to the $\phi_{H^0_u}$ mass due to the exchange of KK gaugino modes $\psi_{\lambda_1}^{(n)}$ with the higgsino state $\psi_{H^0_u}$.
Couplings are described in text.}} 
\end{figure}

\begin{figure}[!hc]
	\centering
	\vspace*{.5cm}
		\includegraphics[width=0.30\textwidth]{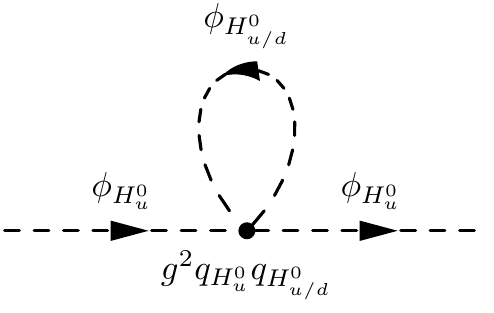}
	\caption{\label{Fig9h} \small{Quadratically divergent quantum corrections at one-loop 
to the $\phi_{H^0_u}$ mass due to self-couplings and couplings with the scalar field $\phi_{H^0_d}$ (described in text).}}
\end{figure}

First, the exchange of ${\rm U(1)}$ gauge boson KK modes contributes ({\it c.f.} Fig.(\ref{Fig6b})), via the gauge couplings derived in Eq.(\ref{AppZpar}),
to the quadratic divergences appearing in the $\phi_{H^0_u}$ mass corrections. This contribution reads as,
\begin{eqnarray}
\mathcal{I}_A &=& \sum_{n=0}^{N} \int\frac{d^4k}{(2\pi)^4} \int dy \ i \ q_{H^0_u}^2 g^{\,2} \, 
[g^{++}_n(y)]^2 \,  \eta^{\mu\nu} \, \delta(y-\pi R_c) 
\frac{-i}{k^2-M^{(n)2}} \left[ \eta_{\mu\nu} 
+ \frac{1-\lambda}{\lambda}\frac{k_{\mu}k_{\nu}}{k^2-{1\over\lambda}M^{(n)2}} \right]
\nonumber\\
&=& q_{H^0_u}^2 g^{\,2} \sum_{n=0}^{N} \int\frac{d^4k}{(2\pi)^4} \frac{[g^{++}_n(\pi R_c)]^2}{k^2-M^{(n)2}} 
\left[ 4 + \frac{1-\lambda}{\lambda-M^{(n)2}/k^2} \right],
\end{eqnarray}
where we have used the 4D effective gauge coupling obtained from Eq.(\ref{AppZpar}). As for the Yukawa sector, the KK
summation and momentum integration are truncated at the 5D cut-off $\Lambda$.
\\ The exchanges of $A_\mu^{(n)}$ states together with the Higgs boson also contribute as drawn in Fig.(\ref{Fig7b}).
Still truncating the sum and integration at $\Lambda$, we obtain [see again Eq.(\ref{AppZpar})]
\begin{eqnarray}
\mathcal{I}'_{A} &=& \sum_{n=0}^{N} \int \frac{d^4k}{(2\pi)^4} \int dy \int dy' \left(-i\ q_{H^0_u} g k^{\mu} \, g^{++}_n(y) \, 
\delta(y-\pi R_c)\right) \left(-i\ q_{H^0_u} g k^{\nu} \, g^{++}_n(y') \, \delta(y'-\pi R_c)\right)  \nonumber\\
&& \qquad\qquad\qquad\qquad\qquad\qquad \times \frac{i}{k^2} \, \frac{-i}{k^2-M^{(n)2}} 
\left[ \eta_{\mu\nu} + \frac{1-\lambda}{\lambda}\frac{k_{\mu}k_{\nu}}{k^2-{1\over\lambda}M^{(n)2}} \right]
\nonumber\\
&=& - q_{H^0_u}^2 g^2 \sum_{n=0}^{N} \int \frac{d^4k}{(2\pi)^4} \frac{[g^{++}_n(\pi R_c)]^2}{k^2-M^{(n)2}} 
\left[ 1 + \frac{1-\lambda}{\lambda-M^{(n)2}/k^2} \right].
\end{eqnarray}
\\ Similarly, the exchanges of $\psi_{\lambda_{1}}^{(n)}$ gaugino states 
(four-component spinorial notation) and higgsinos ($\psi_{H^0_u}$) contribute as in Fig.(\ref{Fig8g}).
Deducing the 4D effective gaugino couplings from the superfield action (\ref{muAction}) [as done exactly in Eq.(\ref{FermLag})], one finds the mass contribution 
\begin{eqnarray}
\mathcal{I}_G &=& - \sum_{n=0}^{N} \int \frac{d^4k}{(2\pi)^4} \int dy \int dy' \left(- \sqrt{2} q_{H^0_u}g g^{++}_n(y) \, \delta(y-\pi R_c)\right)  
\left(\sqrt{2} q_{H^0_u}g g^{++}_n(y') \, \delta(y'-\pi R_c)\right)  \nonumber\\
&& \qquad\qquad\qquad\qquad\qquad\qquad \times \mbox{Tr}\left[ \frac{i (\displaystyle{\not}k+M^{(n)})}{k^2-M^{(n)2}}  
\frac{1-\gamma_5}{2} \frac{i \displaystyle{\not}k}{k^2} \frac{1+\gamma_5}{2} \right]
\nonumber\\
&=& - 4 q_{H^0_u}^2g^2 \sum_{n=0}^{N} \int\frac{d^4k}{(2\pi)^4} \frac{[g^{++}_n(\pi R_c)]^2}{k^2-M^{(n)2}}.
\end{eqnarray}
This loop contribution carries a minus sign in front of the sum due to the Fermi-Dirac statistics. 
Since there is no SUSY breaking, the gauge boson modes have the same 
KK masses and wave functions as their fermionic superpartners. 
\\ Other divergences arise from the Higgs exchange itself; see Fig.(\ref{Fig9h}). The preliminary 4D result of Eq.(\ref{HBiCoupl}) 
and Eq.(\ref{HQuarCoupl}) allow us to write down these two mass corrections, respectively, as:
\begin{eqnarray}
\mathcal{I}_{H_d^0} &=& \int\frac{d^4k}{(2\pi)^4} \left( -i q_{H_u^0}q_{H_d^0}g^2 p^2 G_5^{g^{++}}(p^2;\pi R_c,\pi R_c) \right)  \frac{i}{k^2} 
\nonumber\\ &\to & q_{H_u^0}q_{H_d^0}g^2 \int\frac{d^4k}{(2\pi)^4} \frac{[g^{++}_0(\pi R_c)]^2}{k^2}  
= \frac{q_{H_u^0}q_{H_d^0}g^2}{2 \pi R_c} \int\frac{d^4k}{(2\pi)^4} \frac{1}{k^2} ,
\label{IHd}
\\
\mathcal{I}_{H_u^0} &=& \int\frac{d^4k}{(2\pi)^4} \left( - i q_{H_u^0}^2g^2 p^2 G_5^{g^{++}}(p^2;\pi R_c,\pi R_c) \right) \frac{i}{k^2}
+ \int\frac{d^4k}{(2\pi)^4} \left( - i q_{H_u^0}^2g^2 k^2 G_5^{g^{++}}(k^2;\pi R_c,\pi R_c) \right) \frac{i}{k^2}
\nonumber\\ &\to &
\frac{q_{H_u^0}^2g^2}{2 \pi R_c} \int\frac{d^4k}{(2\pi)^4} \frac{1}{k^2}
+ q_{H_u^0}^2g^2 \int\frac{d^4k}{(2\pi)^4} G_5^{g^{++}}(k^2;\pi R_c,\pi R_c)
\nonumber\\ &= &
\frac{q_{H_u^0}^2g^2}{2 \pi R_c} \int\frac{d^4k}{(2\pi)^4} \frac{1}{k^2}
+ q_{H_u^0}^2g^2 \sum_{n=0}^{N} \int\frac{d^4k}{(2\pi)^4} \frac{[g^{++}_n(\pi R_c)]^2}{k^2-M^{(n)2}}.
\label{IHu} 
\end{eqnarray}
In Eq.(\ref{IHd}), the limit $p^2\to 0$ has been performed (using Eq.(\ref{5Dprop})) since the effective coupling 
involved in Fig.(\ref{Fig9h}) confers to this loop diagram a tadpole-form contribution; indeed the effective diagram in
Fig.(\ref{Fig9h}) may be obtained by summing the
two diagrams of Fig.(\ref{SelfDrawnBis}) after joining the two $\phi_{H^0_d}$ legs and affecting respectively to the $\phi_{H^0_d}$, $\Sigma^{(n)}$,
$\phi_{H^0_u}$ fields the momentum $k^2$ (integrated loop momentum), $p^2=0$ and $0$ (chosen external momentum).   
In Eq.(\ref{IHu}), the limit $p^2\to 0$ has also been performed since the effective coupling 
involved in Fig.(\ref{Fig9h}) for an internal $\phi_{H^0_u}$ field also
confers to this loop diagram a tadpole-form contribution. Here there is even an additional contribution from the exchange of $\Sigma^{(n)}$ with the same
momentum $k^2$ as in the loop. All this can be seen from generating Fig.(\ref{Fig9h}) by summing the
three diagrams of Fig.(\ref{SelfDrawn}) after joining the two upper $\phi_{H^0_u}$ legs in each diagram.   
We have also truncated the KK summation by the 5D cut-off $\Lambda$ 
(like the integration), in the 5D propagator $G_5^{g^{++}}(k^2;\pi R_c,\pi R_c)$, and then inverted it with the integral. 

\begin{figure}[!hc]
	\centering
	\vspace*{.5cm}
		\includegraphics[width=0.30\textwidth]{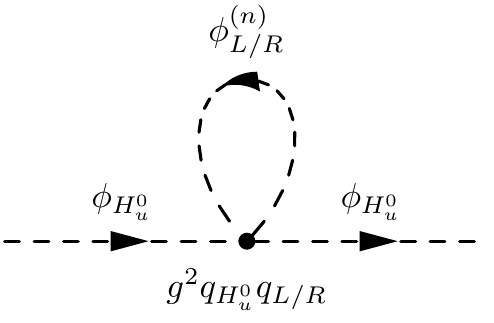}
	\caption{\label{Fig10s} \small{Quadratically divergent quantum corrections at one-loop 
to the $\phi_{H^0_u}$ mass due to gauge couplings with KK scalar superpartners $\phi_{L/R}^{(n)}$ (described in text).}}
\end{figure}

\begin{figure}[!hc]
	\centering
	\vspace*{.5cm}
		\includegraphics[width=0.30\textwidth]{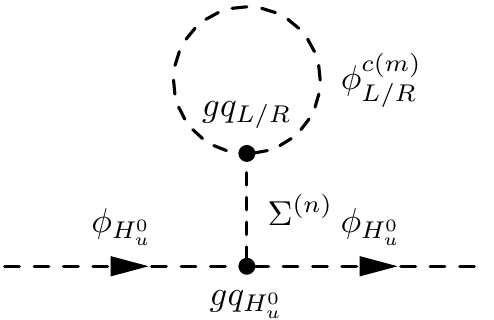}
	\caption{\label{Fig11c} \small{Quadratically divergent quantum corrections at one-loop 
to the $\phi_{H^0_u}$ mass due to the exchange of KK $(--)$ scalar superpartners $\phi_{L/R}^{c(m)}$ and KK $(--)$ scalar modes $\Sigma^{(n)}$
(couplings discussed in text).}} 
\end{figure}

The last contributions to quadratic divergences generated by gauge couplings are the scalar exchanges in Fig.(\ref{Fig10s})-(\ref{Fig11c}). 
In the same way as just above, the 4D D-term coupling previously obtained in Eq.(\ref{HDtCoupl}) leads to the corresponding mass corrections:
\begin{eqnarray}
\mathcal{I}_{L/R} &=& 
\sum_{n=0}^{N} \int\frac{d^4k}{(2\pi)^4} \left( - i q_{L/R} q_{H_u^0} g^2 \int dy [f^n_{L/R}(y)]^2 p^2 G_5^{g^{++}}(p^2;y,\pi R_c) \right)\frac{i}{k^2-m_{L/R}^{(n)2}}   
\nonumber\\ &\to &
\frac{q_{L/R} q_{H_u^0} g^2}{2 \pi R_c} \sum_{n=0}^{N} \int\frac{d^4k}{(2\pi)^4} \int dy [f^n_{L/R}(y)]^2 \frac{1}{k^2-m_{L/R}^{(n)2}}
\nonumber\\ &= &
\frac{q_{L/R} q_{H_u^0} g^2}{2 \pi R_c} \sum_{n=0}^{N} \int\frac{d^4k}{(2\pi)^4} \frac{1}{k^2-m_{L/R}^{(n)2}}.
\label{ILR} 
\end{eqnarray}
in the limit $p^2 \to 0$ and using the orthonormalization condition for wave functions.
\\ Similarly, the 4D effective couplings of two $\phi^{c(m)}_{L/R}$ [defined in Eq.(\ref{PhiDecompBis})] 
to two Higgs bosons -- induced by $\Sigma^{(n)}$ exchanges -- can be directly derived from the scalar field Lagrangian (\ref{ScaFieLagPrim}),
by using relation (\ref{UsefulRelat}), and lead to the respective tadpole contributions of Fig.(\ref{Fig11c}):
\begin{eqnarray}
\mathcal{I}^c_{L/R} = \sum_{m=1}^{M} \int\frac{d^4k}{(2\pi)^4} 
\left(q_{L/R} q_{H_u^0} g^2 \int dy [f^{cm}_{L/R}(y)]^2 \sum_{n=1}^{N} M^{(n)2} \frac{g_n^{++}(y) \ i \ g_n^{++}(\pi R_c)}{p^2-M^{(n)2}} \right) 
\frac{i}{k^2-m_{L/R}^{(m)2}}, \nonumber
\end{eqnarray}
reminding that $f_{L/R}^{cm}(y)=f^{--}_m(c_{L/R};y)$. 
Applying now the completeness relation and the equality $M^{(n)2}/(p^2-M^{(n)2})=\{p^2/(p^2-M^{(n)2})\}-1$, 
one can recast this contribution into the expression
\begin{eqnarray}
\mathcal{I}^c_{L/R} & =& - \sum_{m=1}^{M} \int\frac{d^4k}{(2\pi)^4} 
\left( q_{L/R} q_{H_u^0} g^2 \int dy [f^{cm}_{L/R}(y)]^2 p^2 G_5^{g^{++}}(p^2;y,\pi R_c) \right) \frac{1}{k^2-m_{L/R}^{(m)2}}
\nonumber\\ &\to &
- \sum_{m=1}^{M} \int\frac{d^4k}{(2\pi)^4}  \frac{q_{L/R} q_{H_u^0} g^2}{2 \pi R_c} \int dy [f^{cm}_{L/R}(y)]^2 \frac{1}{k^2-m_{L/R}^{(m)2}} 
\nonumber\\ &=& - \frac{q_{L/R} q_{H_u^0} g^2}{2 \pi R_c} \sum_{m=1}^{M} \int\frac{d^4k}{(2\pi)^4} \frac{1}{k^2-m_{L/R}^{(m)2}} 
\label{reIcLR} 
\end{eqnarray}
in the limit $p^2 \to 0$. 

At this stage where all quadratically divergent contributions have been estimated, 
a useful check is to take the 4D SUSY limit: $k^2/m_{L/R}^{(n)2} \ll 1$ and $k^2/M^{(n)2} \ll 1$ [$n\geq1$] at the zero$th$ order. Doing so,
one recovers indeed the whole quadratically divergent mass correction for the gauge coupling sector of the 4D SUSY theory:
$$
\mathcal{I}_{A} + \mathcal{I}'_{A} + \mathcal{I}_G + \mathcal{I}_{H_u^0} 
+ \mathcal{I}_{H_d^0} + \mathcal{I}_{L} + \mathcal{I}_{R} + \mathcal{I}^c_{L} + \mathcal{I}^c_{R} 
\to q_{H_u^0} \Big ( q_{H_u^0}+q_{H_d^0}+q_L+q_R \Big ) g_{4D}^2 \int\frac{d^4k}{(2\pi)^4} \frac{1}{k^2},
$$
with $g_{4D} = g/\sqrt{2\pi R_c}$ as already defined. Indeed, in 4D SUSY the quadratically divergent 
Higgs mass corrections cancel each other only if the anomaly cancellation
condition $q_{H_u^0}+q_{H_d^0}+q_L+q_R = 0$ is verified, recalling that the quadratic divergence 
cancellation is induced by the simultaneous presence of SUSY and a gauge symmetry (which relies on the absence of
Adler-Bardeen-Jackiw anomalies \cite{anomaly} originating from triangular loops of fermions).

In the present RS SUSY context with localized Higgs fields, we thus have first to wonder what is the global 5D 
anomaly cancellation condition. The contributions to triangular loops, of the fermions belonging to the chiral matter 
superfields $\Phi_L$ and $\Phi_L^{--}$, should vanish due to the vectorial nature of the 5D theory 
[i.e. presence of $\Phi_L^{--}$].  
Same comment holds for the $\Phi^c_L$ and $\Phi_L^{c--}$ superfields. The orbifolding apparently spoils this 5D vectorial
nature [i.e. no zero-mode for $\Phi_L^{--}$] but the anomaly cancellation is recovered in the matter sector through 
certain tree level contributions (see e.g. Ref.~\cite{Ritesh}) induced by the Chern-Simons term (see Ref.~\cite{warpCS} 
for the case of warped orbifolds) together with mixings generated by the St\"{u}ckelberg term 
(for instance, see Ref.~\cite{Stuck}). For intervals in $AdS_5$, anomalies might lead to some constraints 
on the consistent effective field theory description \cite{BenBis}. Finally, one must recall here that the whole 
5D anomaly cancellation condition includes the anomaly cancellation condition of the low-energy 4D chiral theory whose role is to insure the
anomaly cancellation for the contributions of the zero-modes and possibly fields of the considered model confined on 3-branes (in other words for
all states except the KK excitations).

Coming back to our 5D SUSY model, we see that one gets (whatever are the KK masses)
$$
\mathcal{I}_{A} + \mathcal{I}'_{A} + \mathcal{I}_G + \mathcal{I}_{H_u^0} 
+ \mathcal{I}_{H_d^0} + \mathcal{I}_{L} + \mathcal{I}_{R} + \mathcal{I}^c_{L} + \mathcal{I}^c_{R} 
= q_{H_u^0} \Big ( q_{H_u^0}+q_{H_d^0}+q_L+q_R \Big ) \frac{g^2}{2 \pi R_c} \int\frac{d^4k}{(2\pi)^4} \frac{1}{k^2},
$$
so that the quadratic divergence cancellation for the Higgs mass in the gauge sector (and hence in all sectors) is guaranteed by the
4D condition $q_{H_u^0}+q_{H_d^0}+q_L+q_R=0$ for the chiral zero-modes and 4D higgsinos (localized on the TeV-brane) 
which is part of the 5D anomaly cancellation condition.  
On the other side, the vectorial nature of the 5D SUSY theory, which induces the cancellation of 5D anomalies, is also
responsible for the cancellation of the KK $\phi^{(n)}_{L/R}$ contributions to the Higgs mass quadratic divergences   
(entering Eq.(\ref{ILR})) with the KK $\phi^{c(m)}_{L/R}$ contributions (see Eq.(\ref{reIcLR})). This cancellation results
from a compensation KK level by KK level and remains thus true for any 5D cut-off value. 
\\ As a general conclusion (the present analysis being not based on arguments restricted to warped geometries), in higher-dimensional 
SUSY models,
the quadratic divergence cancellation in the Higgs mass is insured by the higher-dimensional anomaly cancellation condition
(as occurs in 4D with the difference that the higher-dimensional anomaly condition can be more complex than the 4D one since it
may include the adjustment of the Chern-Simons term to restore the vectorial behavior
\footnote{This Chern-Simons term has anyway no effects on the quadratic divergences of the Higgs mass.}).

We finish this part by a comment, for completeness. 
While the condition $q_{H_u^0}+q_{H_d^0}+q_L+q_R=0$
is issued from anomalies of type ${\rm U(1)}-$Gravity$-$Gravity, there is a second condition, namely
$q^3_{H_u^0}+q_{H_d^0}^3+q^3_L+q^3_R= 0$, coming from the cubic ${\rm U(1)-U(1)-U(1)}$ anomalies. Let us mention here the third
condition on ${\rm U(1)}$ charges in our toy model: $q_{H_u^0}+q_L+q_R= 0$ related to the existence of a Yukawa interaction for
$\Phi_{L/R}$ (see Eq.(\ref{muAction})). Our motivation for writing an interaction of this type was clearly to consider all the quadratically divergent
contributions to the Higgs mass and in particular those involving Yukawa couplings (discussed above).

\subsection{Sfermion mass splitting}
\label{sfermionpart}

\subsubsection*{Phenomenological framework}

We first describe in more details the phenomenological framework of 
this Section \ref{sfermionpart}. We consider the class of SUSY breaking scenario where (soft) 
squark/slepton masses appear in the bulk and on the boundaries [see SUSY breaking classification of Section \ref{model}]. 
The model studied is the warped 5D pMSSM. We do not compute numerically the heavy KK state mixing effects since we focus on 
dominant low-energy structural effects on scalar couplings induced by the SUSY breaking scenario discussed below.  

Let us thus discuss what are the favored geometrical setups concerning the SUSY breaking scalar mass locations. 
First, to have a generic approach, we assume that all bulk sfermions have additional (i.e. SUSY breaking) 5D mass terms of course invariant under the 
$\mathbb{Z}_{2}$ parity: those are also taken of the type shown in Eq.(\ref{BulScalMass}) but with other $c$ parameters. We have shown in Appendix \ref{SolutionFreeScal}
that adding such masses is equivalent to introduce new 5D scalar parameters, say $c_{\tilde f_{L/R}}$, completely independent from the fermion 
(or superfield) parameters $c_{f_{L/R}}$ and thus affects the scalar localizations. 
In analogy with RS flavor models for fermions, we will generally make the hypothesis that the first generations of sfermions are typically
localized towards the UV boundary (large $c_{\tilde f_{L/R}}$) whereas last families are rather located near the IR boundary (small $c_{\tilde f_{L/R}}$). 
Remaining general, the first scalar generations have also SUSY breaking masses localized on the two boundaries and the 
large soft masses on the Planck-brane are thus not reduced by wave functions overlaps. These Planckian masses mimic $(-+)$ BC 
so that the first generation sfermion zero-modes decouple from the low-energy theory as occurs in the model of Ref.~\cite{PartSusyRS}.  
This class of scenario represents thus a realization of partially split SUSY models which allow to improve the situation \cite{Split} with respect to 
Flavor Changing Neutral Currents (FCNC).
In contrast, the last sfermion generations have soft masses on the TeV-brane which are not suppressed by wave function overlaps. These 
soft masses enlarge the parameter space that we will explore. 
Having typically small $c_{\tilde f_{L/R}}$ for last sfermion families is also an
attractive possibility as it leads to sfermion masses which are mainly generated by the Yukawa interactions after EWSB,
as for SM fermions, and to specific collider signatures as discussed in the following subsections. 
\\ For the considered case
$c_{\tilde f_{L/R}}<c_{f_{L/R}}$, the Yukawa-like couplings of these last generation sfermions to Higgs bosons [first coupling matrix in Eq.(\ref{HCouplMat})] are increased 
-- modulo a square and a Higgs rotation angle/VEV -- relatively to the effective 4D fermion 
Yukawa couplings [see Eq.(\ref{equiv4DyukF})] which are themselves equal to the 4D SUSY scalar Yukawa couplings (forgetting small KK corrections).  
A same comparison holds for the interactions proportional to $\mu_{eff}$ [second coupling matrix of Eq.(\ref{HCouplMat})].  
Since the RS SUSY scenario considered increases globally sfermion couplings to Higgs bosons compared to the usual 4D SUSY case, 
which brings new significant contributions to the (s)quark triangular loop of the gluon-gluon fusion mechanism for Higgs production 
at hadron colliders 
[= main production channel] \cite{AbdPLB} \footnote{KK contributions to this gluon-gluon fusion mechanism have been studied e.g. in 
Ref.~\cite{RSfusion}.},
one might wonder whether such a model is not excluded by present experimental searches \cite{SusyHiggsTev} at Tevatron Run II for 4D SUSY Higgs fields
(or estimated results from SM Higgs searches \cite{SmHiggsTev} e.g. in the decoupling limit).  
First it must be remarked that the Tevatron production cross sections and decay branching ratios as well as various theoretical uncertainties have been recently 
re-evaluated for SUSY Higgs bosons \cite{SmHiggsErr}, also that 4D SUSY Higgs searches
were not exhaustive in the parameter space exploration \cite{SusyHiggsTev} and that other effects could appear in 5D SUSY Higgs productions.
Secondly, the obtained lower limit of the lightest neutral Higgs (noted $h$) 
mass is not so close to the 4D SUSY theoretical upper bound which furthermore can be enhanced in the 5D SUSY 
context \cite{Gautam} enlarging the allowed mass range.   
In fact, the main 4D SUSY contribution to the gluon-fusion mechanism is the stop exchange in the loop, due to its large Yukawa coupling.
In the above RS SUSY setup, the only additional new contribution is the sbottom exchange as we assume the decoupling of the first two generations 
of squarks to avoid constraints from the $K^0-\bar K^0$ mixing (one could even assume the decoupling of the squark doublet $\tilde Q_L$ and sbottom singlet $\tilde b_R$).
Since we take masses $m_{\tilde b} \sim 10^2$ GeV (from little hierarchy arguments), the sbottom Yukawa couplings are of the same order as the (s)top
Yukawa interactions and hence comparable Higgs production rates are expected in 4D versus RS SUSY, which is realistic. Concerning sleptons, we 
assume that only the first generation decouples since these non-colored scalar fields only affect 4D SUSY Higgs searches through the radiative Higgs decay 
into two photons, a decay representing only one of the various channels investigated. Going into the details of the calculation of the Higgs production/decays, 
different sfermion exchanges might suppress each other by destructive interferences of the triangular loops \cite{AbdPLB} 
depending on the signs of various 4D effective couplings over parameter ranges, and some more freedom might even arise from 5D SUSY model building.

Concerning bulk gauginos, their 4D soft masses are taken effectively i.e. without specifying the higher-dimensional geometry  
(as we do not study possible 4D/5D SUSY differences in that sector). 
For example these masses could be localized on the TeV-brane -- without suffering from large overlap suppression like
light generations.

\subsubsection*{Charge and color breaking minima}

Before presenting numerical results, we need to discuss another SUSY aspect which is significantly
modified in this RS context:
the Charge and Color Breaking (CCB) minima. 
We will show that no drastic constraints arise on the parameter space and 
even that no SUGRA-like scenario -- in the sense where trilinear soft scalar terms are proportional to the Yukawa
coupling constants (reducing effectively the trilinear scale $A$) -- needs to be assumed [as usually required in 4D SUSY] 
in order to satisfy those constraints.

In 4D mSUGRA, the constraints coming from imposing the absence of CCB global minima for the potential of squark/slepton  
VEV's read typically as \cite{CCB} (see \cite{NMSSMccb} for the NMSSM case), 
taking the example of the usually dangerous selectron direction, 
\begin{eqnarray}
({\cal Y}_{4D}^e A^e)^2 < 3 \Big ( (\tilde m^e_R)^2  + (\tilde m^e_L)^2 + \hat m_d^2 \Big ) ({\cal Y}_{4D}^e)^2 
\label{4DCCB} 
\end{eqnarray}
where $\tilde m^e_L$, $\tilde m^e_R$ denote the selectron soft masses and $\hat m_d^2=m_{H_d}^2+\mu^2$
with $m_{H_d}$ being the down Higgs soft mass. ${\cal Y}_{4D}^e \sim 10^{-5}$ 
is the Yukawa coupling constant for the electron and $A^e \sim 10^2$ GeV. 
Since the Yukawa couplings simplify each other in the above inequality, there are often not written in 
the literature, but here we keep them for the 5D discussion below. The general conclusion on the 4D case is that  
the CCB constraints, as illustrates the one above, 
remain respected if the soft parameters $A$, $\tilde m$, $m_{H_d}$ and also $\mu$ 
are all of order the EWSB scale (which is compatible with both the Higgs fine-tuning considerations 
and electroweak potential minimization relations).

For mSUGRA models in a warped background, the CCB constraint (\ref{4DCCB}) is replaced by (neglecting the KK
mixing corrections), 
\begin{eqnarray}
({\cal Y}^e A^e e^{-\sigma(\pi R_c)} f_{L}^{0} f_{R}^{0})^2 
< 3 \Big ( \tilde m^e_R (f_{R}^{0} e^{-\sigma(\pi R_c)})^2  + \tilde m^e_L (f_{L}^{0} e^{-\sigma(\pi R_c)})^2 
+ \hat m_d\vert_{eff}^2 \Big ) ({\cal Y}^e f_{L}^{0} f_{R}^{0})^2
\label{RSCCB} 
\end{eqnarray}
as deduced from the form of 4D Higgs couplings (\ref{HCouplMat}) -- of type ${\cal Y}^2$ and $A$ -- and the form
of 4D soft scalar masses on TeV-brane in Eq.(\ref{WholeMassMat}). 
Here $f_{L/R}^{0}=f^{++}_0(c_{\tilde e_{L/R}};\pi R_c)$; $A^{e} {\cal Y}^{e} = {\cal O}(1)$ 
(as discussed in the next subsection on slepton masses in mSUGRA) and $\hat m_d\vert_{eff}^2=m_{H_d}^2+\mu_{eff}^2$
[see the discussion on $\mu_{eff}$ in Section \ref{model}]. Now, the RS CCB criteria of type 
Eq.(\ref{RSCCB}) simplifies to
\begin{eqnarray}
( A^e e^{-\sigma(\pi R_c)} )^2 
< 3 \Big ( \tilde m^e_R (f_{R}^{0} e^{-\sigma(\pi R_c)})^2  + \tilde m^e_L (f_{L}^{0} e^{-\sigma(\pi R_c)})^2 
+ \hat m_d\vert_{eff}^2 \Big ) 
\label{RSCCBsimple} 
\end{eqnarray}
where $A^e e^{-\sigma(\pi R_c)}\sim 10^2$ GeV so that this condition is as natural
as in 4D mSUGRA for effective soft masses at the EWSB scale, which is the case since $\hat m_d\vert_{eff}\sim 10^2$ GeV 
and e.g. $\tilde m_R (f_{R}^{0} e^{-\sigma(\pi R_c)})^2 \sim k^2 e^{-2\sigma(\pi R_c)} \sim (10^2$~GeV$)^2$
[$\tilde m_R \sim k$ from discussion below Eq.(\ref{WholeMassMat}) and see Eq.(\ref{ZeroWaveFerm})
for the wave function order of magnitude] as confirmed by all the values obtained in next subsections.

In an RS framework with a SUSY breaking not specifically of the type SUGRA, the CCB constraint of 
Eq.(\ref{RSCCB}) simply becomes 
\begin{eqnarray}
(A^e e^{-\sigma(\pi R_c)} f_{L}^{0} f_{R}^{0})^2 
< 3 \Big ( \tilde m^e_R (f_{R}^{0} e^{-\sigma(\pi R_c)})^2  + \tilde m^e_L (f_{L}^{0} e^{-\sigma(\pi R_c)})^2 
+ \hat m_d\vert_{eff}^2 \Big ) ({\cal Y}^e f_{L}^{0} f_{R}^{0})^2
\label{RSCCBgeneric} 
\end{eqnarray}
with now $A^e = {\cal O}(1)$ being dimensionless [{\it c.f.} Eq.(\ref{WholeMassMat})]. After simplification, it reads as
\begin{eqnarray}
(A^e e^{-\sigma(\pi R_c)})^2 
< 3 \Big ( \tilde m^e_R (f_{R}^{0} e^{-\sigma(\pi R_c)})^2  + \tilde m^e_L (f_{L}^{0} e^{-\sigma(\pi R_c)})^2 
+ \hat m_d\vert_{eff}^2 \Big ) ({\cal Y}^e)^2 \ ,
\label{RSCCBGenSimple} 
\end{eqnarray}
a condition which is also systematically fulfilled in orders of magnitudes since 
$({\cal Y}^e)^{-1} e^{-\sigma(\pi R_c)} \sim k e^{-\sigma(\pi R_c)}\sim 10^2$ GeV.
Hence, we conclude that within RS SUSY the CCB constraint is generically satisfied,  
with respect to the orders of magnitude, even without assuming a SUGRA-like breaking.

Strictly speaking, the CCB induced conditions must be imposed at the energy scale $Q \sim \langle~\tilde f~\rangle$ but the
running of soft parameters in the RS SUSY framework is beyond our scope.

There exists a related kind of bound which originates from forbidding true minima along
scalar potential directions Unbounded From Below (UFB) \cite{UFB} ({\it c.f.} \cite{NMSSMccb} for the NMSSM). 
The condition for such minima not
to be deeper than the standard EWSB minimum implies the following bound, $m_0/m_{1/2} \ > \ {\cal O}(1)$ 
(in case one assumes universal soft terms). This typical UFB bound should also be quite easily satisfied 
in an RS SUSY context, applying it for simplicity on the 4D effective soft masses.

Finally, one must recall that the actual relevance of these CCB and UFB bounds is not entirely evident in general
because even if e.g. an existing CCB minimum is deeper than the standard EWSB one, it can be acceptable if the 
tunneling rate out of the standard minimum is small relatively to the age of the universe. The various analyses in 
literature lead to the conclusion that these tunneling rates are often quite small (the original paper
is the third one of Ref.~\cite{CCB}). The relevance of 
the CCB and UFB bounds
is thus model-dependent as it depends on cosmology, and in particular on which minimum we drop after inflation 
\cite{COSMOccb}.

\subsubsection*{Numerical results for stop masses: a first discrimination test}

First, we investigate the heavy quark superpartner (namely the stop denoted $\tilde{t}$) sector. The flavor context of the pMSSM, considered in this 
paper, will be reminded in Section \ref{Higgspart} where all flavors are potentially involved.
In the present analysis, we propose a test to discriminate the 4D pMSSM w.r.t. 5D warped pMSSM by looking at the different ways to generate stop masses in these two setups. To be general, we consider here within 4D and 5D SUSY the case where the soft breaking parameters are chosen effectively, only constrained from the experimental data. 

\begin{figure}[!hc]
\centering	
\vspace*{1.5cm}	
\includegraphics[width=0.43\textwidth]{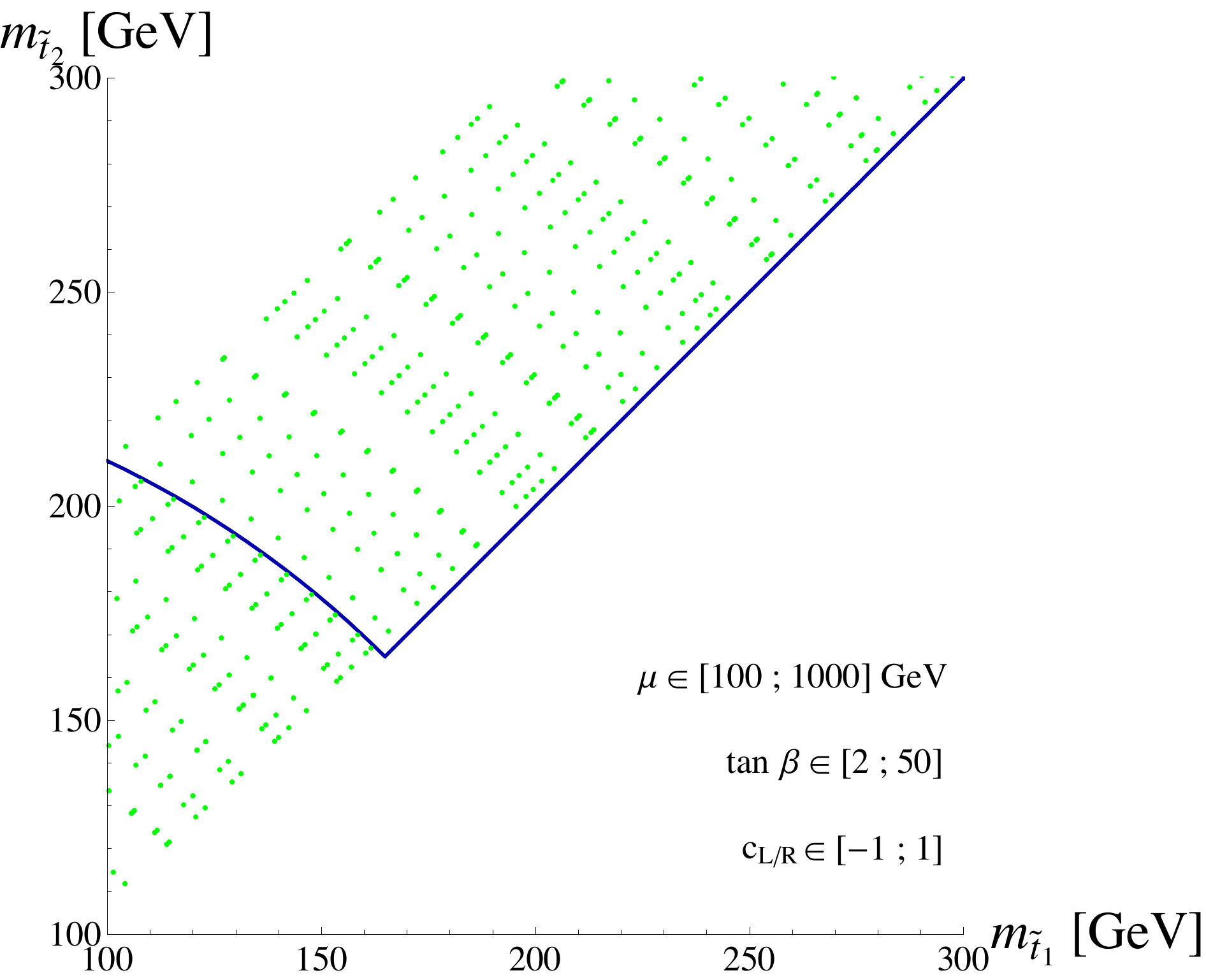}
\hspace*{.5cm}
\includegraphics[width=0.43\textwidth]{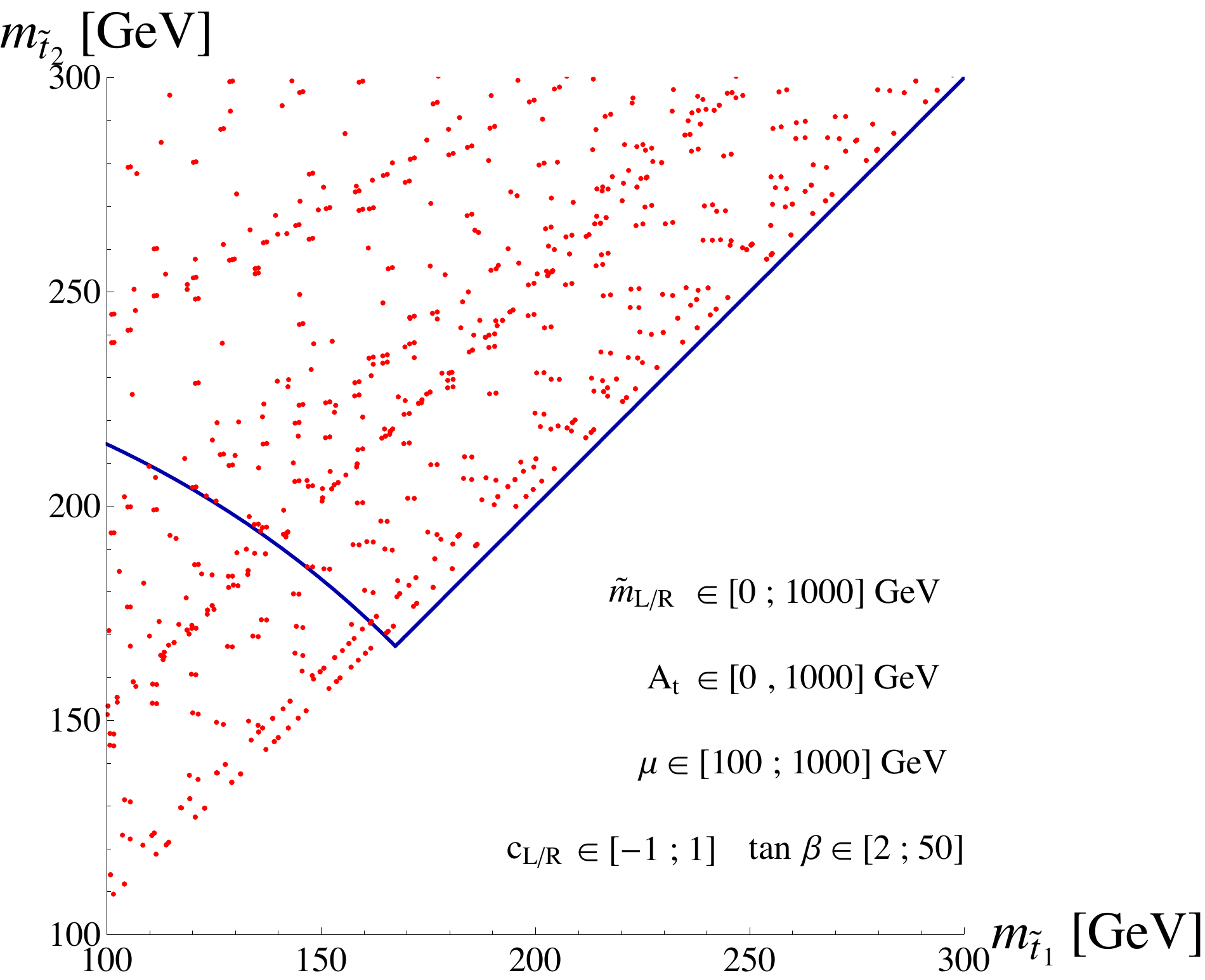}
\caption{\label{mst1mst2} \small{Allowed regions in the plan $m_{\tilde t_2}$ versus $m_{\tilde t_1}$ (in GeV)
within 4D SUSY [demarcated by the blue lines] and RS SUSY [green/red points]. 
The plot on the left side corresponds to the situation where there are no more SUSY breaking terms on the TeV-brane (for the 5D case). 
The thick blue lines represent a lower limit on $m_{\tilde t_2}$ obtained analytically. The parameters are scanned in the intervals indicated on the plot.  
$c_{L/R}=c_{\tilde t_{L/R}}$ are the 5D stop parameters. In the RS case, the so-called effective parameter 
$\mu_{eff}$ is the equivalent of the $\mu$ parameter in 4D SUSY
[{\it c.f.} Eq.(\ref{mueff})]. The interval indicated on 
$\tilde m$ ($A_t$) corresponds to the scan interval for the
soft stop masses $\tilde m_{L/R}$ (trilinear coupling $A_t$) in 4D SUSY and $\tilde{m}_{L/R}|_{eff} = \sqrt{\tilde m_{L/R}ke^{-2k\pi R_c}}$ 
($A_t |_{eff} = A_t ke^{-k\pi R_c}$) in the RS SUSY case.}}
\end{figure}

\begin{figure}[!hc]
\centering	
\vspace*{1.5cm}	
\includegraphics[width=0.43\textwidth]{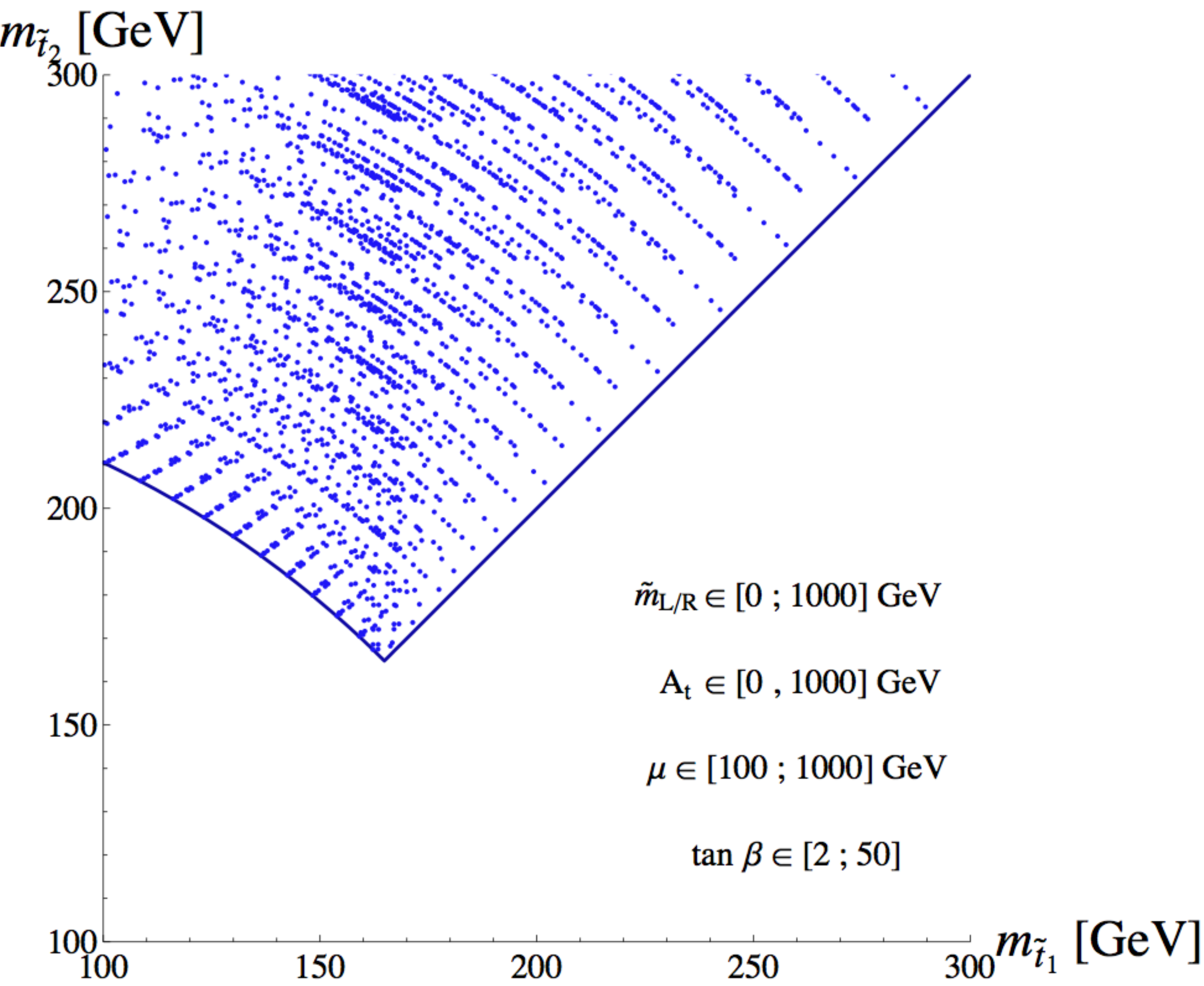}
\caption{\label{mst1mst2MSSM} \small{Allowed regions in the plan $m_{\tilde t_2}$ versus $m_{\tilde t_1}$ (in GeV)
within 4D SUSY. These regions have been obtained for parameters scanned in the intervals indicated on the plot itself.}}
\end{figure}

In Fig.(\ref{mst1mst2}), we have represented the domains possibly explored in 4D or warped SUSY in the plan of the two stop mass eigenstates. 
The blue lines represent the lower limit on $m_{\tilde t_2}$ obtained in 4D SUSY (see discussion below).
The two plots showed correspond to scans of the fundamental parameters in the case where SUSY breaking occurs in the bulk (see Appendix \ref{SolutionFreeScal}), 
and for the right plot, soft SUSY breaking terms (stop masses and trilinear couplings) have been added on the TeV-brane.
For the left plot, the only parameters entering the scan are the SUSY parameters $\tan\beta$, $\mu$ (or $\mu_{eff}$ [see Eq.(\ref{mueff})]), the 5D 
stop parameters $c_{\tilde t_{L/R}}$ and the soft masses 
$\tilde m_{L/R}$ (for 4D SUSY), whereas for the right scan we have further considered the 4D trilinear coupling $A_t$ as well as the TeV-brane SUSY breaking parameters 
$\tilde{m}_{L/R}|_{eff}^2 = \tilde{m}_{L/R}ke^{-2k\pi R_c}$ and $A_t|_{eff} = A_tke^{-k\pi R_c}$ appearing in the stop mass matrix of warped SUSY models (as seen
from Eq.(\ref{WholeMassMat}) and Eq.(\ref{ZeroWaveFerm})). The $A_t|_{eff}$ term is localized on the TeV-brane due to the Higgs localization.
In case of the absence of soft stop mass terms on the TeV-brane (left plot), the stop mass is generated in particular
by its coupling with the Higgs boson -- which can itself be increased through the stop-Higgs wave function overlap controlled by the SUSY breaking stop mass term in the bulk. 
\\ÊThe parameters are scanned over the following ranges, $\tan\beta \in [2;50]$, $\mu, \mu_{eff} \in [100;1000]$ GeV, $c_{\tilde t_{L/R}} \in [-1,1]$, 
$\tilde{m}_{L/R}, \tilde{m}_{L/R}|_{eff} \in [0;1000]$ GeV and $A_t, A_t|_{eff} \in [0;1000]$ GeV. The interval on $\tan\beta$ is conservative given present constraints 
\cite{DjouadiReviewII,TevWebPage}. The $\mu$ or $\mu_{eff}$ ($|c_{\tilde t_{L/R}}|$) 
interval is justified by its required order of magnitude at the EWSB scale (around $\sim 1$) as discussed in Section \ref{model}. 
The $\tilde{m}_{L/R}$ range is motivated by the usual effective scale of 4D SUSY breaking scenarios limiting the Higgs mass fine-tuning and
the $\tilde{m}_{L/R}|_{eff}$ range is motivated by the energy scale orders in RS: $\tilde{m}_{L/R}|_{eff} = (\tilde{m}_{L/R}ke^{-2k\pi R_c})^{1/2} 
\sim (k^2e^{-2k\pi R_c})^{1/2} \sim 10^2$ GeV [see discussion below Eq.(\ref{mueff})]. Finally, The chosen $A_t$ interval is based on usual 4D SUSY breaking scenarios and
the considered $A_t|_{eff}$ range comes from the scales characteristic of RS: $A_t|_{eff} = A_tke^{-k\pi R_c}
\sim 1 \times ke^{-k\pi R_c} \sim 10^2$ GeV [see again below Eq.(\ref{mueff})].  
In order to have a consistent comparison between 4D and 5D models, we have chosen the same ranges in both setups. 
The scans in the plots are not performed on negative values of the $A_t$
(or $A_t|_{eff}$) neither of the $\mu$ (or $\mu_{eff}$) terms but taking the opposite signs does not change the conclusions.
The last remark on the scans is that having chosen slightly larger parameter ranges -- but remaining with the same orders of magnitude as above 
since those are physically motivated -- would have not affected significantly the obtained numerical results presented here.

Let us now explain the differences arising between the 4D and 5D models in plots of Fig.(\ref{mst1mst2}). For that purpose, we need to start from
the stop mass matrix structure. The stop mass matrix within the RS scenario has been derived in Eq.(\ref{WholeMassMat}). Moving to
4D SUSY, we recall the general form of the stop mass matrix in the $\{ \tilde{t}_L , \tilde{t}_R \}$ basis,
\begin{eqnarray}
\label{eq:4Dstopmassmatrix}
\mathcal{M}_{\tilde{t}\tilde{t}}^2\vert_{\mbox{4D SUSY}} &=& \left(\begin{array}{cc}
{ m^2_{t} + Q_Z^{t_L} \cos2\beta\,m_Z^2 + \tilde{m}_L^2 } & { A_{t} - \frac{\mu\,m_{t}}{\tan\beta} } \\
{ A_{t} - \frac{\mu\,m_{t}}{\tan\beta} } & { m^2_{t} - Q_Z^{t_R} \cos2\beta\,m_Z^2 + \tilde{m}_R^2 }
\end{array}\right)
\end{eqnarray}
where $Q_Z^{t_L} \equiv {1\over2}-{2\over3}\sin^2\theta_W$, $Q_Z^{t_R} \equiv -{2\over3}\sin^2\theta_W$ and $m_t$ is the top quark mass. 
One can note at this level that the SM top mass entering this mass matrix (and taken in agreement with 
recent Tevatron measurements \cite{EWWGwebPage}) is larger than the experimental lower bound on the stop mass, $m_{\tilde t_1}>95.7$ GeV 
\cite{PDG,TevWebPage}. 
The mass matrix (\ref{eq:4Dstopmassmatrix}) can be diagonalized by $2\times2$ orthogonal matrices, resulting in the following mass eigenvalues for the stop 
eigenstates $\tilde{t}_{1,2}$,
\begin{eqnarray}
\label{eq:stopeigenmass}
m_{\tilde{t}_{1,2}}^2 &=& {1\over2}\Bigg( 2m_t^2 + {1\over2}\cos2\beta\,m_Z^2 + \tilde{m}_L^2 + \tilde{m}_R^2
\nonumber\\
&& \,\, \mp \sqrt{  \Big( (Q_Z^{t_L}+Q_Z^{t_R})\cos2\beta\,m_Z^2+\tilde{m}_L^2-\tilde{m}_R^2 \Big)^2 + 4\left(A_t - \frac{\mu\,m_t}{\tan\beta}\right)^2 } \,\Bigg) .
\end{eqnarray}

For 4D SUSY, in Fig.(\ref{mst1mst2}), we observe that $m_{\tilde{t}_2}$ has a lower limit (blue line) depending on $m_{\tilde{t}_1}$: this limit is due to the structure of
matrix (\ref{eq:4Dstopmassmatrix}) and it has thus been possible to obtain it analytically. 
Indeed, from Eq.(\ref{eq:stopeigenmass}), summing the two squared masses, we get
\begin{eqnarray}
\label{eq:summstopmass}
m_{\tilde{t}_{2}}^2 + m_{\tilde{t}_{1}}^2 = 2m_t^2 + {1\over2}\cos2\beta\,m_Z^2 + \tilde{m}_L^2 + \tilde{m}_R^2 .
\end{eqnarray}
For arbitrary large soft masses, $\tilde{m}_L^2$ and $\tilde{m}_R^2$, there is no constraint on how high can $m_{\tilde{t}_2}$ be compared to $m_{\tilde{t}_1}$ 
as illustrated in Fig.(\ref{mst1mst2}). We are thus interested in how low can $m_{\tilde{t}_2}$ be w.r.t. $m_{\tilde{t}_1}$ in the low soft mass region (where it appears 
on the plot to exist a 
non-trivial lower limit on $m_{\tilde{t}_2}$). Obviously, one must have $m_{\tilde{t}_2} \geq m_{\tilde{t}_1} \gtrsim 100 \mbox{ GeV}$ from  
definition together with the experimental constraint on stop searches mentioned above. 
The $m_{\tilde{t}_2}$ lowest value thus corresponds to $\tilde{m}_{L/R} = 0$ together with a vanishing $\tilde{t}_{L}-\tilde{t}_{R}$ mixing, resulting in $m_{\tilde{t}_2}^{lowest} = m_{\tilde{t}_1} = \sqrt{m_t^2 + {1\over4}\cos2\beta\,m_Z^2} \sim m_t$. When there is a non-vanishing mixing, for $m_{\tilde{t}_1} \leq m_{\tilde{t}_2}^{lowest}$, we have from Eq.(\ref{eq:summstopmass})
\begin{eqnarray}
\label{eq:blueline}
m_{\tilde{t}_{2}} \geq \sqrt{2m_t^2 + {1\over2}\cos2\beta\,m_Z^2 - m_{\tilde{t}_{1}}^2}
\quad\mbox{for}\quad
m_{\tilde{t}_1} \leq \sqrt{m_t^2 + {1\over4}\cos2\beta\,m_Z^2}
\end{eqnarray}
which corresponds to the first branch of the blue line in both plots of Fig.(\ref{mst1mst2}) -- 
fitting perfectly the lower limit of the 4D SUSY scan point domain that we have also generated for checking [it is shown on Fig.(\ref{mst1mst2MSSM})]. 
The second branch, as mentioned above, simply reveals the constraint
\begin{eqnarray}
m_{\tilde{t}_2} \geq m_{\tilde{t}_1}
\quad\mbox{for}\quad
m_{\tilde{t}_1} \geq \sqrt{m_t^2 + {1\over4}\cos2\beta \, m_Z^2}.
\end{eqnarray}
The minimum $m_{\tilde{t}_2}$ value is thus
\begin{eqnarray}
m_{\tilde{t}_2}^{lowest} = m_{\tilde{t}_1} 
\quad\mbox{for}\quad
m_{\tilde{t}_1} = \sqrt{m_t^2 + {1\over4}\cos2\beta \, m_Z^2} \ \sim \ m_t.
\label{min4Dmstop}
\end{eqnarray}
Note that this result is a quite general result for 4D SUSY models.

Having determined explicitly the 4D SUSY structural limit on $m_{\tilde{t}_2}$ as a function of $m_{\tilde{t}_1}$, 
we now turn our attention to the results in the context of RS SUSY shown in Fig.(\ref{mst1mst2}). Looking at those two plots, we observe an important difference between 
the 4D and 5D SUSY models; $m_{\tilde{t}_2}$ can now reach smaller values (for a given $m_{\tilde{t}_1}$ value) than those in 4D SUSY [limited by the blue line] and is now only constrained by 
\begin{eqnarray}
m_{\tilde{t}_2} \geq m_{\tilde{t}_1} \gtrsim 100 \mbox{ GeV}.
\label{min5Dmstop}
\end{eqnarray}
This difference with the 4D case can be understood as follows. The 4D SUSY conservative constraint of Eq.(\ref{min4Dmstop}) is changed in RS to the global constraint  
$m_{\tilde{t}_2} \gtrsim {\cal Y} \hat v_u f^{++}_0(c_{\tilde t_L};\pi R_c) f^{++}_0(c_{\tilde t_R};\pi R_c)$ as deduced from Eq.(\ref{WholeMassMat}). If there were no SUSY breaking in the bulk, one would have $c_{\tilde t_{L/R}} = c_{t_{L/R}}$ and by consequence ${\cal Y} \hat v_u f^{++}_0(c_{\tilde t_L};\pi R_c) f^{++}_0(c_{\tilde t_R};\pi R_c) = m_t$ 
[neglecting the KK (s)top mixing effect as already mentioned] so that the constraint (\ref{min4Dmstop}) would be recovered. However, when SUSY is broken in the bulk, 
$c_{\tilde t_{L/R}}$ are now effectively free parameters that can lead to either larger or smaller values of ${\cal Y} \hat v_u f^{++}_0(c_{\tilde t_L};\pi R_c) f^{++}_0
(c_{\tilde t_R};\pi R_c)$ relatively to $m_t$. In particular, the lower limit in Eq.(\ref{eq:blueline}) materialized by the first branch in plots is relaxed down to the bound 
(\ref{min5Dmstop}) 
in RS SUSY, thus explaining the RS scanned points going beyond the blue line. 

Hence a measurement of $m_{\tilde{t}_2}$ at high energy colliders in the lower-left region of the RS scans shown here (below the blue line) 
would greatly disfavor 4D SUSY models in their minimal form while constituting possible signatures of RS SUSY scenarios. Such a discrimination test should be already possible 
before a future precision ILC physics, at LHC, given the experimental accuracies expected on squark mass reconstructions. Indeed, an uncertainty of $\sim 5\% - 10\%$ on stop
masses should be reachable at LHC \cite{TDRatlas} which is clearly sufficient for the potential
test suggested here, given the large stop mass deviations in RS with respect to the pure 4D SUSY scenario 
shown in Fig.(\ref{mst1mst2}). 
This typically expected accuracy of $\sim 5\% - 10\%$ on stop masses corresponds to illustrative examples studied in Ref.~\cite{TDRatlas} 
but of course the exact performance would rely on effectively observed events at LHC as well as on the realized SUSY model/parameters chosen by nature.
Note finally that for the values taken throughout this paper, $m_{\tilde \nu_1}>120$ GeV and $m_{\tilde \chi^0_1}>150$ GeV, 
the conservative experimental lower bound on the stop mass is exactly $m_{\tilde t_1}>95.7$ GeV \cite{PDG,TevWebPage} 
so that the region only accessible in RS SUSY on Fig.(\ref{mst1mst2}) is not yet excluded and could be 
revealed by an LHC discovery.

The last comment -- which is interesting for understanding the formal RS SUSY construction and will prove to be useful for Section \ref{stopart} -- 
is about the difference between the two plots of Fig.(\ref{mst1mst2}).
In the first scenario (left plot) where breaking only occurs in the bulk, while, as discussed, $m_{\tilde{t}_2}$ can reach any values $\gtrsim 100 \mbox{ GeV}$, on the other side we 
observe that the scan reaches a maximum value of $m_{\tilde{t}_2}$ depending on $m_{\tilde{t}_1}$. This is due to the chosen usual range for the $c_{\tilde t_{L/R}}$ parameters 
spanning from $-1$ to $+1$, together with the absence of soft mass terms on the TeV-brane. Indeed, in the absence of such mass terms, high values of $m_{\tilde{t}_2}$ are
limited by the diagonal Yukawa-type mass contributions and thus by the largest allowed overlap of the stop wave functions with the Higgs boson, an overlap being limited from
above by the minimal $c_{\tilde t_{L/R}}$ values. In contrast, the scan is not limited from above for the second RS scenario (right plot) due to the presence of soft mass terms
on the TeV-brane (and even of trilinear scalar couplings).

\subsubsection*{Numerical results for smuon masses in mSUGRA}

In this part, we consider within 4D SUSY the subcase of the mSUGRA scenario where the soft breaking parameters have universal values at the Grand 
Unified Theory (GUT) scale,
$m_0$, $m_{1/2}$, $A_0$, the whole trilinear scalar couplings $A_0 {\cal Y}$ being proportional to Yukawa coupling constants ${\cal Y}$. 
The two other input (SUSY) parameters are $\tan\beta$ and $sign(\mu)$. The soft masses $\tilde{m}_{L/R}$ will be run here 
\footnote{We use the standard RGE evolution to the weak scale at one-loop for the studied smuon masses.} from their GUT value $m_0$
down to low-energy, where we study collider physics, and the obtained values included into the mass matrix form (\ref{eq:4Dstopmassmatrix}).  
To adopt a general approach, we will take the low-energy $\vert A \vert$ parameter to be between zero and the TeV scale. This maximum scale is justified, 
in 4D mSUGRA, by the order of effective SUSY breaking scales and by the condition of absence of CCB minima [see typically Eq.(\ref{4DCCB})].

In a minimal RS version of the mSUGRA scenario, the set of parameters entering the scalar mass matrices can be taken as follows.
First, the parameter $\mu_{eff}$ (see Eq.(\ref{mueff})), which enters the mass matrix form (\ref{WholeMassMat}), is taken at the $\mu$ value one gets in 4D mSUGRA
since it is the equivalent parameter. The possible kind of differences arising between 4D and 5D models in the running necessary to derive the $\mu$ parameter 
is not studied here; this is a potential source of additional differences that could lead to new tests for distinguishing between 4D and 5D SUSY. Similarly, 
for the diagonal scalar mass matrix elements, we take the same values as the ones obtained from the 4D mSUGRA running, motivated by the fact that we 
will focus on differences between 4D and 5D mSUGRA arising in the off-diagonal mass matrix elements. Last but not least, in a mSUGRA like scenario, one 
would have a new [compared to Eq.(\ref{WholeMassMat})] trilinear coupling constant 
$A {\cal Y} = {\cal O}(1)$ with $A \sim (1/{\cal Y}) \sim k$ to not introduce new scales. The quantity $A\vert_{eff}=A {\cal Y} k e^{-k\pi R_c}\sim$ TeV ($A=0$
could also be an acceptable scale) appearing now in
the $A$ terms of Eq.(\ref{WholeMassMat}), namely in $A {\cal Y} e^{-k\pi R_c} \hat v_u f^0_Lf^0_R$, is taken equal numerically to the 4D SUSY breaking parameter 
$A$ introduced just above, neglecting once again possible differences arising in the 5D running.

Let us now emphasize 4D versus 5D differences in the smuon mass matrix (studying the example of the smuon is motivated by 
the theoretical framework described above and the experimental performances discussed below) within this context.
\\ In 4D mSUGRA, the off-diagonal entry to squared smuon mass $A_{\mu} {\cal Y}_{\mu} \hat v_d$, driven by the muon mass ($m_\mu$), is typically around $10^2$ GeV$^2$ at most  
which leads to a $\bar {\tilde \mu}_L \tilde \mu_R$ smuon mass term induced by the Higgs VEV much smaller than the diagonal soft mass
terms $\tilde m^2_{L} \bar {\tilde \mu}_L \tilde \mu_L$ and $\tilde m^2_{R} \bar {\tilde \mu}_R \tilde \mu_R$ [see the matrix form (\ref{eq:4Dstopmassmatrix})]. 
Indeed, one should typically have 
$\tilde m^2_{L,R} \gtrsim 10^4$ GeV$^2$ so that smuon mass eigenvalues are not excluded by the conservative current experimental 
lower bound $m_{\tilde \mu_1}>94$ GeV \cite{PDG,TevWebPage}. 
\\ The other off-diagonal $\bar {\tilde \mu}_L \tilde \mu_R$ contribution to the smuon mass is at most $m_\mu \mu \tan\beta \sim 5 \ 10^3$ GeV$^2$,  
for extremely optimized parameter values, so that this contribution remains also systematically 
smaller than $\tilde m^2_{L,R} \gtrsim 10^4$ GeV$^2$. Hence the two Left-Right mass mixing terms are limited in 4D mSUGRA,
relatively to the complete diagonal elements of the $2\times 2$ smuon squared mass matrix.

In the RS version of mSUGRA, one has a trilinear-induced mass term $A_{\mu} {\cal Y}_{\mu} e^{-k\pi R_c} \hat v_d f^0_Lf^0_R$ where the $f^0_{L/R}$ wave function 
values at $y=\pi R_c$ are not constrained from above by the muon mass because $f_{L/R}^{0}=f^{++}_0(c_{\tilde \mu_{L/R}};\pi R_c)$ whereas the
muon mass is controlled by independent fermionic 5D parameters $c_{\mu_{L/R}}$ in the present SUSY breaking scheme.    
\\ Similarly, the off-diagonal Left-Right smuon squared mass $\mu_{eff} {\cal Y}_{\mu} \hat v_d \tan\beta f^0_Lf^0_R$ of Eq.(\ref{WholeMassMat}) can take benefit from large factors from the
scalar wave function overlaps with the Higgs brane,
while $\tan\beta$ takes comparable values as in 4D SUSY and ${\cal Y}_{\mu} \sim 1/k$ is compensated by $f^0_Lf^0_R \propto k$ ({\it c.f.} Eq.(\ref{ZeroWaveFerm})).
Therefore, in contrast to the 4D mSUGRA case, the off-diagonal mixing smuon mass terms are not constrained by the muon mass and can thus get  
higher values.

\begin{figure}[!hc]
\centering
\vspace*{1.5cm}	
\includegraphics[width=0.47\textwidth]{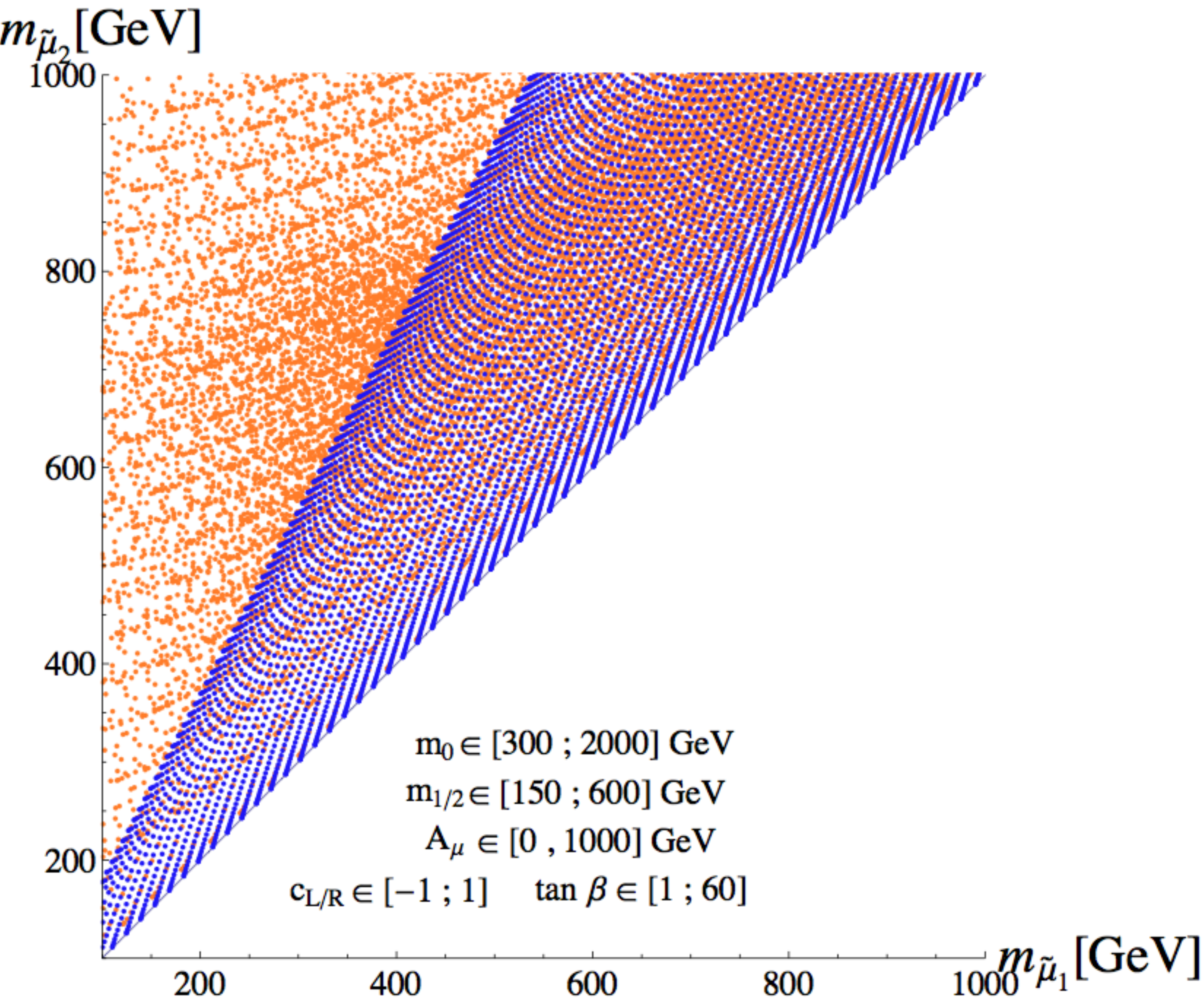}
\caption{\label{SUGRA} \small{Points obtained in the plan $m_{\tilde \mu_2}$ versus $m_{\tilde \mu_1}$ (in GeV), for an mSUGRA type scenario  
within 4D SUSY [blue points] and RS SUSY [orange points], from a scan performed in the intervals indicated on the plot.   
The interval indicated on $A_\mu$ corresponds to the range for the soft 
trilinear coupling $A_\mu$ in 4D SUSY and for the $A_\mu|_{eff}$ effective dimension-one parameter defined 
in RS SUSY (see text). $c_{L/R}=c_{\tilde \mu_{L/R}}$ denote now the 5D smuon parameters.}}
\end{figure}

In consequence, the Left-Right mass mixing for smuon masses can reach higher amounts within RS mSUGRA than in 4D mSUGRA.
In turn (having a universal mass $m_0$ tends to have a configuration with identical diagonal squared masses), 
the splitting between the mass eigenvalues $m_{\tilde \mu_1}$ and $m_{\tilde \mu_2}$ can be much larger as illustrates Fig.(\ref{SUGRA}).
\\ In order to obtain this figure, we have scanned the input parameters over these ranges: $\tan\beta \in [1;60]$, $m_0 \in [300;2000]$ GeV, $m_{1/2} \in [150;600]$ GeV,
$c_{\tilde t_{L/R}} \in [-1,1]$ and $A_\mu, A_\mu|_{eff} \in [0;1000]$ GeV (see above discussion). The interval on $\tan\beta$ corresponds to the domain allowed
by the Higgs potential minimization conditions in mSUGRA \cite{DjouadiReviewII}. The choice of $m_0$ and $m_{1/2}$ ranges is based on the constraints coming
from SUSY searches at colliders combined with the requirement that the Lightest Supersymmetric Particle (LSP) is not the tau lepton superpartner (the stau) 
\cite{TevWebPage,firstLHCexclusion}; taking higher $m_0$ and $m_{1/2}$ values does not modify significantly the scan presented in Fig.(\ref{SUGRA}). 
The output smuon mass ranges (i.e. the shown plot domain in Fig.(\ref{SUGRA})) is motivated by the order of magnitude of usual effective 4D SUSY breaking scale 
-- near the EWSB scale -- allowing to protect the Higgs boson against too dramatic mass fine-tuning. 
The scans here are not done for negative values of the $A_\mu$
(or $A_\mu|_{eff}$) terms and the $\mu$ (or $\mu_{eff}$) terms are neither taken negatively but choosing the opposite signs would not change the present conclusions.
Besides, only values in the range $[100;1000]$ GeV have been kept for the $\vert \mu \vert$ (or $\vert \mu_{eff} \vert$) quantity; 
this is imposed by the orders of experimental bounds and the Higgs potential minimization conditions already mentioned. 
The last remark, as before, is that having chosen slightly larger parameter ranges -- staying with the same orders of magnitude  
physically motivated -- would have not affected significantly the obtained numerical results shown in this part. 
\\ In conclusion, if a $m_{\tilde \mu_2}$ measurement is obtained at ILC or even at LHC at a value larger -- including the
experimental uncertainty in this comparison -- than the 4D SUSY upper limit appearing on Fig.(\ref{SUGRA})
for a given $m_{\tilde \mu_1}$ value (assumed to be measured also),
this result would rule out the 4D mSUGRA scenario [at least in its simplest form] and constitute a good indication for an RS mSUGRA model.  
Indeed all $m_{\tilde \mu_2}$ values above the upper 4D mSUGRA limit can be reached within RS mSUGRA as shown by the scan over parameters.   
Such a discrimination should be already possible at LHC given the accuracies expected on slepton mass reconstructions:
a $\sim 5\% - 10\%$ uncertainty is reasonable to expect \cite{TDRatlas} and clearly sufficient in a large part of parameter space (at high $m_{\tilde \mu_2}$) 
for the proposed potential test -- given the large mass splitting reachable 
theoretically in RS mSUGRA [illustrated in Fig.(\ref{SUGRA})]. 
This typical uncertainty of $\sim 5\% - 10\%$ on the smuon mass corresponds to illustrative examples studied in Ref.~\cite{TDRatlas} 
but of course this performance will be quite model/parameter-dependent and can be improved by combining different related mass measurements.
Note that the present experimental bound $m_{\tilde \mu_1}>94$ GeV \cite{PDG,TevWebPage} is respected on the plot of Fig.(\ref{SUGRA}).

\subsection{$H$ boson decays}
\label{Higgspart}

The phenomenological framework of this Section \ref{Higgspart} is the same as the one described in Section \ref{sfermionpart}. 
We have seen that in this RS framework the interactions between Higgs bosons
and sfermions can be significantly increased with respect to the 4D SUSY case. This is mainly due to the fact that the 5D $c_{\tilde f_{L/R}}$ parameters involved in these Higgs interactions 
are quite free (more precisely $c_{\tilde f_{L/R}}$ only affect squark/slepton masses) in RS SUSY whereas the same Higgs interactions are fixed by SM Yukawa coupling constants (related to 
SM fermion masses) in 4D SUSY. In this section, we will look at the effects of these increases on Higgs decay branching ratios. 
The analysis emphasis will be put on the example of the heaviest neutral Higgs boson $H$ as, kinematically, the lightest $h$ field (being heavier than $\sim 91$ GeV from LEP results 
but smaller than $\sim 140$ GeV from SUSY Higgs structure \cite{DjouadiReviewII}) cannot decay into pairs of on-shell sfermions (having lower experimental limits of 
$\sim 10^2$ GeV \cite{PDG,TevWebPage}) 
and hence does not feel optimal RS effects, except of course if the theoretical upper limit on its mass $m_h$ can really be sufficiently relaxed in warped SUSY \cite{Gautam}. 
The pseudo-scalar field $A$ and charged Higgs boson $H^\pm$ can similarly have increased decay widths into sleptons, as we will show it occurs for $H$ in RS SUSY.

Technically, the branching ratio formulas for Higgs bosons can be found in \cite{DjouadiReviewII} together with EW and NLO radiative corrections. We have included the  
leading corrections involved: the $b$ quark running mass, the radiative corrections to the neutral Higgs boson masses $m_h$ and $m_H$ as well as the 
corrections to the trilinear Higgs coupling $\Delta\lambda_{Hhh}$ within the $\epsilon$ approximation (see Ref.~\cite{DjouadiReviewII}, Section 1.3.3, for details).
\\ These branching ratios depend on the various Higgs couplings;
in RS SUSY, the $H$ boson couplings to SM fermions, Higgs/gauge bosons and higgsinos/gauginos are taken as in 4D SUSY 
(since heavy KK mixing/exchange effects are neglected as mentioned above) 
while the $H$ couplings to squarks/sleptons are deduced from stop couplings in Eq.(\ref{HCouplMat}) -- and rotation to the sfermion mass basis.
\\ The higher-dimensional parameters entering these effective $H$ couplings to squarks/sleptons are $c_{\tilde q_{L/R}}$, $c_{\tilde \ell_{L/R}}$ (with absolute values around unity),  
$\mu_{eff}$ [see Eq.(\ref{mueff})] and $A|_{eff} = Ake^{-k\pi R_c} \sim$ TeV which appears in the scalar coupling matrix of Eq.(\ref{HCouplMat}).
The TeV scale parameter $\mu_{eff}$ ($A|_{eff}$) will be taken numerically as $\mu$ ($A$) in 4D SUSY.
Similarly, the effective quantity $\tilde{m}_{L/R}|_{eff} = (\tilde{m}_{L/R}ke^{-2k\pi R_c})^{1/2} \sim$ TeV that shows up in scalar mass matrices  
of type (\ref{WholeMassMat}) is taken numerically approximately equal to the 4D soft mass $\tilde{m}_{L/R} \sim$ TeV [see Eq.(\ref{eq:4Dstopmassmatrix})]. 
We denote as usual the effective 4D soft gaugino masses for the bino and wino respectively $M_1$ and $M_2$.

\begin{figure}[!hc]
\centering	\vspace*{1.5cm}
\includegraphics[width=0.48\textwidth]{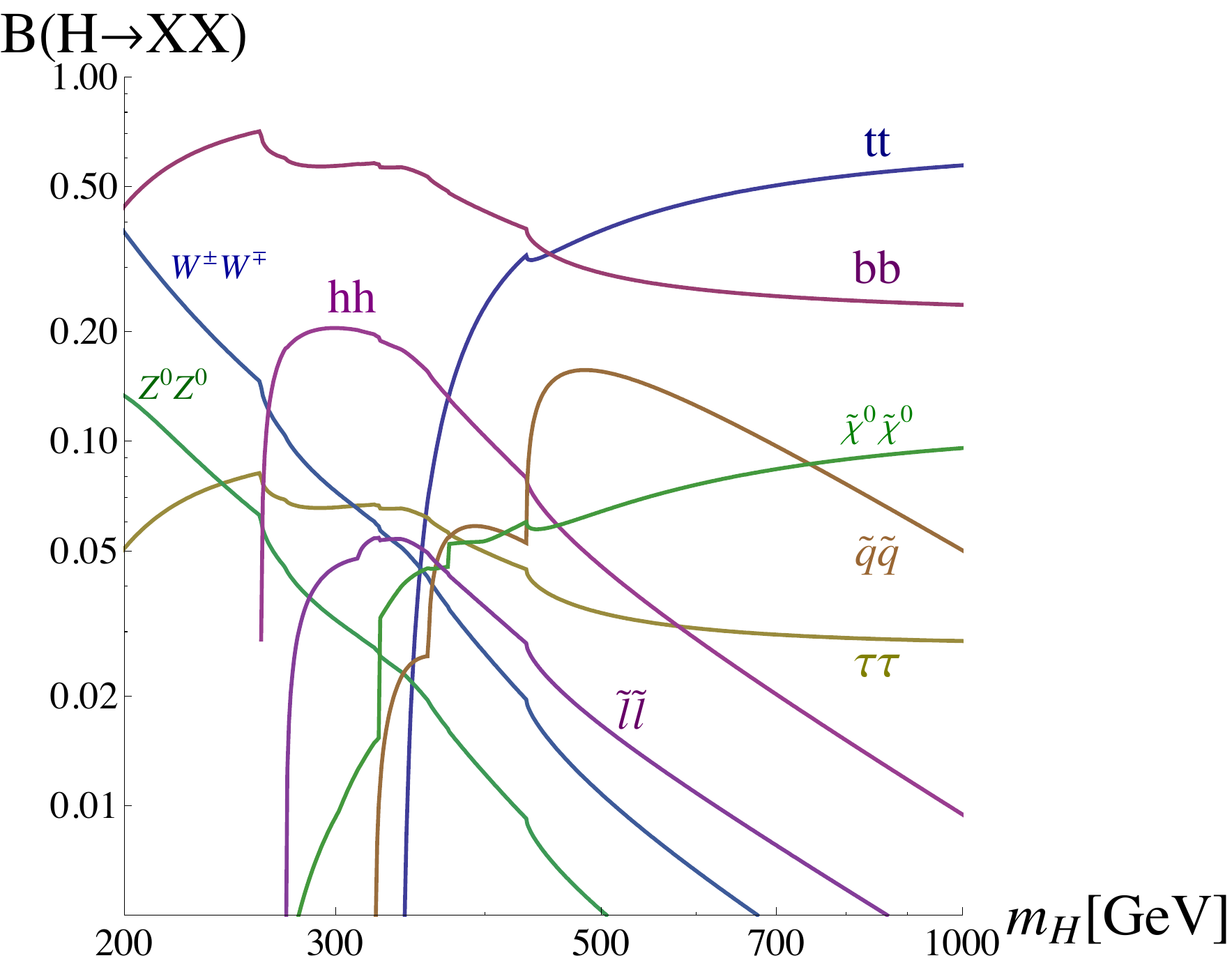}
\includegraphics[width=0.48\textwidth]{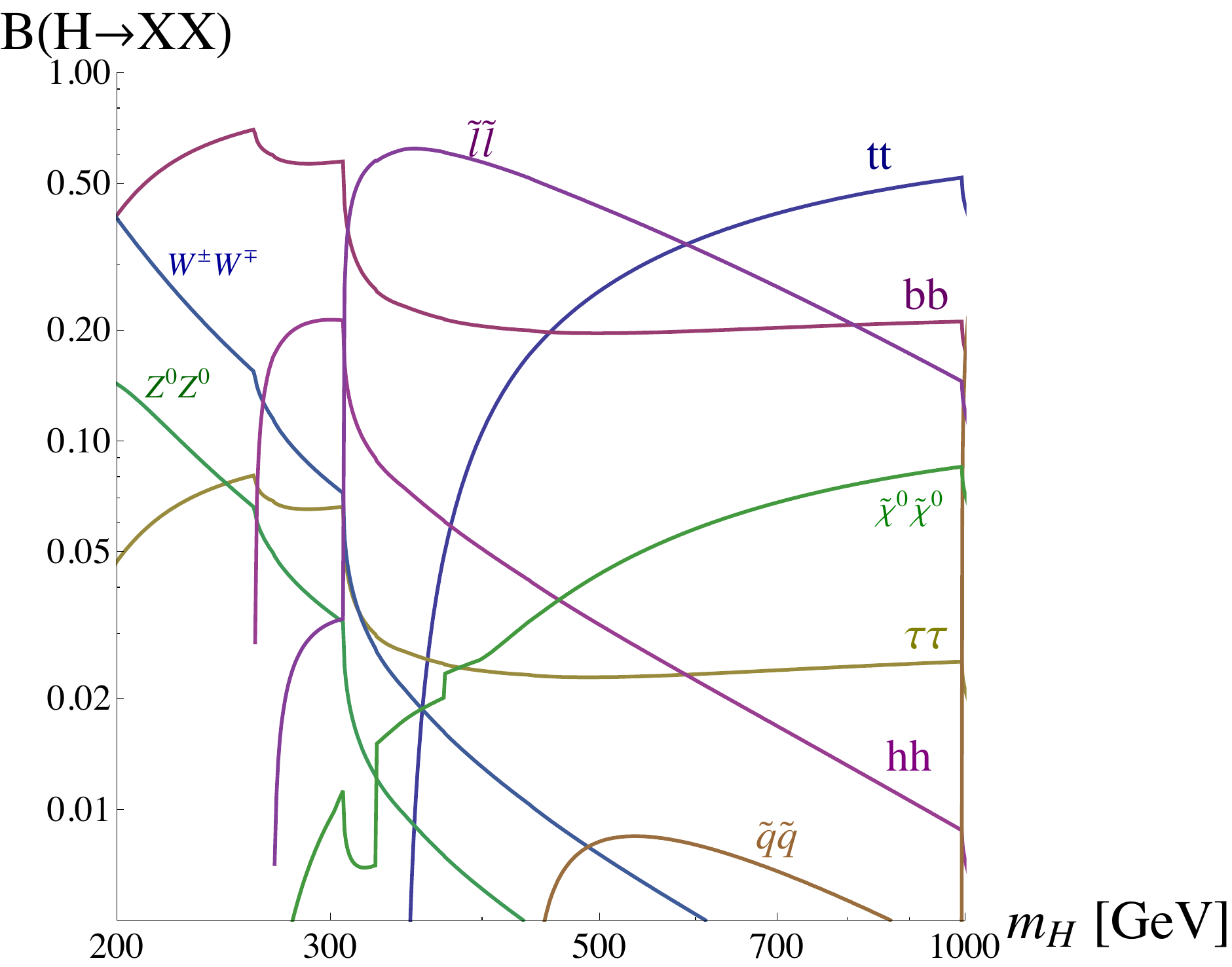}
\caption{\label{HdecSLvTT} 
\small{Branching ratios of the $H$ boson decays as a function of its mass $m_H$ (in GeV) within pure SUSY [left] and RS SUSY [right]. The types of final states are indicated directly on the plot; all kinematically allowed channels for neutralinos (among $\bar{\tilde{\chi}}^0_i\tilde{\chi}^0_j$, with $i,j=\{1,2,3,4\}$) and squarks/sleptons are summed.
The $\tau$ lepton channel is included but not the ones with branching ratios below $\sim 10^{-2}$ in this $m_H$ range (like the triangular-loop induced decays into photons and gluons).
The values for the gaugino sector parameters in the two plots are: $\mu = \mu_{eff} = 170\mbox{ GeV}$, $\tan\beta = 6$ and $M_1 = 160$ GeV, 
$M_2 = 1000\mbox{ GeV}$, leading to
the neutralino masses $m_{\tilde{\chi}^0_1} \simeq 128\mbox{ GeV}$, $m_{\tilde{\chi}^0_2} \simeq 174\mbox{ GeV}$ and lightest chargino mass $m_{\tilde{\chi}^{\pm}_1} \simeq 167\mbox{ GeV}$. 
The other common parameters are taken at $A_t|_{eff}=A_t = A_b = A_{\nu} = A_{\ell^\pm} = -500\mbox{ GeV}$. 
The soft masses in 4D SUSY are: $\tilde{m}_L^{q} = 170\mbox{ GeV}$, $\tilde{m}_R^{q} = 1000\mbox{ GeV}$ for the squarks and $\tilde{m}_L^{\ell} = 180\mbox{ GeV}$, 
$\tilde{m}_R^{\ell} = 1000\mbox{ GeV}$ for the sleptons. 
For this set of parameters, the smallest mass eigenvalues obtained are:  
$m_{\tilde{\nu}_{\mu1}} \simeq m_{\tilde{\nu}_{\tau1}} \simeq 169\mbox{ GeV}$ for sneutrinos, 
$m_{\tilde{\mu}_1} \simeq m_{\tilde{\tau}_1} \simeq 186\mbox{ GeV}$ for charged sleptons, 
$m_{\tilde{t}_1} \simeq 216\mbox{ GeV}$ for the stop and $m_{\tilde{b}_1} \simeq 179\mbox{ GeV}$ for the sbottom quark. 
In RS SUSY, the effective soft masses are: $\tilde{m}_L^{q}|_{eff} = 275\mbox{ GeV}$, $\tilde{m}_R^{q}|_{eff} = 1000\mbox{ GeV}$, $\tilde{m}_L^{\ell}|_{eff} = 105\mbox{ GeV}$, 
$\tilde{m}_R^{\ell}|_{eff} = 1000\mbox{ GeV}$ leading to $m_{\tilde{\nu}_{\mu1}} \simeq m_{\tilde{\nu}_{\tau1}} \simeq 135\mbox{ GeV}$,
$m_{\tilde{\mu}_1} \simeq m_{\tilde{\tau}_1} \simeq 154\mbox{ GeV}$ and $m_{\tilde{t}_1} \simeq m_{\tilde{b}_1} \simeq 218\mbox{ GeV}$   
while the 5D parameters are $c_{\tilde q_{L}}=c_{\tilde q_{R}} = 0.2$ for squarks and $c_{\tilde \ell_{L}}=c_{\tilde \ell_{R}} = -0.5$ for sleptons (for all flavors).}}
\end{figure}

In Fig.(\ref{HdecSLvTT}), we show the branching ratios for the main decay channels of the $H$ boson as a function of its mass
for a given set of parameters. Both 4D and 5D SUSY scenarios are represented.
The tree level Higgs mass $m_H$ (like $m_h$ and the neutral Higgs mixing angle $\alpha$) 
depends on the parameter $\tan \beta$ and on $m_A$ (pseudo-scalar mass) which has been varied (above $\sim 200$ GeV) to span the $m_H$ interval in  
Fig.(\ref{HdecSLvTT}). The modifications of radiative corrections due to heavy KK modes are expected to raise Higgs masses by an amount of at most
${\cal O}(10)$ GeV \cite{Gautam} so that Fig.(\ref{HdecSLvTT}) would not be significantly affected by those.
The heavy Higgs mass and $\tan \beta$ ranges considered in Fig.(\ref{HdecSLvTT}) and in the following Higgs branching 
ratio plots, namely $m_H \in [200,1000]$ GeV and $\tan \beta \in [6,30]$
\footnote{More precisely, we will only take $\tan \beta =6$ and $\tan \beta = 30$ in order to illustrate the main
characteristic behaviours.}, are clearly in agreement with the several
constraints coming from direct charged/neutral Higgs boson searches at LEP \cite{DjouadiReviewII} 
or Tevatron Run II \cite{TevWebPage} within a SUSY framework. The modifications of these constraints due to
KK mode effects should not be significant here for $M_{KK} \gtrsim 3$ TeV ({\it c.f.} Section \ref{model}).
\\ For the sake of convenience, in this section, we only write the sets of parameter values in the captions of Higgs branching ratio plots. 
But let us give here some general comments about these sets chosen in the plots. First, the branching ratios are shown for either positive or negative values of the $A$
(or $A|_{eff}$) and $\mu$ (or $\mu_{eff}$) terms as these signs do not affect the main ratio behaviors. Secondly, 
all the superpartner mass eigenvalues considered respect their last conservative/combined experimental lower limits derived from direct SUSY searches 
at colliders \cite{PDG,TevWebPage} \footnote{The mass constraints obtained recently at the LHC in non-SUGRA scenarios assume specific conditions \cite{spec}, like a gravitino LSP
\cite{gravLSP}, a gluino mass smaller \cite{SQsupGL} or equal \cite{SQeqGL} to squark ones, that we do not assume here.}
some of which we quote here for comparison: $m_{\tilde{\chi}^0_1} > 120\mbox{ GeV}$, $m_{\tilde{\chi}^0_2} > 116\mbox{ GeV}$, $m_{\tilde{\chi}^{\pm}_1} > 164\mbox{ GeV}$,
$m_{\tilde{\nu}_{1}} > 120\mbox{ GeV}$, $m_{\tilde{\ell}_1} > 107\mbox{ GeV}$, $m_{\tilde{t}_1} > 95\mbox{ GeV}$ and $m_{\tilde{b}_1} > 89\mbox{ GeV}$. 
In order to compare the Higgs couplings via its branching ratios, we have chosen the parameters so that the superpartner mass eigenvalues are approximately identical in the 4D and 5D pMSSM 
(KK corrections might be up to a few percents and are irrelevant in this analysis); this is also motivated by the present philosophy of developing tests of discrimination between the two SUSY 
scenarios (4D versus 5D) for a situation where light SUSY particles would have been discovered [and thus their masses at least approximately estimated]. Note also that
the lightest neutralino is systematically the LSP in our choices of parameters so that $\tilde \chi_1^0$ represents the potential candidate for dark matter as in usual 4D SUSY theories.   
In order to minimize the corrections to EW observables, we have further imposed $A_t = A_b$, $A_{\nu} = A_{\ell}$,  
$\tilde{m}^u_{L/R} = \tilde{m}^d_{L/R}$ for soft squark masses and $\tilde{m}^\nu_{L} = \tilde{m}^e_{L}$ for the slepton ones (in the RS case as well). Finally, 
the sfermion mass matrices as well as soft trilinear scalar couplings are taken universal for all families and diagonal in the flavor basis (which guarantees the
absence of FCNC at tree level) as motivated by the SUSY flavor problem. Note that these hypotheses imply the absence of flavor mixing in the scalar sector which
allows a better identification of the smuon and stop studied in Sections \ref{sfermionpart} and \ref{stopart}.
We also assume that all phases in the soft SUSY breaking potential are zero to eliminate all new sources of CP-violation. Those assumptions are characteristic
of the pMSSM \cite{pMSSM}.

For simplifying the discussion on this first Fig.(\ref{HdecSLvTT}), we have chosen large effective TeV-brane soft masses $\tilde m^{q,\ell}_{R},\tilde m^{q,\ell}_{R}\vert_{eff}$  
so that the heaviest sfermion mass eigenvalues become large enough to close the associated channels $H \to \tilde f_1\tilde f_2, \tilde f_2\tilde f_2$.
We see on Fig.(\ref{HdecSLvTT}) that starting from a 4D SUSY regime where the $W^\pm$ boson, top quark and lightest neutral Higgs channels are dominant 
(the top dominance is due, in particular, to the low $\tan \beta$ regime increasing the top-Higgs coupling), and then moving to the RS case with small $c_{\tilde \ell_{L/R}}$ 
values, the slepton channel is globally increased and even becomes dominant at intermediate $m_H$ values. This slepton channel has
a width decreasing with the third power of the $H$ mass which explains its branching behavior on the plot. 
The first reason for having larger slepton branching ratios in that RS case is that the slepton-Higgs couplings are favored by  
the large wave function overlaps between sleptons and the localized Higgs boson induced by the small $c_{\tilde \ell_{L/R}}$ values taken. The
other reason, of same kind, is the choice of large $c_{\tilde q_{L/R}}$ reducing the squark overlaps with Higgs bosons relatively to the 4D SUSY case. 
Indeed, one can clearly observe, by comparing the two plots of Fig.(\ref{HdecSLvTT}), the decrease of the branching ratio $B(H\to\bar{\tilde{q}}\tilde q)$ in RS
which leads to an increase of $B(H\to\bar{\tilde{\ell}}\tilde \ell)$.

\begin{figure}[!hc]
\centering  \vspace*{1.5cm}	
\includegraphics[width=0.48\textwidth]{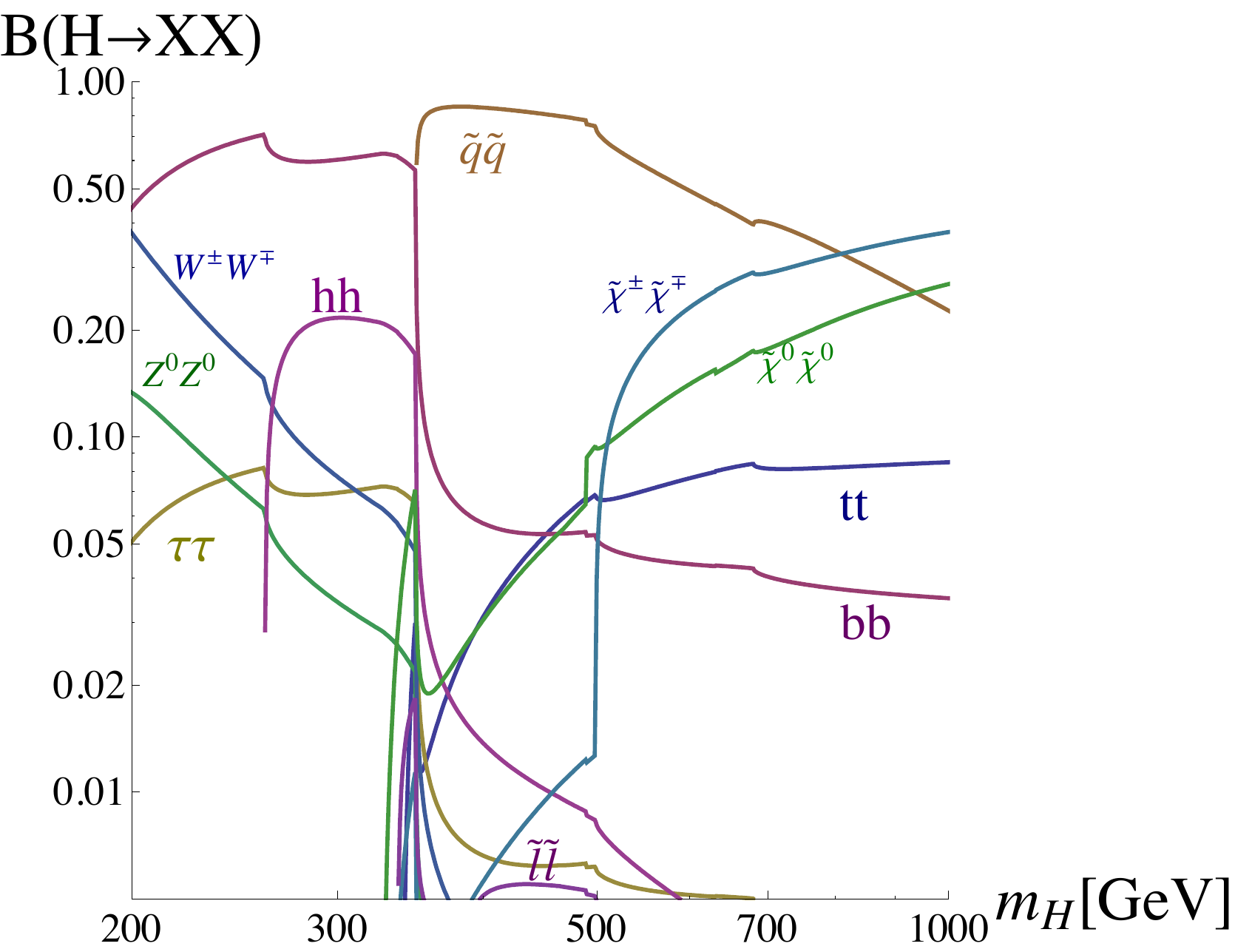}	
\includegraphics[width=0.48\textwidth]{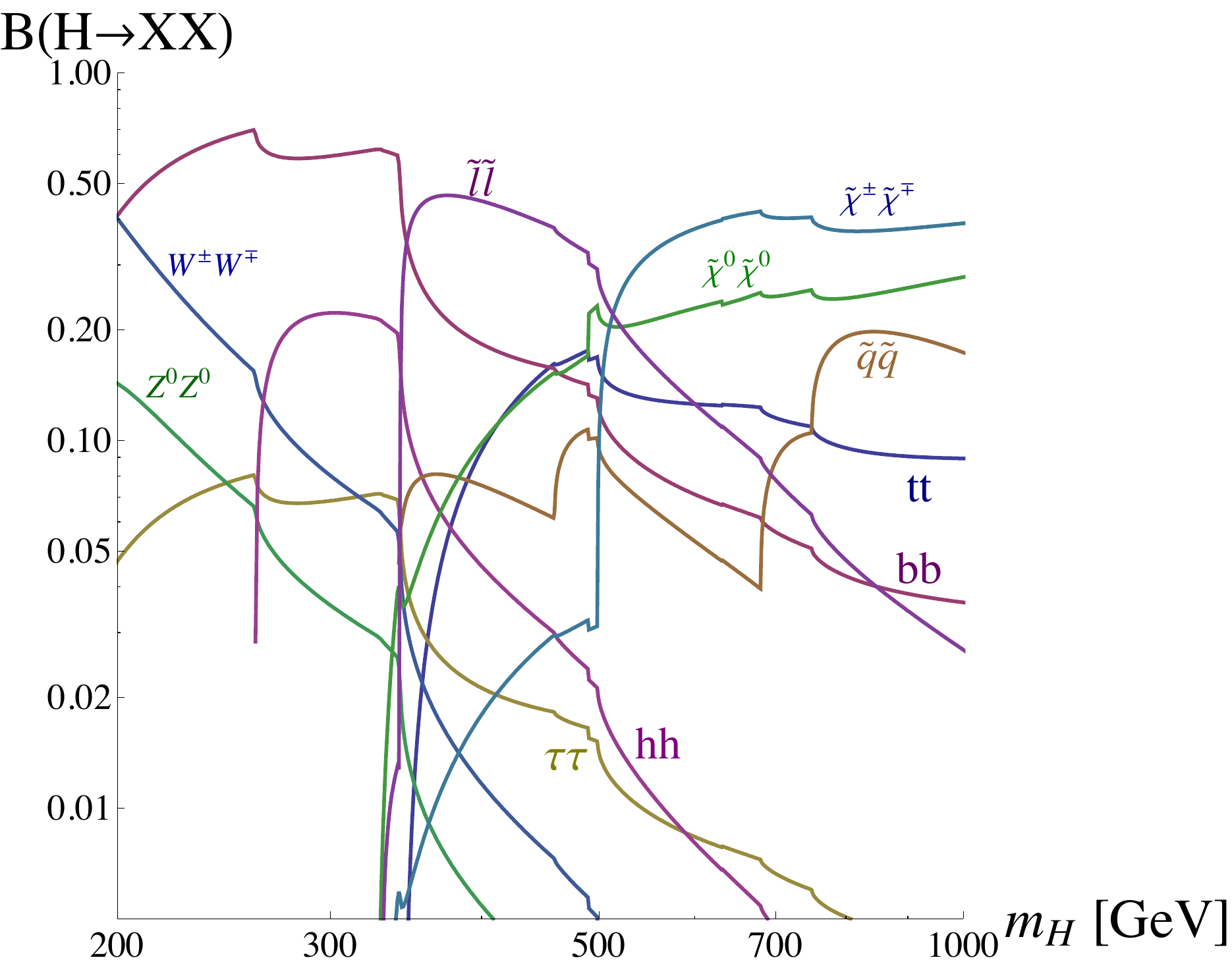}
\caption{\label{HdecSLvSQ} \small{Branching ratios of the $H$ boson decays as a function of its mass $m_H$ (in GeV) within pure SUSY [left] and RS SUSY [right]. 
The types of final states are indicated directly on the plot; all kinematically allowed channels for neutralinos (among $\bar{\tilde{\chi}}^0_i\tilde{\chi}^0_j$, with $i,j=\{1,2,3,4\}$),
charginos (among $\tilde{\chi}^\pm_k\tilde{\chi}^\mp_l$, with $k,l=\{1,2\}$) and squarks/sleptons are summed, as in Fig.(\ref{HdecSLvTT}). 
The values for the gaugino sector parameters are: $\mu = \mu_{eff} = 300\mbox{ GeV}$, $\tan\beta = 6$, $M_1 =600$ GeV and $M_2 = 190\mbox{ GeV}$ leading to
$m_{\tilde{\chi}^0_1} \simeq 163\mbox{ GeV}$, $m_{\tilde{\chi}^0_2} \simeq 305\mbox{ GeV}$ and $m_{\tilde{\chi}^{\pm}_1} \simeq 164\mbox{ GeV}$. 
The other common parameters are $A_t|_{eff}=A_t = A_b = -200\mbox{ GeV}$ and $A_\nu|_{eff}= A_{\nu} = A_{\ell^\pm} = -300\mbox{ GeV}$. 
The soft masses in 4D SUSY are: $\tilde{m}_{L}^{q} = 210\mbox{ GeV}$, $\tilde{m}_{R}^{q} = 220\mbox{ GeV}$ and $\tilde{m}_{L/R}^{\ell} = 180\mbox{ GeV}$ which give rise to  
$m_{\tilde{\nu}_{\mu1}} \simeq m_{\tilde{\nu}_{\tau1}} \simeq 169\mbox{ GeV}$, $m_{\tilde{\mu}_1} \simeq m_{\tilde{\tau}_1} \simeq 175\mbox{ GeV}$, 
$m_{\tilde{t}_1} \simeq 175\mbox{ GeV}$ and $m_{\tilde{b}_1} \simeq 206\mbox{ GeV}$. 
In RS SUSY, the effective soft masses are: $\tilde{m}_L^{q}|_{eff} = 500\mbox{ GeV}$, $\tilde{m}_R^{q}|_{eff} = 260\mbox{ GeV}$, $\tilde{m}_L^{\ell}|_{eff} = 125\mbox{ GeV}$ 
and $\tilde{m}_R^{\ell}|_{eff} = 1000\mbox{ GeV}$ giving $m_{\tilde{\nu}_{\mu1}} \simeq m_{\tilde{\nu}_{\tau1}} \simeq 165\mbox{ GeV}$,
$m_{\tilde{\mu}_1} \simeq m_{\tilde{\tau}_1} \simeq 171\mbox{ GeV}$, $m_{\tilde{t}_1} \simeq 171$ GeV and $m_{\tilde{b}_1} \simeq 229\mbox{ GeV}$
while the 5D parameters are $c_{\tilde q_{L}}=c_{\tilde b_{R}} = 0$, $c_{\tilde t_{R}} = 0.3$ and $c_{\tilde \ell_{L/R}}= -0.5$.}}
\end{figure}

We now present in Fig.(\ref{HdecSLvSQ}) the branching ratios for the main decay channels of the $H$ boson as a function of $m_H$ 
for another set of parameters [specified in the caption]. The soft masses $\tilde m^{q}_{R}$, in particular, are smaller than in the previous figure so that     
the stop masses are smaller and channels into the heaviest state $\tilde q_2$ open up, leading to a dominant squark channel at intermediate $m_H$ 
in the 4D SUSY case. The $M_{1,2}$ parameters are also typically lower than in Fig.(\ref{HdecSLvTT}), so that channels into charginos open up and neutralino
channels have larger phase space factors, which modifies the branching profiles rendering in particular the various gaugino channels 
dominant above $m_H\simeq 800$ GeV in 4D SUSY.
We see as well on Fig.(\ref{HdecSLvSQ}) that compared to this new 4D SUSY regime with alternatively dominant squark and bottom channels,   
in RS the slepton channel rates can again be greatly enhanced (and still become dominant at intermediate $m_H$) for small enough $c_{\tilde \ell_{L/R}}$ values
increasing the slepton-Higgs couplings. 
This enhancement of $B(H\to\bar{\tilde{\ell}}\tilde \ell)$ is also correlated with the large $c_{\tilde q_{L/R}}$ values allowing to have small squark-Higgs couplings 
relatively to 4D SUSY, and in turn to have small amounts of $B(H\to\bar{\tilde{q}}\tilde q)$. We note from the RS plot that $B(H\to\bar{\tilde{\ell}}\tilde \ell)$
reaches values a bit smaller than $\sim 50\%$ whereas it was above $\sim 60\%$ in Fig.(\ref{HdecSLvTT}), which is
due to the $\tilde m^{\ell}_{L}\vert_{eff}$ value taken larger here.

\begin{figure}[!hc]
\centering   \vspace*{1.5cm}	
\includegraphics[width=0.48\textwidth]{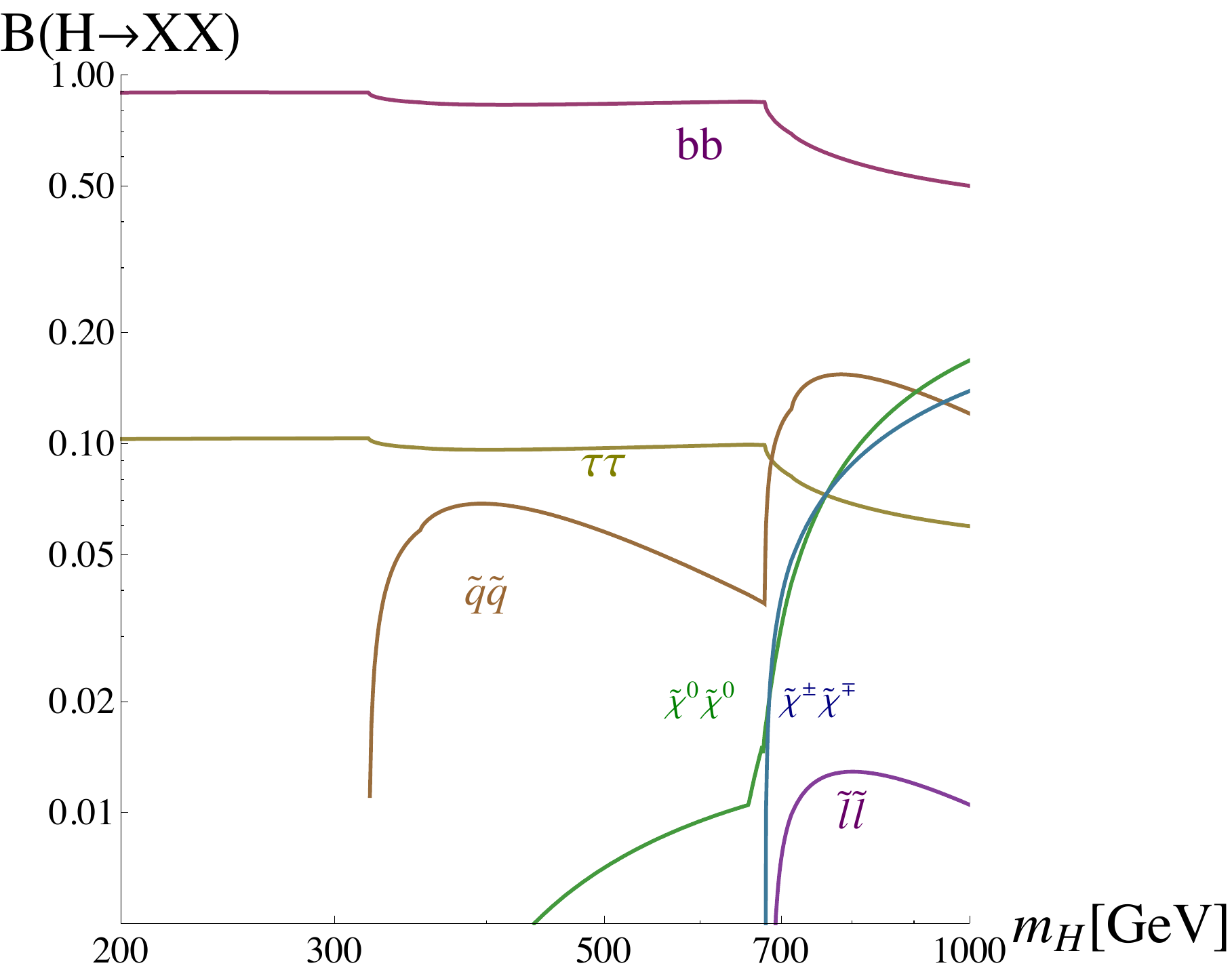}	
\includegraphics[width=0.48\textwidth]{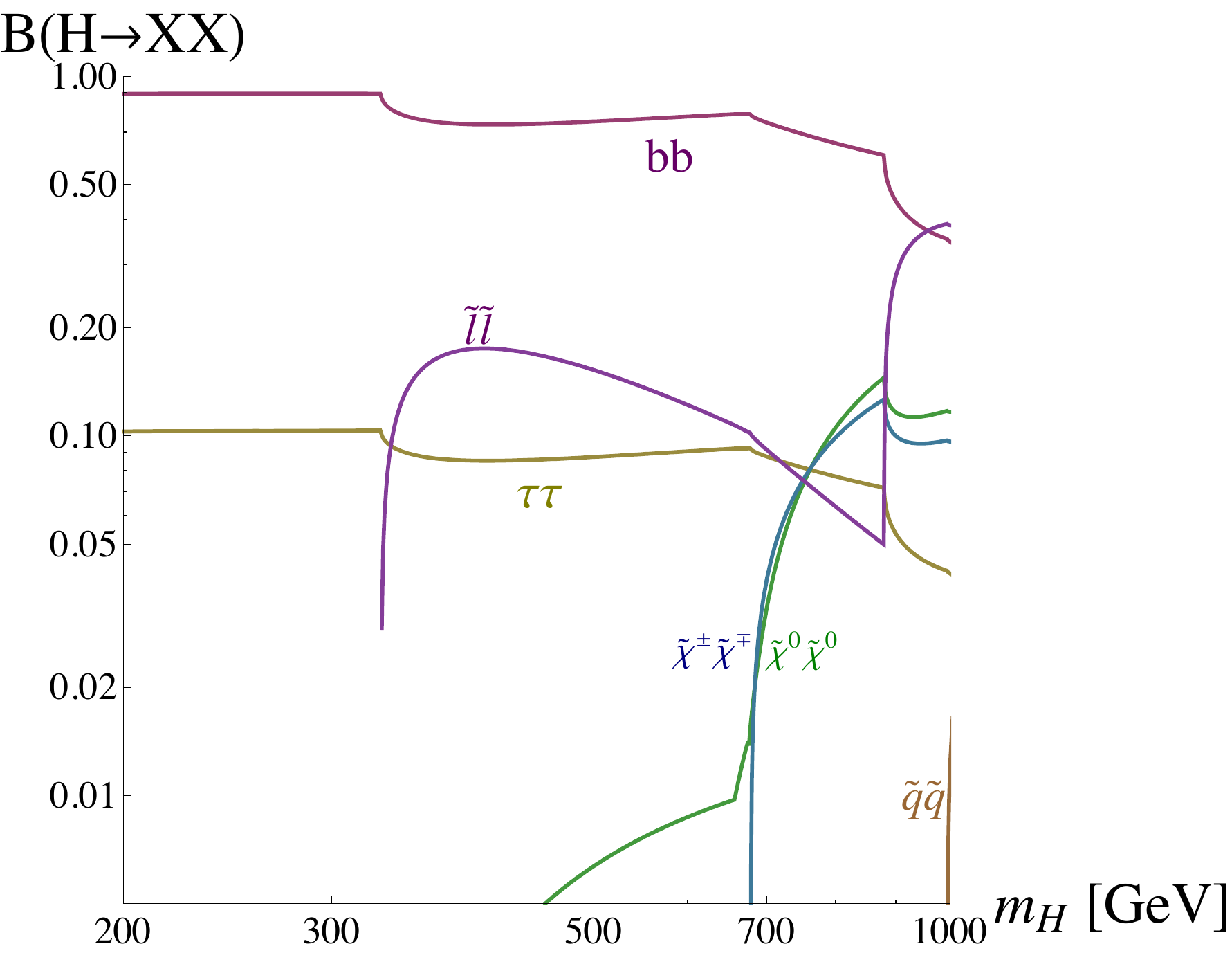}
\caption{\label{HdecSLvBB} \small{Branching ratios for the $H$ boson decays as a function of $m_H$ (in GeV) within pure SUSY [left] and RS SUSY [right]. 
The types of final states are indicated directly on the plot; all kinematically allowed channels for neutralinos,
charginos and squarks/sleptons are summed, as in previous figures. 
The values for the gaugino sector parameters are: $\mu = \mu_{eff} = 170\mbox{ GeV}$, $\tan\beta = 30$, $M_1 = M_2 = 500\mbox{ GeV}$ leading to
$m_{\tilde{\chi}^0_1} \simeq 157\mbox{ GeV}$, $m_{\tilde{\chi}^0_2} \simeq 176\mbox{ GeV}$ and $m_{\tilde{\chi}^{\pm}_1} \simeq 164\mbox{ GeV}$. 
The trilinear couplings read as $A_t|_{eff}=A_t = A_b= - A_{\nu} = - A_{\ell^\pm} = -500\mbox{ GeV}$. 
The 4D soft masses are: $\tilde{m}_{L}^{q} = 170\mbox{ GeV}$, $\tilde{m}_{R}^{q} = 500\mbox{ GeV}$, $\tilde{m}_{L}^{\ell} = 180\mbox{ GeV}$ 
and $\tilde{m}_{R}^{\ell} = 500\mbox{ GeV}$ which give 
$m_{\tilde{\nu}_{\mu1}} \simeq m_{\tilde{\nu}_{\tau1}} \simeq 168\mbox{ GeV}$, $m_{\tilde{\mu}_1} \simeq m_{\tilde{\tau}_1} \simeq 185\mbox{ GeV}$, 
$m_{\tilde{t}_1} \simeq 160\mbox{ GeV}$ and $m_{\tilde{b}_1} \simeq 177\mbox{ GeV}$. 
In RS SUSY, the effective soft masses are: $\tilde{m}_L^{q}|_{eff} = 275\mbox{ GeV}$, $\tilde{m}_R^{q}|_{eff} = 1000\mbox{ GeV}$, 
$\tilde{m}_L^{\ell}|_{eff} = 130\mbox{ GeV}$ and 
$\tilde{m}_R^{\ell}|_{eff} = 500\mbox{ GeV}$ giving $m_{\tilde{\nu}_{\mu1}} \simeq m_{\tilde{\nu}_{\tau1}} \simeq 172\mbox{ GeV}$,
$m_{\tilde{\mu}_1} \simeq m_{\tilde{\tau}_1} \simeq 165\mbox{ GeV}$ and $m_{\tilde{t}_1} \simeq m_{\tilde{b}_1} \simeq 219\mbox{ GeV}$
while the 5D parameters are $c_{\tilde q_{L/R}} = 0.2$ and $c_{\tilde \ell_{L/R}}= -0.5$.}}
\end{figure}

We present in Fig.(\ref{HdecSLvBB}) the branching ratios for the main $H$ decay channels as a function of $m_H$ in the high $\tan\beta$ regime.
We see on this figure that starting from such a 4D SUSY regime where the bottom channel is extremely dominant 
and going to the RS case, the Higgs-to-slepton branching ratio is largely enhanced -- once again due to the small $c_{\tilde \ell_{L/R}}$ values --  
reaching significant levels 
around $20\%$ and even dominantly at $\sim 40\%$ for a large $m_H$ (thanks to the $H\to\bar{\tilde{\ell_1}}\tilde \ell_2$ opening). 
This $B(H\to\bar{\tilde{\ell}}\tilde \ell)$ enhancement is also due to the large $c_{\tilde q_{L/R}}$ values which tend to suppress   
$B(H\to\bar{\tilde{q}}\tilde q)$ relatively to the 4D SUSY configuration. The reasons why the $H$ decay width into sleptons cannot be comparable 
in size to the one for bottom quark final states, at $m_H \lesssim 800$ GeV in RS, are the large $\tan\beta$, phase space and color factor.

\begin{figure}[!hc]
\centering \vspace*{1.5cm}
\includegraphics[width=0.49\textwidth]{BR_mssm_2.pdf}
\includegraphics[width=0.49\textwidth]{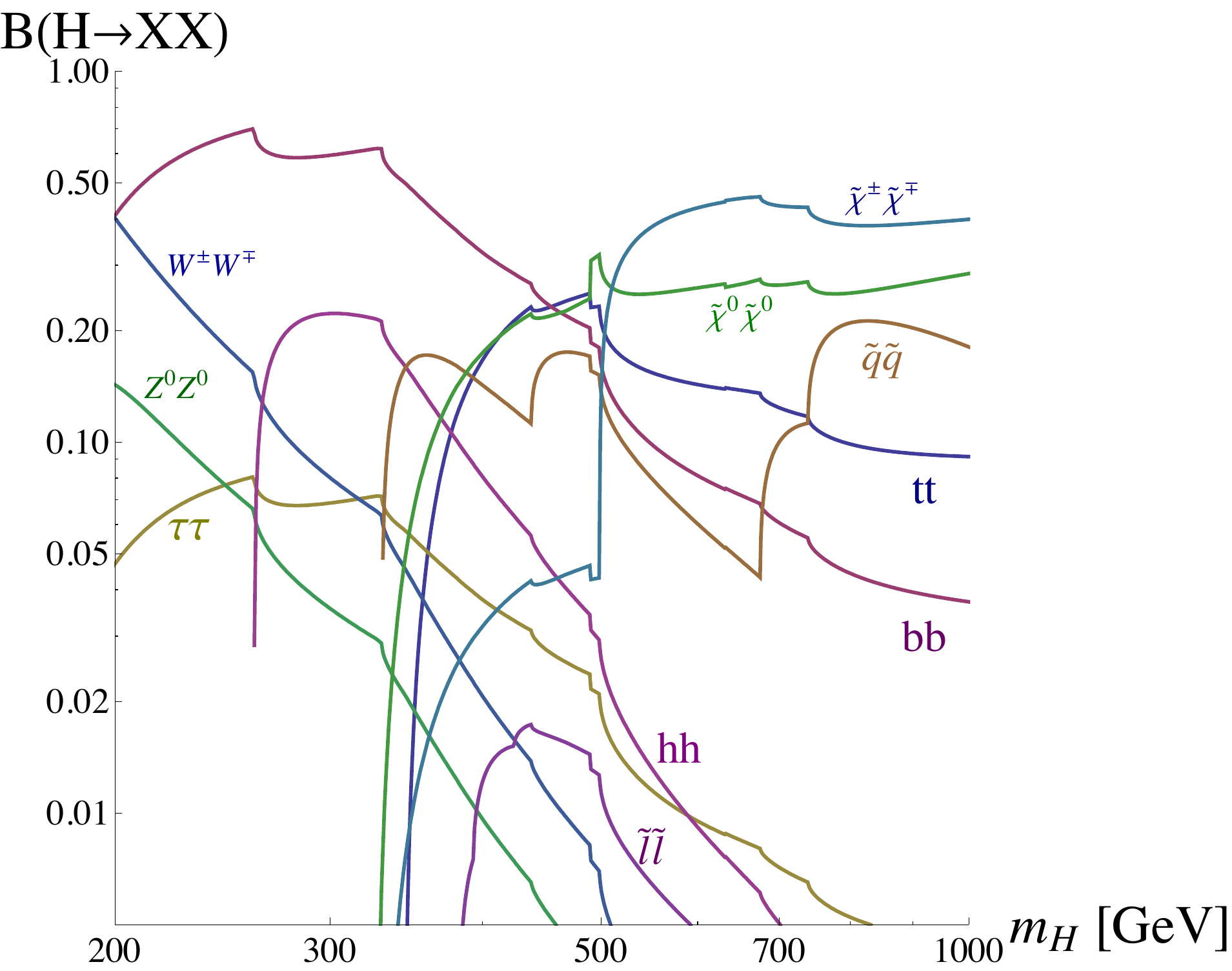}
\caption{\label{HdecSLEPTON} \small{Branching ratios for the $H$ boson decays as a function of $m_H$ (in GeV) within pure SUSY [left] and RS SUSY [right]. 
The types of final states are indicated directly on the plot. The parameters in the higgsino/gaugino sector, the scalar trilinear couplings 
and the soft masses in 4D SUSY are the same as in Fig.(\ref{HdecSLvSQ}) so that the left plot here is identical to the left plot of Fig.(\ref{HdecSLvSQ}).  
In RS SUSY, the effective soft masses are: $\tilde{m}_L^{q}|_{eff} = 500\mbox{ GeV}$, $\tilde{m}_R^{q}|_{eff} = 250\mbox{ GeV}$, 
$\tilde{m}_L^{\ell}|_{eff} = 1000\mbox{ GeV}$ and 
$\tilde{m}_R^{\ell}|_{eff} = 800\mbox{ GeV}$ giving $m_{\tilde{\nu}_{\mu1}} \simeq m_{\tilde{\nu}_{\tau1}} \simeq 197\mbox{ GeV}$,
$m_{\tilde{\mu}_1} \simeq m_{\tilde{\tau}_1} \simeq 169\mbox{ GeV}$, $m_{\tilde{t}_1} \simeq 165$ GeV and $m_{\tilde{b}_1} \simeq 219\mbox{ GeV}$
while the 5D parameters are $c_{\tilde q_{L}}=c_{\tilde b_{R}} = 0$, $c_{\tilde t_{R}} = 0.3$ and $c_{\tilde \ell_{L/R}} = +0.49$.}} 
\end{figure}

Of course in the different configuration of relatively high $c_{\tilde \ell_{L/R}}$ values, the slepton-Higgs couplings, and in turn the associated Higgs channels, 
are not significantly increased with respect to the 4D pMSSM, as illustrated in Fig.(\ref{HdecSLEPTON}) in a low $\tan\beta$ example; note that
in the RS case the sneutrino channel opens up at a larger $m_H$ due to sneutrinos being a bit heavier than in the 4D case. 
The main difference between the two plots of Fig.(\ref{HdecSLEPTON}) is the decrease of the squark channel in RS due to still quite 
large $c_{\tilde q_{L/R}}$ parameters (for smaller $c_{\tilde q_{L/R}}$ one would recover branching profiles similar to those in 4D SUSY). 
This leads in particular to an increase of $B(H\to\tilde{\chi}^{\pm}\tilde{\chi}^{\mp},\bar{\tilde{\chi}}^0\tilde{\chi}^0)$. Finally, for a better understanding
of Fig.(\ref{HdecSLEPTON}), we remark that in RS  
the slepton masses are mainly generated through large TeV-brane soft masses (reduced to the EWSB scale by wave function overlap factors) since 
the masses induced by Yukawa-type couplings to the Higgs VEV are suppressed by the present high $c_{\tilde \ell_{L/R}}$ values. For comparison, 
within 4D SUSY, the soft masses in Fig.(\ref{HdecSLEPTON}) are smaller (roughly at the EWSB scale) since those are not suppressed by overlap factors,
while the masses induced by Yukawa coupling constants are negligible due to the tiny lepton masses.

Let us summarize this part on Higgs decays and conclude. In order to maximize the $H$ couplings to sleptons in 4D SUSY,
the $A$ terms have been taken of the order of the TeV, and not at zero. The slepton masses have also been taken close to their lower experimental
limits to maximize the branching ratios of the slepton channel in 4D SUSY. 
However we have found that even in these optimal cases the slepton channel branchings cannot reach amounts above $\sim 5\%$ 
(in agreement with Ref.~\cite{DjouadiReviewII}) while 
important slepton branchings can arise in RS -- where $B(H\to\bar{\tilde{\ell}}\tilde \ell)$ can reach up to $\sim 60\%$
(even larger ratios are accessible for instance by decreasing again $c_{\tilde \ell_{L/R}}$ w.r.t. present values, or by introducing a Right-handed (s)neutrino). 
Besides, the other way to try to increase the slepton channel branching ratio in 4D SUSY is to decrease the other branching ratios: for that purpose, 
we have explored -- keeping trilinear couplings at the TeV scale -- 
the main typical domains of 4D SUSY (breaking) parameter space characterized by different types of dominant $H$ decay channels.
The conclusion was still that the slepton channel rate can never be as important as it can be in RS. Furthermore, the case of a dominant slepton channel
can occur in RS (see e.g. Fig.(\ref{HdecSLvSQ})) whereas it is impossible in 4D SUSY. 
This is due to the fact that in 4D SUSY slepton-Higgs couplings are severely constrained by the small Yukawa coupling constants of SM leptons. 
Therefore the experimental observation of either a dominant slepton channel or a slepton channel with say e.g. $B(H\to\bar{\tilde{\ell}}\tilde \ell)\simeq 50\%$ 
would exclude the 4D pMSSM and indicate the possible existence of a warped version of the pMSSM.
Such a potential test should in principle be doable at colliders; a four-momentum reconstruction of slepton cascade decays based on 
measured slepton masses could allow to identify the decay channel 
$H\to\bar{\tilde{\ell}}\tilde \ell$ which is necessary to estimate its branching ratio -- or (it would be sufficient here) a lower bound on this ratio. 
This kinematics approach being experimentally challenging, one could also use the clean leptonic event topology to identify the slepton channel. 
Indeed, the decay $H\to\bar{\tilde{\ell}}\tilde \ell$ would lead to final states with an higher lepton multiplicity due to the decay $\tilde \ell\to\ell\tilde{\chi}^0_i$ 
(recall that $\tilde{\chi}^0_1$ is the LSP).

\subsection{Stop pair production  at ILC}
\label{stopart}

In this part we consider the particularly clean environment of the future leptonic collider, the ILC, and we focus on the squark pair production which only occurs
through s-channel exchanges. In 4D SUSY, as is well-known, the stop can be specially light due to its large Left-Right mass mixing terms favored by strong Yukawa 
coupling constants. We will first provide the theoretical tools needed for computing the more general cross section of the sfermion pair production through neutral SM gauge
boson exchanges, namely $e^+ e^- \rightarrow \tilde{f}_i \overline{\tilde{f}}_j$ ($i,j =\{1,2\}$ labeling the mass eigenstates), in the 4D pMSSM. 
Then, after a discussion on the KK squark mixing effects, we will analyze numerically the obtained cross sections in the two RS SUSY breaking frameworks 
discussed previously (see the discussion on Fig.(\ref{mst1mst2})): with and without brane soft terms. 
Note that we concentrate here on the stop $\tilde{t}_{1,2}$ states and do not specify the structure of the sector made of the first generations of sfermions.

\subsubsection*{Formal cross section}

Let us describe the cross section for the reaction $e^+ e^- \rightarrow \tilde{f}_i \overline{\tilde{f}}_j$ [$i,j =\{1,2\}$] 
proceeding via the exchanges of the EW neutral gauge bosons -- photon ($\gamma$) and $Z$ boson -- in the s-channel. 
For polarized electron and positron beams, this cross section has the following form at tree level (it can be found with more details in Ref.~\cite{Bartl}) 
\begin{eqnarray}
\label{eq:crosssection}
\sigma(e^+ e^- \rightarrow \tilde{f}_i \overline{\tilde{f}}_j) &=& \frac{\pi \alpha^2 \kappa_{ij}^3}{s^4} \Bigg( 
(Q^{f}_{e.m.})^2 \delta_{ij} (1-\mathcal{P}_- \mathcal{P}_+)
- \frac{Q^{f}_{e.m.} c_{ij} \delta_{ij}}{2 c_W^2 s_W^2} \bigg [v_e(1-\mathcal{P}_-\mathcal{P}_+) - a_e(\mathcal{P}_- - \mathcal{P}_+)\bigg ] {\bf D_{\gamma Z}}
\nonumber\\
&&\qquad\qquad + \frac{c_{ij}^2}{16 s_W^4 c_W^4}\bigg [(v_e^2 + a_e^2)(1-\mathcal{P}_- \mathcal{P}_+) - 2a_ev_e(\mathcal{P}_- - \mathcal{P}_+)\bigg ] {\bf D_{ZZ}} \Bigg)
\end{eqnarray}
where $s_W\equiv\sin\theta_W$, $c_W\equiv\cos\theta_W$, $v_e = 4 s_W^2-1$ and $a_e=-1$ are the vector and axial-vector couplings of the electron to the $Z$ boson 
whereas $Q^{f}_{e.m.}$ is the electric charge of the sfermion supersymmetric partner: $f$.
$\mathcal{P}_{\pm}$ denote the degree of polarization of the $e^{\pm}$ beams, with the convention $\mathcal{P}_{\pm}=-1,\,0,\,+1$ for the Left-polarized, unpolarized, 
Right-polarized $e^{\pm}$ beams, respectively (e.g. $\mathcal{P}_- = -0.9$ means that $90\%$ of the electrons are Left-polarized and the rest is unpolarized). 
$c_{ij}$ is the $Z \tilde{f}_i \overline{\tilde{f}}_j$ coupling matrix (up to a factor $1/c_W$):
\begin{eqnarray}
\label{eq:Zstopcoupling}
c_{ij} = \left(\begin{array}{cc}
{I_{3L}^{f} \cos^2\theta_{\tilde{f}}-Q^{f}_{e.m.} s_W^2} & {-{1\over2}I_{3L}^{f} \sin2\theta_{\tilde{f}}} \\
{-{1\over2}I_{3L}^{f} \sin2\theta_{\tilde{f}}} & {I_{3L}^{f} \sin^2\theta_{\tilde{f}}-Q^{f}_{e.m.} s_W^2}
\end{array}\right)
\end{eqnarray}
where $\theta_{\tilde{f}}$ is the sfermion mixing angle defined by
\begin{eqnarray}
\label{eq:defthetamix}
\tilde{f}_1 \equiv \cos\theta_{\tilde{f}} \, \tilde{f}_L + \sin\theta_{\tilde{f}} \, \tilde{f}_R
\quad,\quad
\tilde{f}_2 \equiv \cos\theta_{\tilde{f}} \, \tilde{f}_R - \sin\theta_{\tilde{f}} \, \tilde{f}_L \quad .
\end{eqnarray}
In the 4D pMSSM, this mixing angle is calculated from the diagonalization of the pMSSM stop mass matrix previously given in Eq.(\ref{eq:4Dstopmassmatrix}). In the 
RS SUSY setup, it is obtained from the mass matrix derived in Eq.(\ref{WholeMassMat}) [possibly including mixings with KK modes]. 
Finally, in Eq.(\ref{eq:crosssection}), $\sqrt{s}$ is the center-of-mass energy, 
$\kappa_{ij} = [(s-m^2_{\tilde{f}_i}-m^2_{\tilde{f}_j})^2-4 m^2_{\tilde{f}_i} m^2_{\tilde{f}_j}]^{1/2}$ and
\begin{eqnarray}
{\bf D_{ZZ}} = \frac{s^2}{(s-m_Z^2)^2+\Gamma_Z^2 m_Z^2} \quad,\quad {\bf D_{\gamma Z}} = \frac{s(s-m_Z^2)}{(s-m_Z^2)^2+\Gamma_Z^2 m_Z^2} \quad ,
\end{eqnarray}
$\Gamma_Z = 2.4952 \pm 0.0023 \mbox{ GeV}$ \cite{PDG} being the full decay width of the $Z$ boson.

\subsubsection*{On the difficulty to distinguish 4D and RS SUSY in custodial realizations}

In our previous studies of Sections \ref{sfermionpart} and \ref{Higgspart}, the KK sfermion mixing effects were sub-leading relatively to the studied structural effects
on scalar masses/couplings. In the RS realization with a gauge custodial symmetry in the bulk, having specific BC KK states (the so-called
`custodians') with theoretically possible low masses 
in the particle spectrum \footnote{FCNC and EW precision tests constrain such masses from below.}, there is a priori a hope that 
KK mixings bring other potential signatures of the warped SUSY version. In this section we thus study the RS version with an implementation of 
the ${\rm SU(2)_L \times SU(2)_R \times U(1)_X}$ gauge custodial symmetry in the bulk (introduced in Section \ref{model}). The custodians are the new fermions
-- without zero-modes -- filling the ${\rm SU(2)_R}$ representations to which matter (MS)SM (super)fields are promoted.
Being in a SUSY context, we thus have to embed the custodians into superfields which leads to the theoretical prediction of the existence of custodian superpartners: 
let us call them the `scustodians' (those are scalar fields). Since we do not consider a SUSY breaking \`a la Scherk-Schwarz, the custodians and their associated
scustodians possess the same BC and hence identical 5D profiles and KK mass spectra. For instance, in complement to the stop squark states $\tilde{t}_{L/R}$ and
to a possible $t'$ custodian (with same electric charge as the SM top quark $t$ of which it can be an ${\rm SU(2)_R}$ partner), 
we would now have associated scustodians that we note $\tilde{t}'_{L/R}$. In RS, it is known that the mixing -- induced by the 
localized Higgs VEV -- of the $t$ quark with e.g. a $t'$ partner is favored since 
the 5D $c_t$ parameter, expected to be rather small for this third generation, tends to decrease the $t'$ mass $m_{(-+)}^{(1)}(c_{t})$ \footnote{Strictly speaking,
$m_{(-+)}^{(1)}(c_{t})$ is the mass of the first KK mode with $(-+)$ (and $(+-)$) BC.} and thus to make it closer to $m_t$. The $t$-$t'$ mixing is expected to be larger than 
the $t$-$t^{(1)}$ mixing [not characteristic of the bulk gauge custodial realization] cause $m_{(-+)}^{(1)}$ can be significantly smaller than $t^{(1)}$ masses  
(= masses of the first KK excitation of SM top quarks which have $(++)$ BC) being above $\sim 3$ TeV typically from EW precision constraints. 
More generally, the small $c_{b,t}$ parameters of the third quark generation 
($b$ and $t$) tend to increase the $b$ and $t$ wave function overlaps with the TeV-brane where is the Higgs, a feature which also favors the $t$-$t'$ or $b$-$b'$ mixing.  
One may expect similar effects in the third generation squark sector (in virtue of the same 5D profiles and KK masses) giving rise possibly to a significant mixing between 
the stop squarks $\tilde{t}_{L/R}$ and some light scustodians $\tilde{t}'_{L/R}$
\footnote{For scalar fields, the $\tilde{t}$-$\tilde{t}'$ mass mixing terms come from the $\mu$ and $A$ mass terms.}. 
In our SUSY breaking scenario where $c_{\tilde t} \neq c_t$, the $c_{\tilde t}$ parameter
would be related to the stop mass only (no more to the precisely-known top quark mass), a freedom that could even reinforce the $\tilde{t}$-$\tilde{t}'$ mixing.   
This mixing can affect the $Z \tilde{t} \overline{\tilde{t}}$ coupling since the $Z$ charge of the $\tilde{t}'$, $Q_Z^{\tilde{t}'}$, can be different from the $\tilde{t}$ 
one, $Q_Z^{\tilde{t}}$ (it is not the case for the photon and gluon couplings). Studying the stop pair production via the $Z$ exchange 
is more appropriate at ILC since the stop pair production is dominated by the gluon exchange at LHC. The reaction $e^+ e^- \rightarrow \tilde{t}_i \overline{\tilde{t}}_j$
has also the interest to not involve the $W$ boson neither gaugino contributions.

However, this task of generating significant corrections to the $Z \tilde{t} \overline{\tilde{t}}$ vertex, through $\tilde{t}$-$\tilde{t}'$ mixings  
and a large difference $Q_Z^{\tilde{t}'}$-$Q_Z^{\tilde{t}}$,  
has revealed itself to be quite difficult to realize for the two following reasons. First, strong theoretical constraints hold on the choices of group representations
and hence on the $Z$ charge of the $\tilde{t}'$, $Q_Z^{\tilde{t}'}=I_{3L}^{\tilde{t}'}-(2/3)s^2_W$, due to the gauge structure of Yukawa couplings which has to be 
invariant under the ${\rm SU(2)_L \times SU(2)_R \times U(1)_X}$ symmetry (and also allowed by the orbifold symmetry). 
For example, scalar couplings between $\tilde{t}$ (Left or Right) and some $\tilde{t}'$ scustodians can only be realized for scustodians with an ${\rm SU(2)_L}$ 
isospin equal to $I_{3L}^{t_{L/R}}\pm1/2$. Having so, their $Q_Z^{\tilde{t}'}$ cannot be very different from $Q_Z^{t_{L/R}}=Q_Z^{\tilde{t}_{L/R}}$. 
Secondly, another (related) origin of 
suppression in the variation of $Q_Z^{\tilde{t}_{L/R}}$ is the compensation between various scustodians, e.g. between effects from a $\tilde{t}-\tilde{t}'$ 
mixing (with $Q_Z^{\tilde{t}'} = Q_Z^t + 1/2$) and a $\tilde{t}-\tilde{t}''$ mixing (with $Q_Z^{\tilde{t}''} = Q_Z^t - 1/2$). 
Moreover, the dominant mixing between the $\tilde{t}_L$ and $\tilde{t}_R$ states tends to reduce the mixing effect of scustodians.
\\ Based on an exploration of the parameter space and models, 
we have found numerically that even for optimal group representations and scustodian masses set at $\sim 10^2$ GeV 
\footnote{Those low masses could maybe even be excluded by FCNC and EW precision constraints.},
the deviations $\delta\sigma_{th}$ induced by the scustodian mixings on the cross section $\sigma(e^+ e^- \rightarrow \tilde{t}_i \overline{\tilde{t}}_j)$ in 4D SUSY 
cannot be significant in regard to the experimental precision on this cross section measurement one can reasonably expect at ILC: 
$\delta\sigma_{exp} \sim 5\%$ \cite{Bartl} ($\sim 1\%$ at most \cite{FRICHARD}).
Hence we must conclude that testing 4D SUSY against RS SUSY, by measuring precisely
the cross section $\sigma(e^+ e^- \rightarrow \tilde{t}_i \overline{\tilde{t}}_j)$, is not realistic.

Since the RS SUSY signature of KK stop (partner) mixing effects on the $\tilde{t}_i \overline{\tilde{t}}_j$ production is too difficult, 
one could think of studying off-shell exchanges or mixing effects of KK gauge bosons 
\footnote{Maybe even KK gaugino effects or KK top mixings.} but then the simpler fermion production, like the top pair production, would probably be more fruitful
to consider as no supersymmetric cascade decays should be reconstructed there.  
Nevertheless, it has already been shown that detecting a resonance tail [our work hypothesis being that no KK (gauge boson) states have been produced on-shell] 
of KK gluon in $t\bar t$ production at LHC appears to be challenging due to small realistic cross sections
\cite{LHCgluonKK} (see Ref.~\cite{LHCbosonKK} for other KK gauge bosons). The indirect effects of such virtual KK mode exchanges in $t\bar t$ production
at ILC are more promising \cite{Sher} but would be polluted by the SUSY background 
($\tilde{t} \bar{\tilde{t}}$ production followed by $\tilde{t}\to t\tilde{\chi}^0_1$, $t\bar t$ at loop level,\dots).

\subsubsection*{Discriminating between RS versions: bulk versus brane SUSY breaking}

In our framework with tree level SUSY breaking squark/slepton masses and trilinear $A$ couplings, where we do not specify the mechanism at the origin
of SUSY breaking (our phenomenological approach is more `bottom-up' like), we will show that the stop pair production at ILC can be studied in order to 
differentiate between different variants of RS SUSY: the scenarios with or without soft scalar mass and trilinear $A$ terms on the TeV-brane. 
\\ The KK (s)fermion mixings cannot be significant in the stop pair production 
-- the scustodian mixing is irrelevant (as shown in previous subsection) and a fortiori the same conclusion holds for usual KK stops -- 
so that such effects are not considered in this part. The heavy (at least at $\sim 3$ TeV) 
KK gauge boson effects are also neglected in favor of the studied low-energy structural effects in scenarios with/without brane soft mass terms.

\begin{figure}[!hc]
\centering \vspace*{1.5cm}
\includegraphics[width=0.325\textwidth]{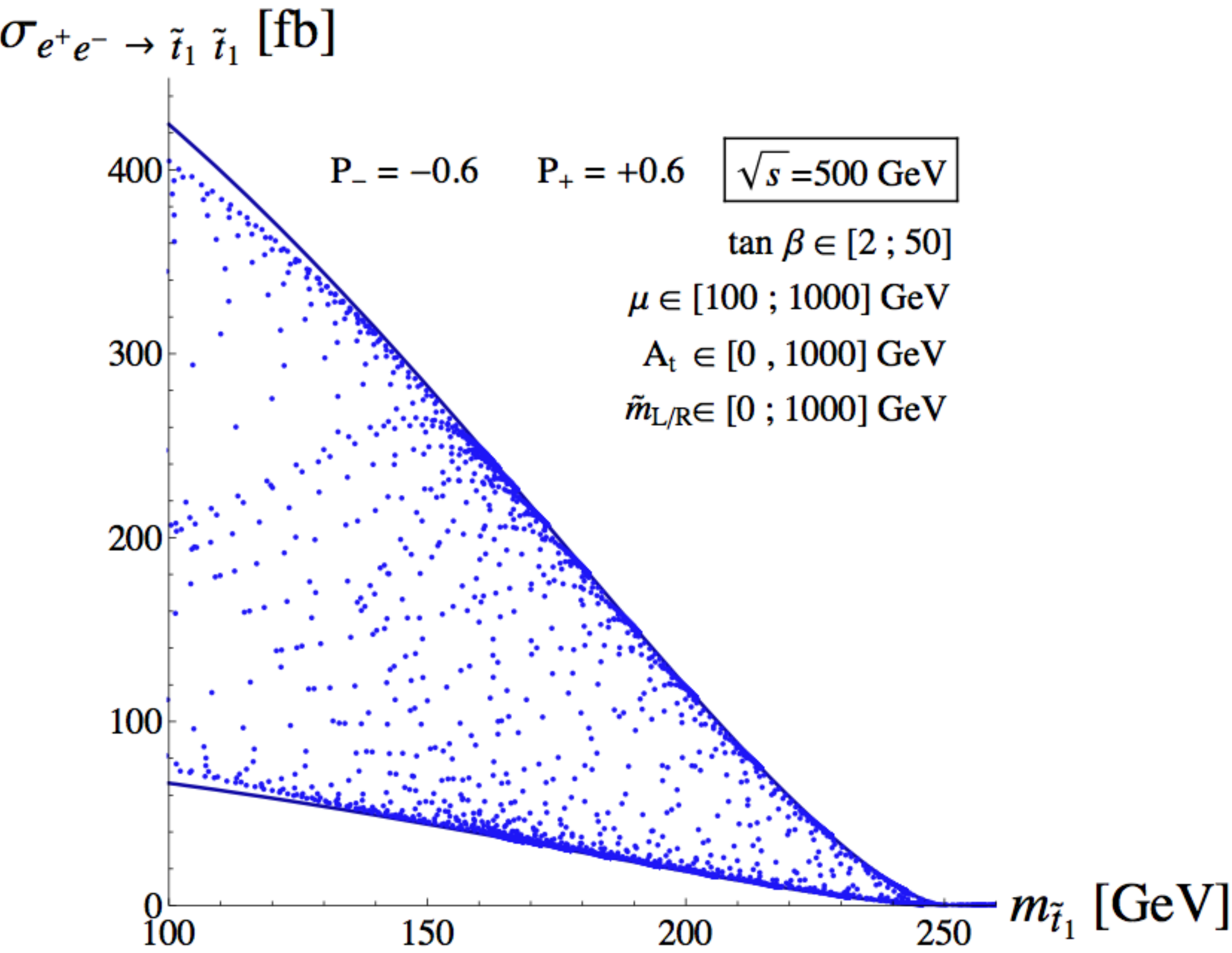}
\includegraphics[width=0.325\textwidth]{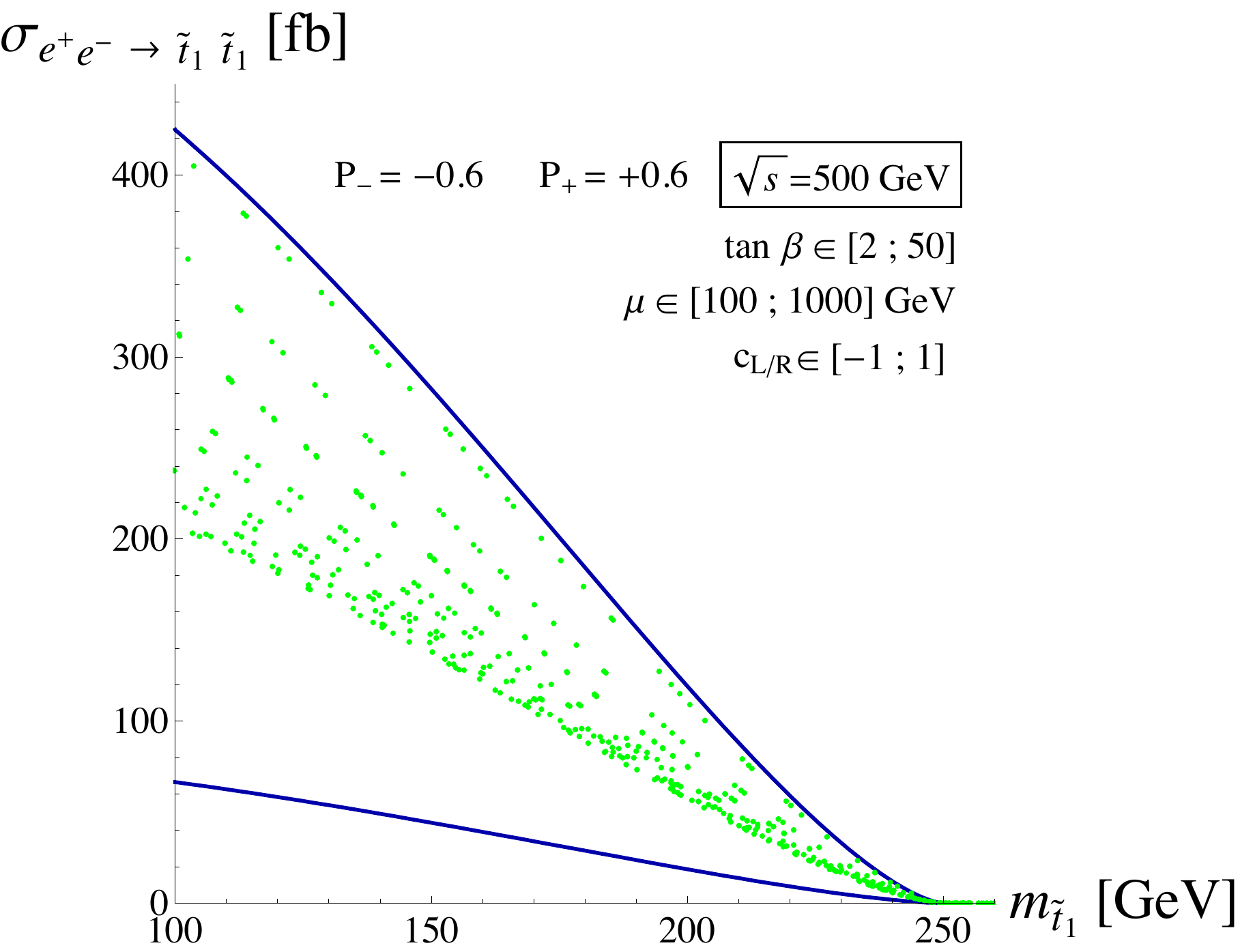}
\includegraphics[width=0.325\textwidth]{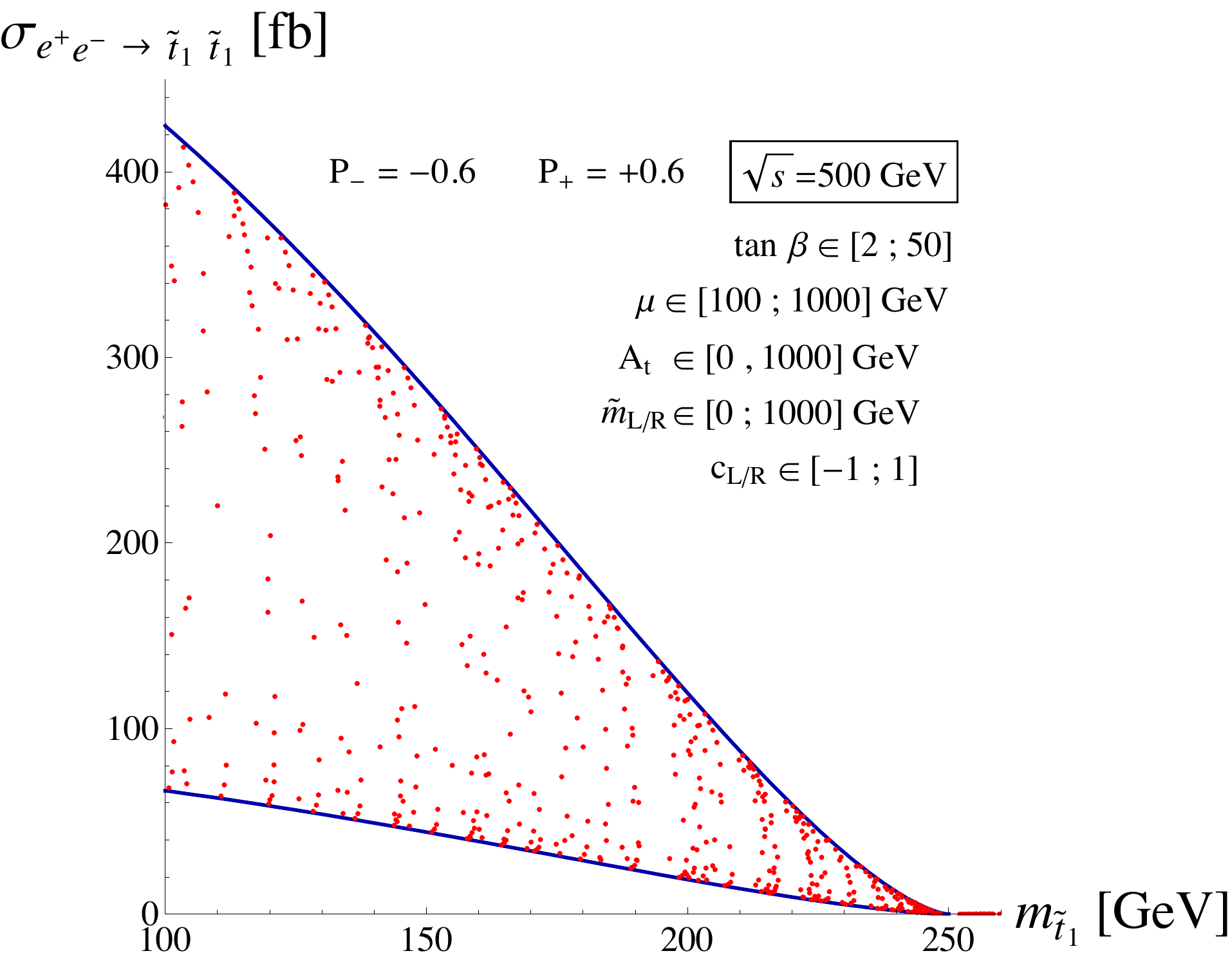}
  	\caption{ \small{Cross section for the reaction $e^+ e^- \rightarrow \tilde{t}_1 \overline{\tilde{t}}_1$ (in $fb$), at a center-of-mass energy 
	$\sqrt{s} = 500 \mbox{ GeV}$  
	and with the polarizations $\mathcal{P}_\pm$ (see Eq.(\ref{eq:crosssection})) indicated on the plot, as a function of $m_{\tilde{t}_1}$ (in GeV) 
	within the 4D pMSSM [left plot], RS pMSSM without brane soft terms [middle] and RS pMSSM with brane soft terms  
	[right]. All points presented here are obtained from a scan inside the ranges given on the plots for the parameters $c_{L/R}=c_{\tilde t_{L/R}}$, $\tan\beta$, 
	$\mu$ (and $\mu_{eff}$ in RS), the soft stop masses $\tilde m_{L/R}$ (and brane masses $\tilde{m}_{L/R}|_{eff}$) and the soft trilinear coupling $A_t$ 
	(and brane coupling $A_t |_{eff}$).
	The limiting thick blue lines correspond to the analytically obtained cross sections for an extremal $Z\tilde{t}_1\overline{\tilde{t}}_1$ coupling: 
	$c_{11} = Q_Z^{t_{R}}$ [lower limit] and $c_{11} = Q_Z^{t_{L}}$ [upper limit].} }
  \label{fig-500-11}
\end{figure}

\begin{figure}[!hc]
\centering \vspace*{1.5cm}
\includegraphics[width=0.325\textwidth]{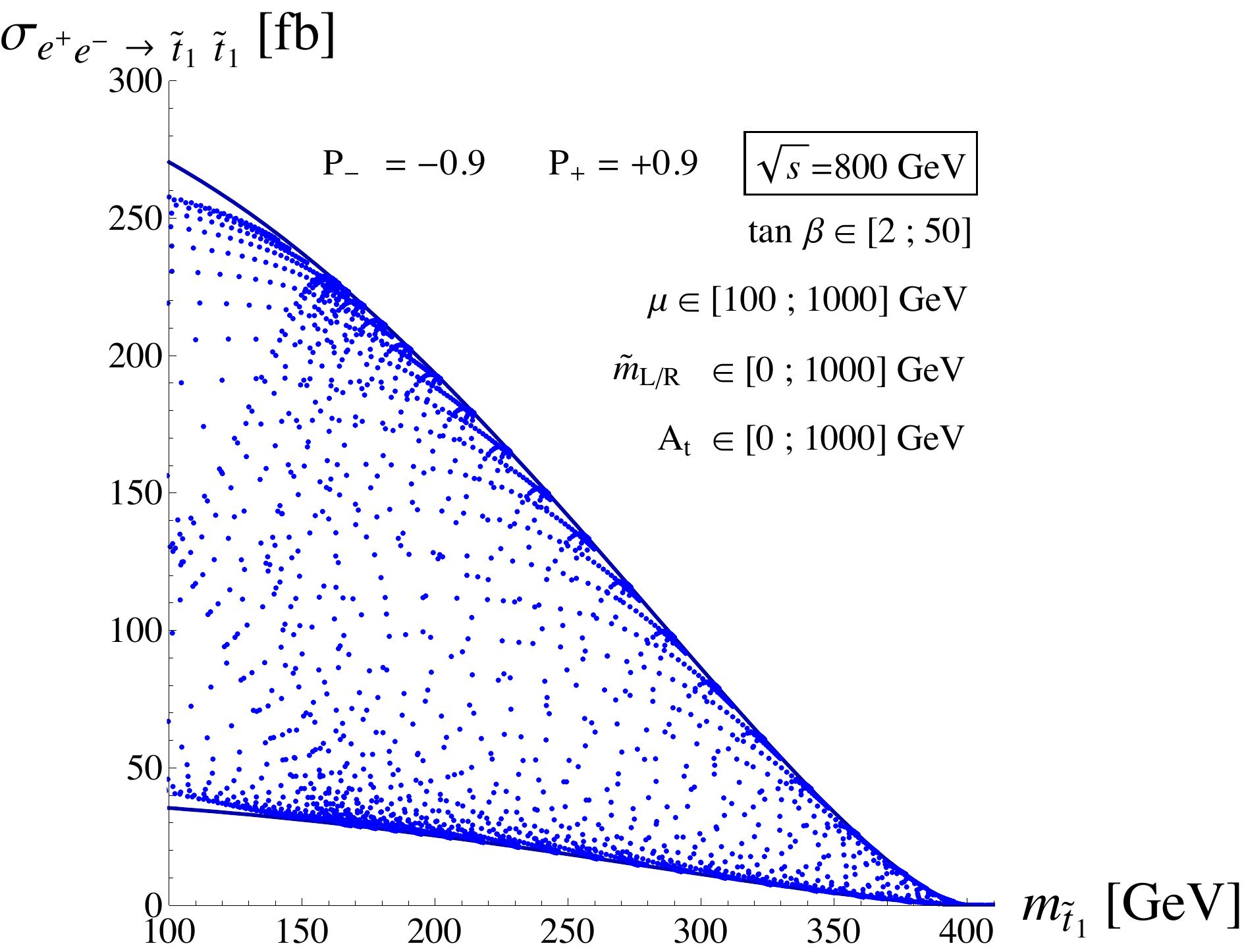}
\includegraphics[width=0.325\textwidth]{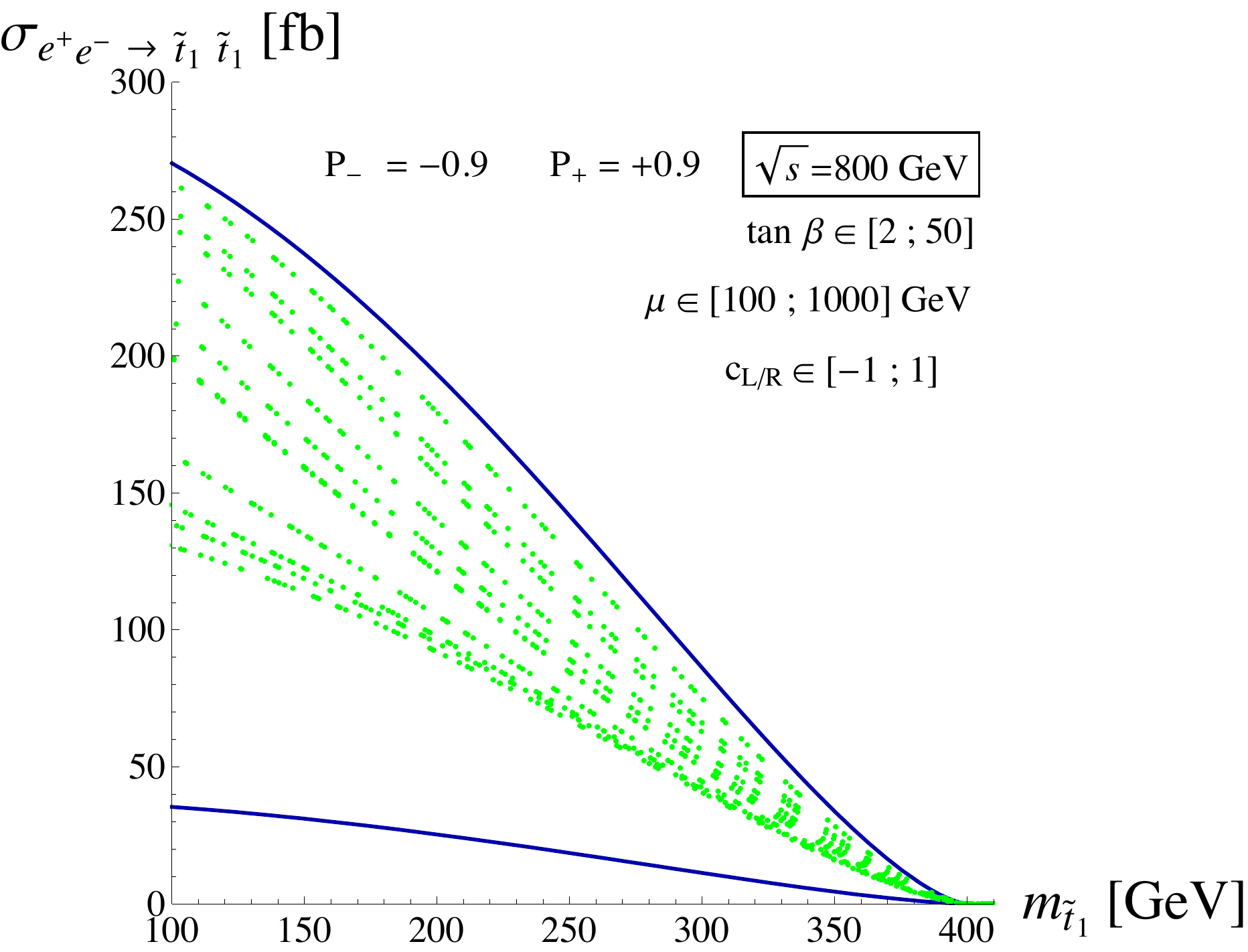}
\includegraphics[width=0.325\textwidth]{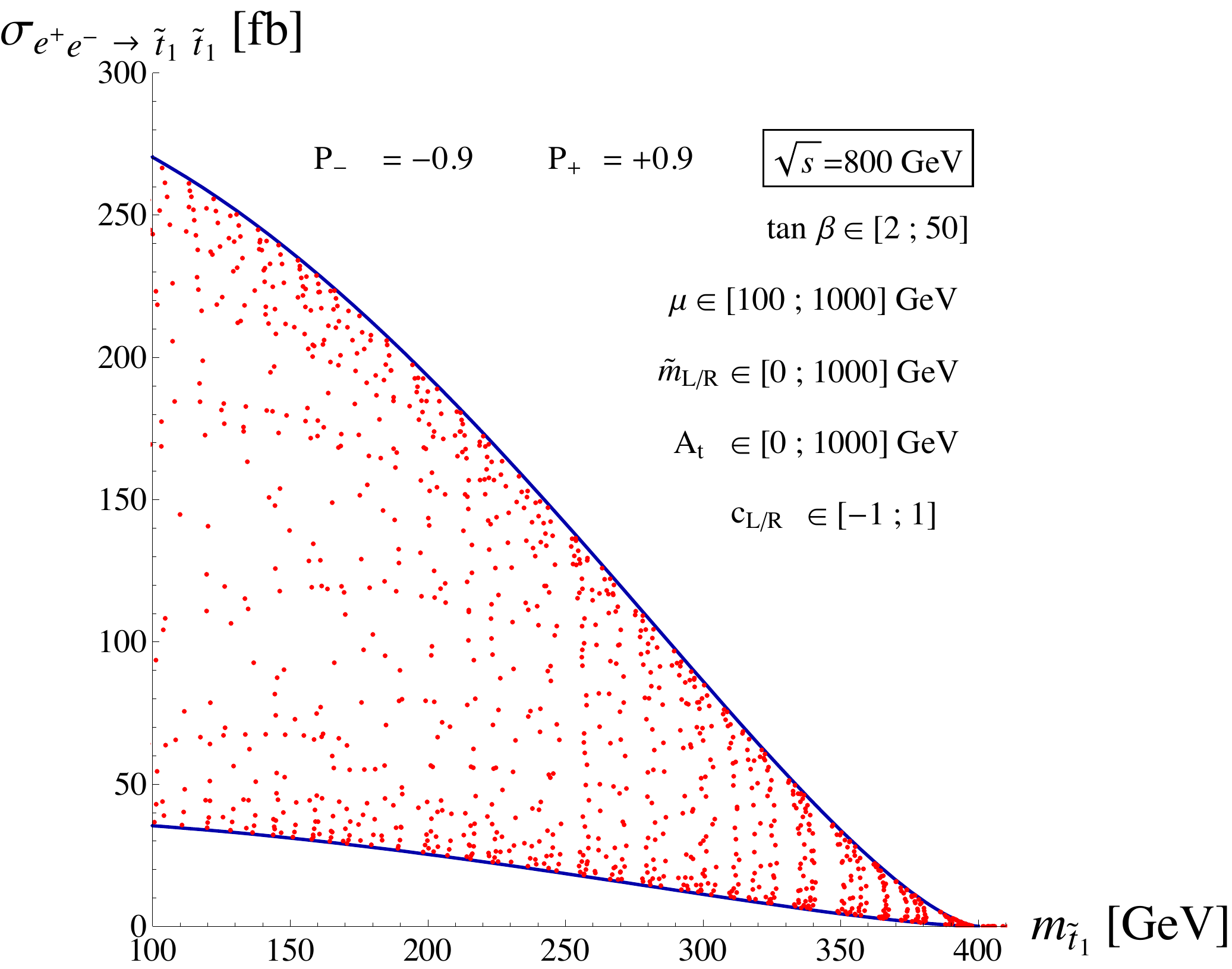}
  	\caption{ \small{Same as in Fig.(\ref{fig-500-11}) for $\sqrt{s} = 800$ GeV and $\mathcal{P}_{\pm} = \pm 0.9$.}}
  \label{fig-800-11}
\end{figure}

In Fig.(\ref{fig-500-11}) we present the cross sections for the reaction $e^+ e^- \rightarrow \tilde{t}_1 \tilde{t}_1$ as a function of the lightest stop mass $m_{\tilde{t}_1}$ within the
4D pMSSM and RS pMSSM with(out) soft terms [i.e. stop masses and trilinear coupling] on the TeV-brane. 
\\ Those results are derived from scans on the parameters 
$c_{\tilde t_{L/R}}$, $\tan\beta$, $\mu$ (or $\mu_{eff}$ [see Eq.(\ref{mueff})]), $\tilde m_{L/R}$ ($\tilde{m}_{L/R}|_{eff} = (\tilde{m}_{L/R}k)^{1/2}e^{-k\pi R_c}$ in RS SUSY)
and $A_t$ (or $A_t|_{eff} = A_tke^{-k\pi R_c}$).
As before, the allowed ranges in our scan for each parameter are $\tan\beta \in [2;50]$, $\mu, \mu_{eff} \in [100;1000]$ GeV, $c_{\tilde t_{L/R}} \in [-1,1]$, 
$\tilde{m}_{L/R}, \tilde{m}_{L/R}|_{eff} \in [0;1000]$ GeV and $A_t, A_t|_{eff} \in [0;1000]$ GeV. 
The experimental constraint on the stop mass, $m_{\tilde t_1}>95.7$ GeV \cite{PDG,TevWebPage}, is respected on the plots. Once again,
the signs of $A_t$ ($A_t|_{eff}$) and $\mu$ ($\mu_{eff}$) do not affect the present analysis.
\\ÊThe scanned points on the three plots of this figure never cross the drawn lines. The reason is that these lower and upper lines, obtained analytically, correspond 
respectively to the cross sections for the extremal $Z\tilde{t}_1\overline{\tilde{t}}_1$ couplings $c_{11} = Q_Z^{t_{R}}$ and $c_{11} = Q_Z^{t_{L}}$ associated 
to $\cos\theta_{\tilde{t}}=0$ and $\cos\theta_{\tilde{t}}=1$, as shown in Eq.(\ref{eq:Zstopcoupling}). We also observe that the obtained regions in the plots end up
at the kinematic limit as expected. 
\\ Finally, Fig.(\ref{fig-500-11}) was obtained at a center-of-mass energy, $\sqrt{s} = 500$ GeV, and for some polarizations of the two beams, 
$\mathcal{P}_{\pm} = \pm 0.6$, which are realistic for an ILC prospect. Changing these experimental inputs, $\sqrt{s}$ and $\mathcal{P}_{\pm}$, 
changes the production amplitudes as well as kinematic limits but these inputs do not affect the main shape of the obtained scans and are thus not really
relevant for our study. This can be observed by comparing Fig.(\ref{fig-500-11}) with Fig.(\ref{fig-800-11}) where the same scans have been done for $\sqrt{s} = 800$ GeV 
and $\mathcal{P}_{\pm} = \pm 0.9$. One can thus freely adjust the center-of-mass energy and beam polarizations to get an optimal ratio of signal over backgrounds.

Now that we have described the way Fig.(\ref{fig-500-11}) and Fig.(\ref{fig-800-11}) were obtained, we can compare the three plots of say Fig.(\ref{fig-500-11}).  
The first conclusion is that the whole shapes of the scans within the 4D and 5D SUSY frameworks do not offer distinctive features (typically 
in both cases the entire domain in-between the upper and lower analytical limits can be filled), forbidding thus to develop a new test of discrimination between 
these two SUSY realizations.   
\\ Comparing the middle and right plots of Fig.(\ref{fig-500-11}), we observe that in the case without brane soft terms only cross section values roughly in the   
higher half of the theoretically allowed region are reached. This means that the relevant $Z$ charge $c_{11}$ of the $\tilde{t}_1$ squark [{\it c.f.} Eq.(\ref{eq:Zstopcoupling})] is 
close to $Q_Z^{t_L}$ 
($c_{11}=Q_Z^{t_L}$ corresponds to the upper limit in scans) and hence that $\tilde{t}_1$ is never too far from being mainly composed by the $\tilde{t}_L$ state
i.e. $\cos\theta_{\tilde{t}}$ is constrained (depending on $m_{\tilde{t}_1}$) to be relatively close to unity accordingly to Eq.(\ref{eq:defthetamix}). 
This result has to be enlightened with the previously explained upper bound (depending on $m_{\tilde{t}_1}$)
on the highest stop mass eigenvalue $m_{\tilde{t}_2}$ exhibited in Fig.(\ref{mst1mst2}) [left plot] 
in the absence of brane soft terms. The presence of an upper bound on $m_{\tilde{t}_2}$ reflects the existence of an upper 
limit on the amplitude of mixing between the two $\tilde{t}_L$ and $\tilde{t}_R$ squark states. 
More formally, the mixing angle $\theta_{\tilde{t}} \in [0,\pi/2]$ in the 4D pMSSM [see Eq.(\ref{eq:defthetamix})] is constrained to a smaller 
range $[0,\theta_{RS}]$ (with $\theta_{RS} < \pi/2$) in the warped pMSSM without brane soft terms. So we recover the conclusion from the middle plot of Fig.(\ref{fig-500-11}). 
Starting from the case of the middle plot of Fig.(\ref{fig-500-11}) and adding in particular a soft trilinear stop coupling on the TeV-brane has the virtue to enlarge the mixing 
between $\tilde{t}_L$ and $\tilde{t}_R$ after EWSB. Doing so, the scanned points now fill the whole allowed theoretical region 
as can be seen on the right plot of Fig.(\ref{fig-500-11}).

\begin{figure}[!hc]
\vspace*{1.5cm}
\includegraphics[width=0.325\textwidth]{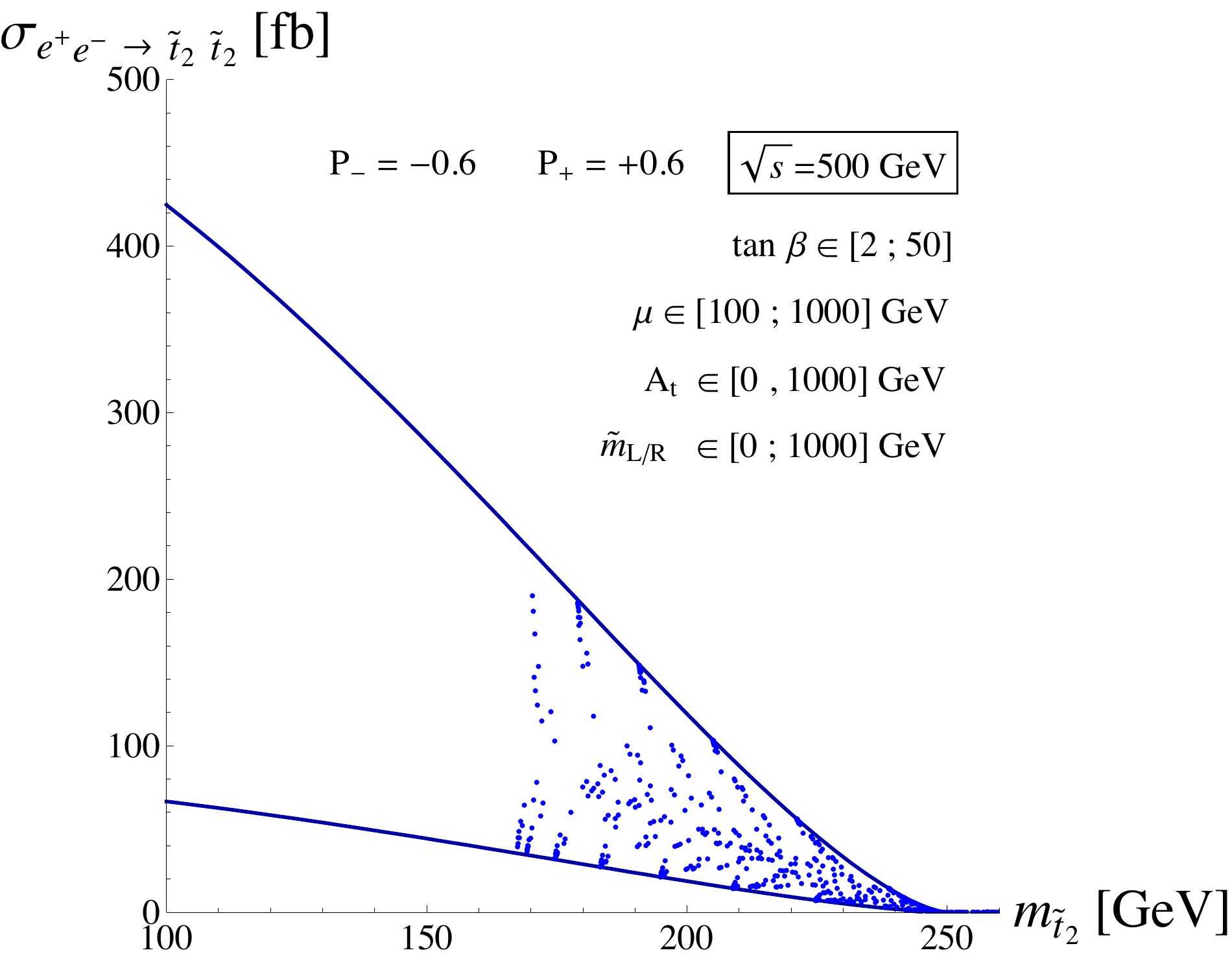}
\includegraphics[width=0.325\textwidth]{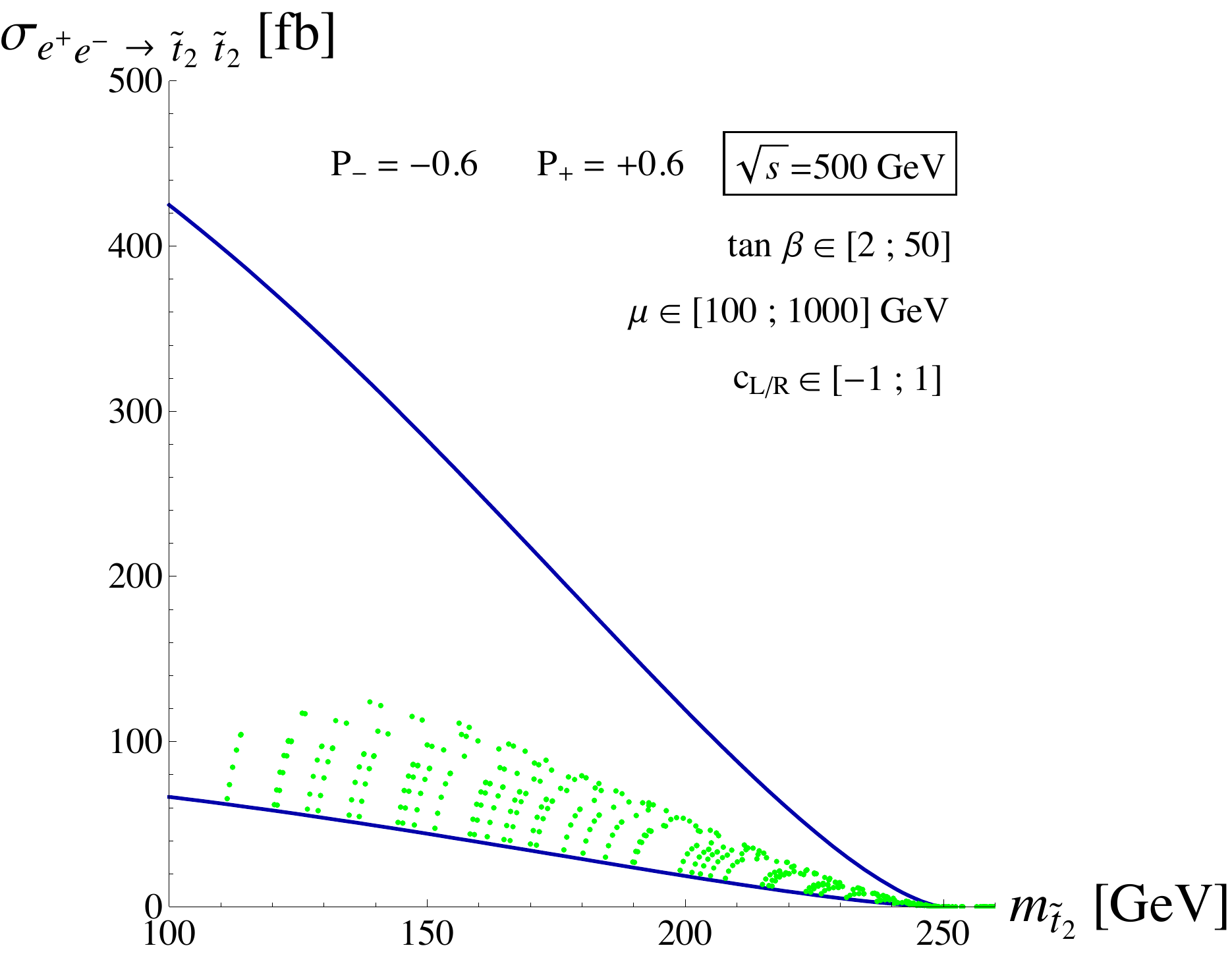}
\includegraphics[width=0.325\textwidth]{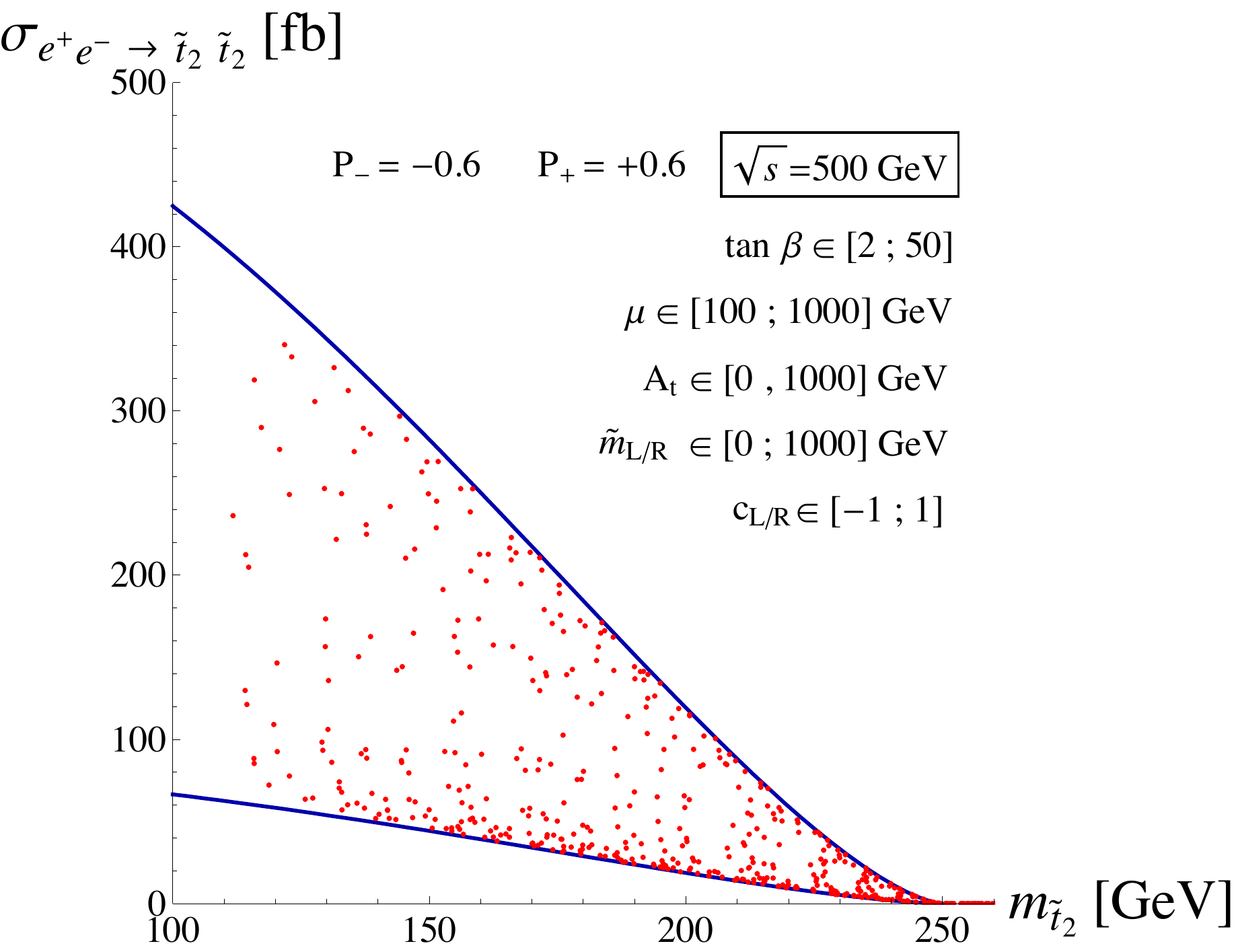}
\caption{ \small{Cross section for the reaction $e^+ e^- \rightarrow \tilde{t}_2 \overline{\tilde{t}}_2$ (in $fb$), at a center-of-mass energy $\sqrt{s} = 500 \mbox{ GeV}$  
	and with the polarizations $\mathcal{P}_\pm$ indicated on the plot, as a function of $m_{\tilde{t}_2}$ (in GeV) 
	within the 4D pMSSM [left plot], RS pMSSM without brane soft terms [middle] and RS pMSSM with brane soft terms  
	[right]. All points presented here are obtained from a scan inside the ranges given on the plots for the parameters $c_{L/R}=c_{\tilde t_{L/R}}$, $\tan\beta$, 
	$\mu$ (and $\mu_{eff}$ in RS), the soft stop masses $\tilde m_{L/R}$ (and brane masses $\tilde{m}_{L/R}|_{eff}$) and the soft trilinear coupling $A_t$ 
	(and brane coupling $A_t |_{eff}$).
	The limiting thick blue lines correspond to the analytically obtained cross sections for an extremal $Z\tilde{t}_2\overline{\tilde{t}}_2$ coupling: 
	$c_{22} = Q_Z^{t_{R}}$ [lower limit] and $c_{22} = Q_Z^{t_{L}}$ [upper limit].} }
  \label{fig-500-22}
\end{figure}

In Fig.(\ref{fig-500-22}), we further illustrate the $e^+ e^- \rightarrow \tilde{t}_2 \overline{\tilde{t}}_2$ reaction as a function of $m_{\tilde{t}_2}$, with $\sqrt{s} = 500 \mbox{ GeV}$. 
The scanned points on plots never cross the drawn lines corresponding  
respectively to the cross sections for the extremal $Z\tilde{t}_2\overline{\tilde{t}}_2$ couplings $c_{22} = Q_Z^{t_{R}}$ and $c_{22} = Q_Z^{t_{L}}$ associated 
to $\cos\theta_{\tilde{t}}=1$ and $\cos\theta_{\tilde{t}}=0$ (see Eq.(\ref{eq:Zstopcoupling})). We observe that in the 4D pMSSM case [left plot], 
there is a theoretical lower limit on $m_{\tilde{t}_2}$ around $m_t$
as we have already described when discussing Fig.(\ref{mst1mst2}) and Fig.(\ref{mst1mst2MSSM})
[see also Eq.(\ref{min4Dmstop})]. 
\\
In the middle plot of Fig.(\ref{fig-500-22}), without brane soft terms, only cross section values in a    
lower part of the theoretical region are reached that is to say that the relevant $Z$ charge $c_{22}$ of the $\tilde{t}_2$ squark 
[{\it c.f.} Eq.(\ref{eq:Zstopcoupling})] is close to $Q_Z^{t_R}$ ($c_{22}=Q_Z^{t_R}$ corresponds to the lower limit in the scan) 
and thus that $\tilde{t}_2$ is never too far from being mainly composed by the $\tilde{t}_R$ state. This is consistent with the analysis of
the middle plot of Fig.(\ref{fig-500-11}) which has revealed that $\tilde{t}_1$ is never too far from being mainly composed by the $\tilde{t}_R$.

\begin{figure}[!hc]
\centering 
\vspace*{1.5cm}
\includegraphics[width=0.325\textwidth]{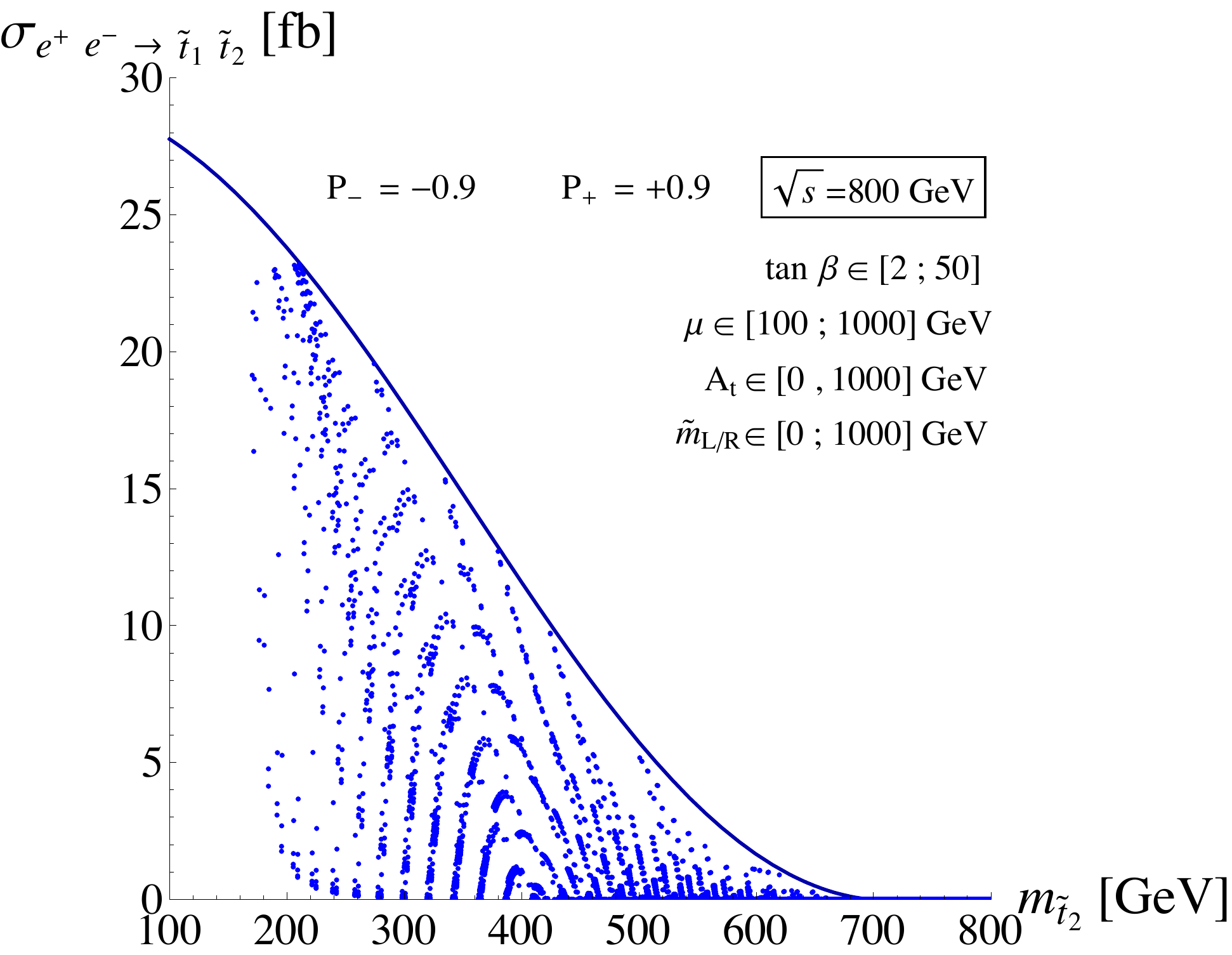}
\includegraphics[width=0.325\textwidth]{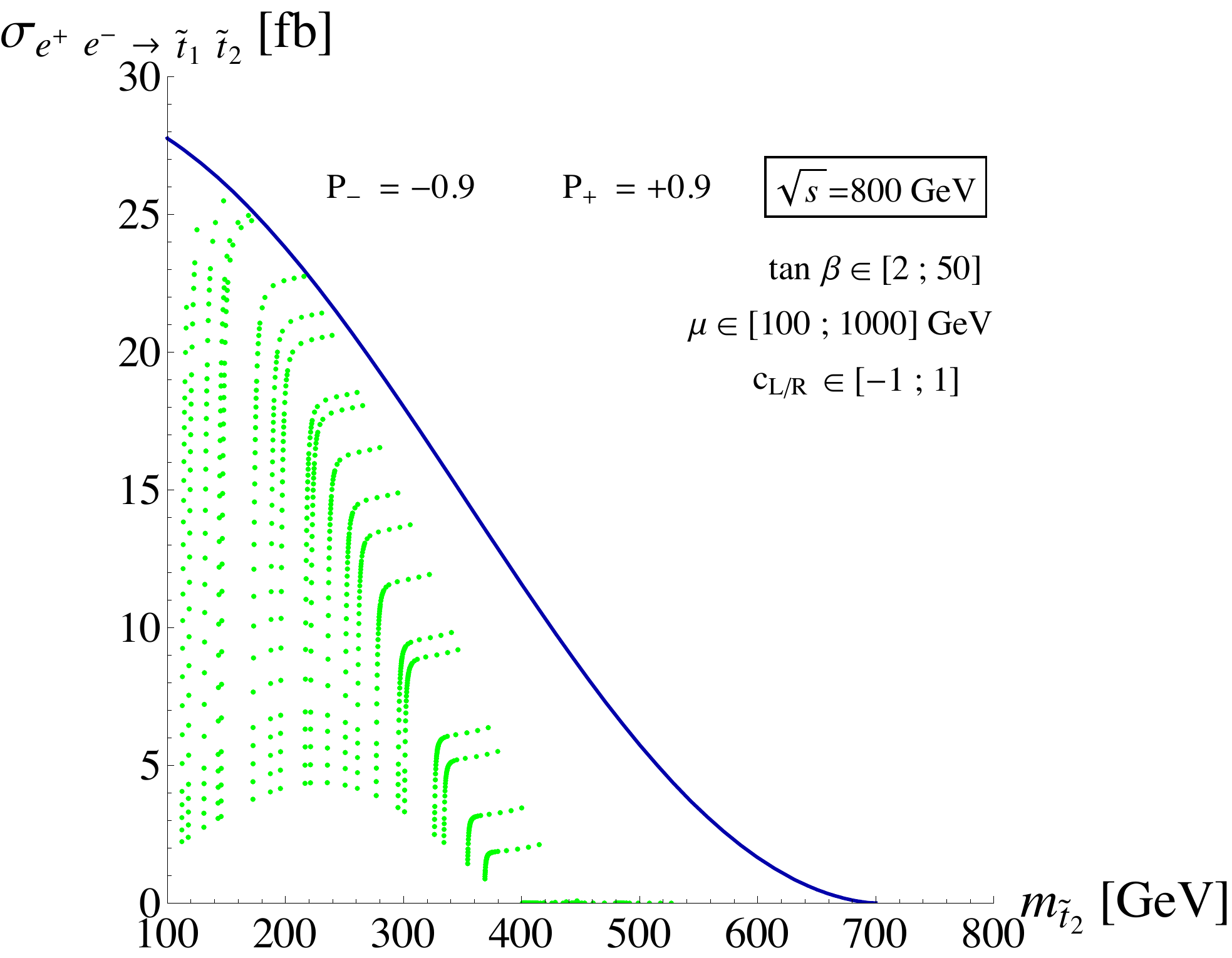}
\includegraphics[width=0.325\textwidth]{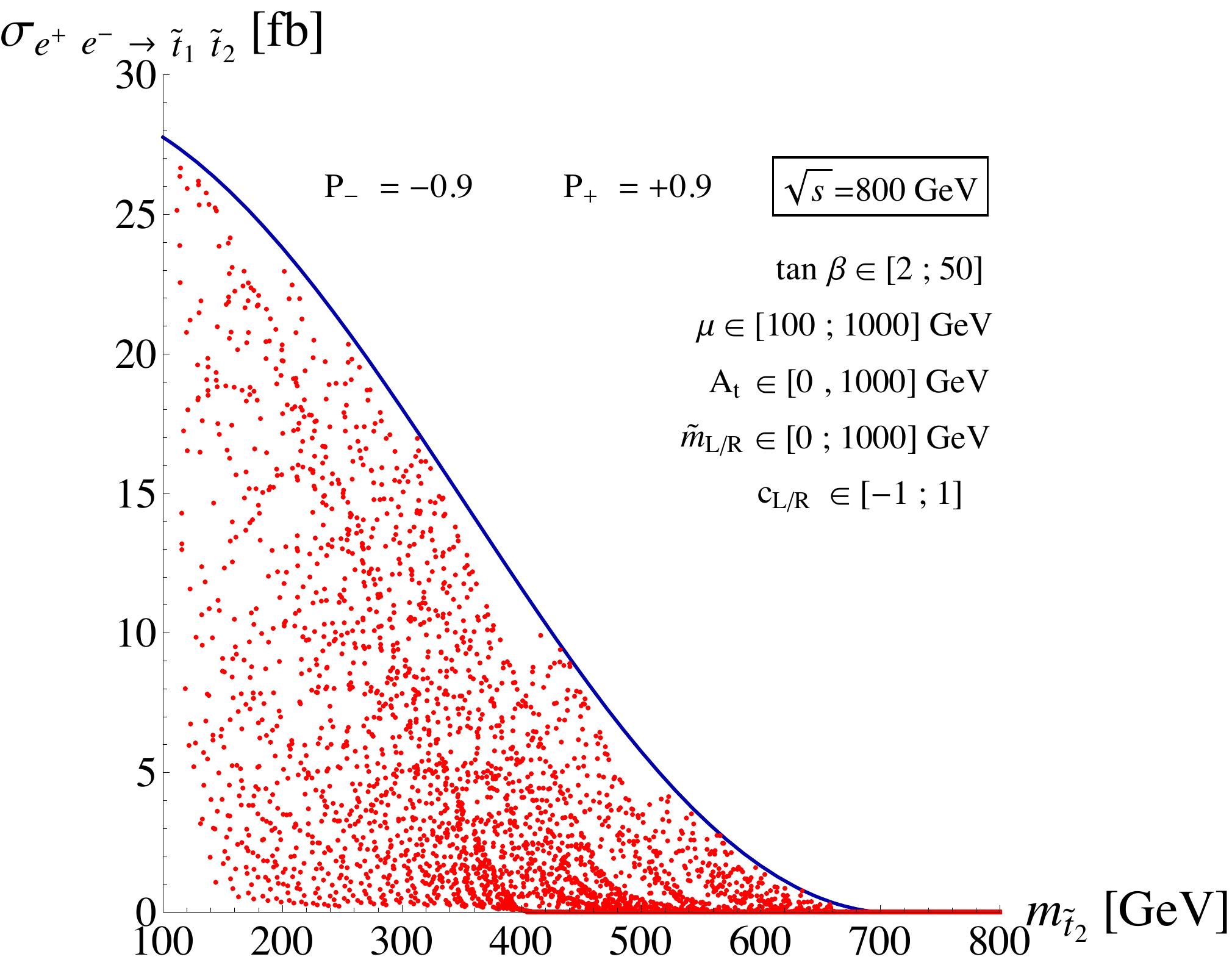}  	
\caption{ \small{Cross section for the reaction $e^+ e^- \rightarrow \tilde{t}_1 \overline{\tilde{t}}_2$ (in $fb$) at a center-of-mass 
       energy $\sqrt{s} = 800 \mbox{ GeV}$  
       as a function of $m_{\tilde{t}_2}$ (in GeV) 
	within the 4D pMSSM [left plot], RS pMSSM without brane soft terms [middle] and RS pMSSM with brane soft terms  
	[right]. All points presented here are obtained from a scan inside the ranges given on the plots for the parameters $c_{L/R}=c_{\tilde t_{L/R}}$, $\tan\beta$, 
	$\mu$ (and $\mu_{eff}$ in RS), the soft stop masses $\tilde m_{L/R}$ (and brane masses $\tilde{m}_{L/R}|_{eff}$) and the soft trilinear coupling $A_t$ 
	(and brane coupling $A_t |_{eff}$).
	The limiting thick blue line corresponds to the analytically obtained cross section for an extremal $Z\tilde{t}_1\overline{\tilde{t}}_2$ coupling: 
	 $c_{12} = -(1/2)\times I_{3L}^{t} =-1/4$.}}
  \label{fig-800-12}
\end{figure}

In Fig.(\ref{fig-800-12}), we finally illustrate the $e^+ e^- \rightarrow \tilde{t}_1 \overline{\tilde{t}}_2$ reaction as a function of $m_{\tilde{t}_2}$, with $\sqrt{s} = 800 \mbox{ GeV}$. 
Now the scanned points on plots never cross the line corresponding  
to the cross section for the extremal $Z\tilde{t}_1\overline{\tilde{t}}_2$ coupling 
$c_{12} = -(1/2)\times I_{3L}^{t} =-1/4$ [unique line drawn on the three plots] associated 
to $\sin2\theta_{\tilde{t}}=1$ (see Eq.(\ref{eq:Zstopcoupling})). 
\\
In the middle plot of Fig.(\ref{fig-800-12}), without brane soft terms, the reason why the cross section values never reach the analytical line at $c_{12}=-1/4$
is that $\sin2\theta_{\tilde{t}}$ never reaches the unity. Indeed, we have seen that, in the warped pMSSM without brane soft terms the mixing angle 
$\theta_{\tilde{t}}$ is constrained from above.

For instance, a measurement of the cross section $\sigma(e^+ e^- \rightarrow \tilde{t}_1 \overline{\tilde{t}}_1)$ and mass $m_{\tilde{t}_1}$ at ILC could allow
to distinguish between the two RS scenarios of SUSY breaking. Indeed, the case where the experimental values of $\sigma(e^+ e^- \rightarrow \tilde{t}_1 
\overline{\tilde{t}}_1)$ and $m_{\tilde{t}_1}$ would fall outside the theoretically predicted domain in the middle plot of Fig.(\ref{fig-500-11}) would exclude
the scenario without soft scalar mass and trilinear $A$ terms on the TeV-brane. Such measurements would be compatible with the other RS version (presence
of brane soft terms) if they belong to the theoretical region in the right plot of Fig.(\ref{fig-500-11}) [otherwise these experimental data would conflict
both with 4D and 5D SUSY].   
\\ The same kind of test could be investigated with measurements of $\sigma(e^+ e^- \rightarrow \tilde{t}_2 \overline{\tilde{t}}_2)$ and $m_{\tilde{t}_2}$ 
[see Fig.(\ref{fig-500-22})] or by reconstructing $\sigma(e^+ e^- \rightarrow \tilde{t}_1 \overline{\tilde{t}}_2)$ and e.g. $m_{\tilde{t}_2}$ 
[see Fig.(\ref{fig-800-12})]. It is interesting to note that performing the three above tests together at ILC could even lead to correlated indications for the RS model 
with brane soft terms. 
\\ Such tests are possible if the predicted theoretical regions have fixed conservative limits. Now we have seen when discussing Fig.(\ref{mst1mst2}) that the theoretical region    
was limited (from above) by the minimal $c_{\tilde t_{L/R}}$ values. It is thus [see above discussion] the case also for the theoretical domains obtained in the middle plots of
Fig.(\ref{fig-500-11}), Fig.(\ref{fig-800-11}), Fig.(\ref{fig-500-22}) and Fig.(\ref{fig-800-12}) -- not speaking about the absolute limits (blue lines) induced by 
the extremal $Z$ couplings. However, in the case of a bulk gauge custodial symmetry, one should impose typically  
$c_{\tilde t_{L/R}} \gtrsim -0.5$ \cite{ADMS} to avoid the existence of too light scustodians which would have been observed at colliders. There would thus be 
fixed limits to the theoretically allowed domains. 
By the way the above constraint, more strict than the one we have imposed: $c_{\tilde t_{L/R}} > -1$, would lead to even more restricted area of spread 
scanned points. 
Furthermore, some information (and hence constraints) can be obtained on the $c_{\tilde t_{L/R}}$ parameters from stop mass related
measurements.   
\\ Concerning the experimental feasibility of such tests, a reasonable precision expected at ILC for the $\sigma(e^+ e^- \rightarrow \tilde{t}_i 
\overline{\tilde{t}}_j)$ measurement is $\delta\sigma_{exp} \sim 5\%$ \cite{Bartl} ($\sim 1\%$ at most \cite{FRICHARD}) while it was shown in Ref.~\cite{Feng}
that $m_{\tilde{t}}$-reconstructions at ILC with an accuracy around one percent can be obtained using $20$ fb$^{-1}$ of data. 
The exact uncertainties reachable at ILC depend on the methods used and SUSY (breaking) parameters. Stating on the above 
uncertainties at the percent level is realistic and sufficient for the realizations of the proposed tests, given the large sizes of theoretically forbidden regions in the
middle plots of Fig.(\ref{fig-500-11})-(\ref{fig-800-12}).

\section{Conclusion}
\label{concludepart}

If the fundamental theory of nature is of string theory kind, the effective low-energy model 
manifesting itself at present and future particle colliders would probably be an higher-dimensional SUSY model.
In this paper, we have studied the theory and phenomenology of such models entering the class of
particularly well-motivated frameworks with a warped space-time.

First, we have derived in a consistent analytical way, within the realistic 5D pMSSM context,
the effective 4D Lagrangian for the brane-Higgs interactions as well as the complete 
sfermion mass matrices (illustrating explicitly the example of the stop quark) induced 
partially by such interactions.

Those theoretical investigations have provided us with the necessary tools to
perform concrete phenomenological studies of the RS SUSY models. In this respect, we have first  
demonstrated that the cancellation of quadratic divergences in the one-loop corrections to the
Higgs mass is deeply related to the 5D anomaly cancellation (similarly to 4D).
We have also found that the quadratic divergence cancellation occurs for any 5D cut-off, which means possibly for
a `truncated' 5D SUSY theory (as it must be due to its non-renormalizable aspect).
In these loop calculations, the accent was also put on the justification of the infinite KK summation 
required in the KK regularization which has open up a rich debate in the literature a decade ago.  
Possible perspectives are the extensions of these calculations to the sfermion masses.
\\ Concerning collider physics, we have tried to answer a crucial question: in the hypothetical
situation where some superpartners
were discovered whereas all KK modes were outside the reach of direct detection (at LHC and/or ILC), 
how one could distinguish experimentally between a pure 4D SUSY scenario and a warped 5D SUSY model ?  
In particular, the virtual effects of KK gluon excitations on the stop pair production
at LHC might certainly be at least as tricky to detect as the similar effects for the top pair final state \cite{LHCgluonKK}.
We have developed series of tests that could be more clear or at least complementary. For instance, we have shown that
the heaviest stop eigenstate can reach lower mass values in RS SUSY than in 4D SUSY.
Other clear 4D/5D SUSY differences, that may be used for data-based discriminations, are those arising from
the slepton mass sector of mSUGRA scenarios where larger mass splittings can occur in 5D setups. 
Finally, in the same philosophy applied to the SUSY Higgs sector, it has been shown in particular that branching ratios
of $H$ decay channels into sleptons can reach dominant levels in the RS pMSSM, a feature absent of the   
conventional 4D SUSY models.
In the future, it will be interesting to explore a related direction: the search at hadron colliders
for the SM Higgs boson -- or the lightest neutral 4D SUSY Higgs field $h$ -- could be affected by new significant
contributions to the gluon-fusion mechanism induced by 5D effects [if not too much constrained by 
present experimental data], effects representing other potential distinctive signatures of 5D SUSY scenarios. 
\\ Our phenomenological 
analysis was end up with the suggestion of complementary methods for pinning up the presence of soft terms on the IR boundary
using stop pair productions at ILC. This task being more related to the discrimination between 
various RS SUSY versions.     
\\
\\ 
\noindent \textbf{Acknowledgments:}  
The authors are grateful to Aldo~Deandrea, Abdelhak~Djouadi, Ulrich~El\-lwanger, Gero~v.~Gersdorff, Christophe Grojean, 
Swarup~K.~Majee, Fran\c{c}ois~Richard and James~D.~Wells for useful discussions as well as to 
Marc~Besan\c{c}on for motivating conversations at early stages of this work. 
We also thank M.~Calvet for her contribution to the manuscript. 
This work is supported by the European network HEPTOOLS and the A.N.R. {\it CPV-LFV-LHC} 
under project \textbf{NT09-508531} as well as 
the A.N.R. {\it TAPDMS} under project \textbf{09-JCJC-0146}.

\newpage 

\newpage

\appendix 
\noindent \textbf{\Large Appendix} \vspace{0.5cm}


\renewcommand{\thesubsection}{A.\arabic{subsection}} 
\renewcommand{\theequation}{A.\arabic{equation}} 
\setcounter{subsection}{0} 
\setcounter{equation}{0} 

\section{The supersymmetric 5D Lagrangian}
\label{lagrangianpart} 

We consider the usual warped space-time based on the 5D metric $G_{AB}$:
\begin{eqnarray}
&& ds^2 = G_{AB}\,dx^A dx^B = e^{-2\sigma(y)}\eta_{\mu\nu}dx^{\mu} dx^{\nu} - dy^2 \quad,\quad y \in [-\pi R_c;+\pi R_c]  \\
&& \sqrt{+G} \equiv \sqrt{\det[G_{AB}]} = \sqrt{(-1)^4(e^{-2\sigma(y)})^4} = e^{-4\sigma(y)} 
\end{eqnarray}
where capital latin letters $A,B$ are 5D Lorentz indices, 
$\eta_{\mu\nu}$ is the 4D flat $(+,-,-,-)$ Minkowski metric
and $x^{\mu}$ denote the usual coordinates ($\mu=0,1,2,3$) while $y$ parametrizes the fifth dimension.

\subsection{Superfield content}
\label{SuperContent}

Writing higher-dimensional SUSY Lagrangians with ordinary $N=1$ 4D superfields only allows to make the $N=1$ 4D SUSY invariance manifest and it prevents 
from explicitly covariant forms. In spite of these limitations, it allows for a more compact form than when using the component fields themselves and also to easily get the 
bulk-boundary couplings. 
We thus adopt this formalism. This approach was first developed for theories with 10 dimensions \cite{4DSuperFA}, then extended to other dimensions 
\cite{4DSuperFB} and to incorporate the radion superfield \cite{4DSuperFC}.

In this Appendix \ref{lagrangianpart}, we derive explicitly in terms of the fields, within the RS SUSY framework with a bulk ${\rm U(1)}$ 
gauge symmetry, the whole Lagrangian encoding the Yukawa and gauge couplings of the two 
$N=1$ 5D (or $N=2$ 4D) bulk hypermultiplets $\{\Phi_L,\Phi_L^{--}\}$ and $\{\Phi^c_L,\Phi_L^{c--}\}$ to two $N=1$ 4D complex chiral superfields localized on the TeV-brane 
called $H_{u}^0$ (as it will play the role of the `up-type' neutral Higgs superfield when extending this toy model to the pMSSM) and $H_{d}^0$ (for `down-type'). 
The $N=1$ chiral superfield $\Phi_L$ 
[the subscript $L$ corresponds to the chirality of the contained fermion field once promoted to a four-component spinor: see just below
\footnote{Note that $L/R$ would be seen as a gauge index within the pMSSM, corresponding to a doublet/singlet under ${\rm SU(2)_L}$.}] 
with Neumann BC at $y=0,\pi R_c$ (noted $(++)$ in the main text) reads as  
\bea
\Phi_L(x^{\mu},y;\theta,\overline{\theta}) &\equiv& \phi_L(x^{\mu},y) + \sqrt{2} \theta \zeta_L(x^{\mu},y) e^{-{1\over2}\sigma(y)} - \theta\theta F_L(x^{\mu},y) 
+ i \theta \sigma^{\mu} \overline{\theta} \partial_{\mu}\phi_L(x^{\mu},y) 
\nonumber\\
&& + \frac{i}{\sqrt{2}} \theta\theta \overline{\theta} \overline{\sigma}^{\mu} \partial_{\mu} \zeta_L(x^{\mu},y) e^{-{1\over2}\sigma(y)} 
- {1\over4} \theta\theta\overline{\theta}\overline{\theta} \partial_{\mu}\partial^{\mu}\phi_L(x^{\mu},y) , 
\label{eq:phiL}
\eea
$\zeta_L$ being the two-component spinorial field, $\phi_L$ its scalar superpartner and $F_L$ the auxiliary field,
whereas its complex conjugated develops according to 
\bea
\overline{\Phi}(x^{\mu},y;\theta,\overline{\theta}) &\equiv& \overline{\phi}_L(x^{\mu},y) + \sqrt{2} \overline{\theta} \overline{\zeta}_L(x^{\mu},y) e^{-{1\over2}\sigma(y)} - \overline{\theta}\overline{\theta} \overline{F}_L(x^{\mu},y) - i 
\theta \sigma^{\mu} \overline{\theta} \partial_{\mu}\overline{\phi}_L(x^{\mu},y)
\nonumber\\
&& + \frac{i}{\sqrt{2}} \overline{\theta\theta} \theta \sigma^{\mu} \partial_{\mu} \overline{\zeta}_L(x^{\mu},y) e^{-{1\over2}\sigma(y)} 
- {1\over4} \overline{\theta}\overline{\theta}\theta \theta \partial_{\mu}\partial^{\mu} \overline{\phi}_L(x^{\mu},y)
\eea
and its charge conjugated superfield with Dirichlet BC at $y=0,\pi R_c$ (noted $(--)$) is
\bea
\Phi^{--}_L(x^{\mu},y;\theta,\overline{\theta}) &\equiv& \phi^c_L(x^{\mu},y) + \sqrt{2} \theta \chi_L(x^{\mu},y) e^{-{1\over2}\sigma(y)} - \theta\theta F_L^c(x^{\mu},y) + i \theta \sigma^{\mu} \overline{\theta} \partial_{\mu}\phi^c_L(x^{\mu},y) 
\nonumber\\
&& + \frac{i}{\sqrt{2}} \theta\theta \overline{\theta}\overline{\sigma}^{\mu}\partial_{\mu}\chi_L(x^{\mu},y) e^{-{1\over2}\sigma(y)} - {1\over4} \theta\theta\overline{\theta}\overline{\theta} \partial_{\mu}\partial^{\mu}\phi^c_L(x^{\mu},y).
\label{eq:phicL}
\eea
We define another superfield $\Phi^c_L$ through its charge conjugated state [this will allow us to introduce only Left-handed chiral superfields, as usually in the 4D pMSSM]:
\bea
\Phi^c_L (x^{\mu},y;\theta,\overline{\theta}) &\equiv& \phi_R(x^{\mu},y) + \sqrt{2} \theta \chi_R(x^{\mu},y) e^{-{1\over2}\sigma(y)} - \theta\theta F_R(x^{\mu},y) + i \theta \sigma^{\mu} \overline{\theta} \partial_{\mu}\phi_R(x^{\mu},y) 
\nonumber\\
&& + \frac{i}{\sqrt{2}} \theta\theta \overline{\theta}\overline{\sigma}^{\mu}\partial_{\mu}\chi_R(x^{\mu},y) e^{-{1\over2}\sigma(y)}  
- {1\over4} \theta\theta\overline{\theta}\overline{\theta} \partial_{\mu}\partial^{\mu}\phi_R(x^{\mu},y) 
\label{eq:phiR}
\eea
and the opposite BC superfield is:
\bea
\Phi_L^{c--} (x^{\mu},y;\theta,\overline{\theta}) &\equiv& \phi^c_R(x^{\mu},y) + \sqrt{2} \theta \zeta_R(x^{\mu},y) e^{-{1\over2}\sigma(y)} - \theta\theta F_R^c(x^{\mu},y) + i \theta \sigma^{\mu} \overline{\theta} \partial_{\mu}\phi^c_R(x^{\mu},y) 
\nonumber\\
&& + \frac{i}{\sqrt{2}} \theta\theta \overline{\theta}\overline{\sigma}^{\mu}\partial_{\mu}\zeta_R(x^{\mu},y) e^{-{1\over2}\sigma(y)}    
- {1\over4} \theta\theta\overline{\theta}\overline{\theta} \partial_{\mu}\partial^{\mu}\phi^c_R(x^{\mu},y).
\label{eq:phicR}
\eea
In our notations, the four-component fermions $\psi_{L/R}$ ($L/R$ indicating the Lorentz chirality) read in terms of the two-component fields as
$$
\psi^t \equiv (\zeta ,\bar \chi), \ \ \ \psi_L^t \equiv (\zeta ,0), \ \ \ \psi^{c \ t} \equiv (\chi,\bar \zeta). 
$$
We consider a ${\rm U(1)}$ gauge symmetry whose $N=1$ 5D (or $N=2$ 4D) gauge supermultiplet is known to have 
the same field content as {\bf one} $N=1$ vector supermultiplet 
with $(++)$ BC, for which we write the decomposition as (in the Wess-Zumino gauge) 
\begin{eqnarray}
V(x^{\mu},y;\theta,\overline{\theta}) &\equiv& \theta \sigma^{\mu} \overline{\theta} A_{\mu}(x^{\mu},y) 
- i \overline{\theta}\overline{\theta} \theta \lambda_1(x^{\mu},y) e^{-{3\over2}\sigma(y)} + i \theta\theta \overline{\theta} 
\overline{\lambda}_1(x^{\mu},y) e^{-{3\over2}\sigma(y)} + {1\over2} \theta\theta \overline{\theta}\overline{\theta} D(x^{\mu},y)
\label{SuperVectorial}
\end{eqnarray}
$A_{\mu}$ denoting the gauge boson, $\lambda_1$ the Weyl gaugino field and $D$ the complex auxiliary field,
plus {\bf one} $(--)$ $N=1$ chiral superfield, that we write 
\begin{eqnarray}
\Omega(x^{\mu},y;\theta,\overline{\theta}) &\equiv& 
{1\over\sqrt{2}} [\Sigma(x^{\mu},y) - i A_5(x^{\mu},y)] -i \sqrt{2} \theta \lambda_2(x^{\mu},y) e^{-{1\over2}\sigma(y)} - \theta\theta F_{\Omega}(x^{\mu},y) 
\nonumber\\
&& + i \theta \sigma^{\mu} \overline{\theta} \partial_{\mu}{1\over\sqrt{2}} [\Sigma(x^{\mu},y) - i A_5(x^{\mu},y)]
+ \frac{1}{\sqrt{2}} \theta\theta \overline{\theta} \overline{\sigma}^{\mu} \partial_{\mu} \lambda_2(x^{\mu},y) e^{-{1\over2}\sigma(y)} 
\nonumber\\
&& - {1\over4} \theta\theta\overline{\theta}\overline{\theta} \partial_{\mu}\partial^{\mu}{1\over\sqrt{2}} [\Sigma(x^{\mu},y) - i A_5(x^{\mu},y)]
\label{SuperVectorialbis}
\end{eqnarray}
$\Sigma$ being a real scalar field.

\subsection{Superfield action}
\label{SuperAction}

The field content described in previous subsection \ref{SuperContent} together with the following action $\mathcal{S}_{5D}$ 
define the toy model analyzed in this part \ref{lagrangianpart}. This action 
is given by $\mathcal{S}_{5D} = \mathcal{S}_{gauge}+ \mathcal{S}_{Higgs}\Big|_{brane} + \mathcal{S}_{matter} $ with,
following the formalism of Ref.~\cite{4DSuperFC},
\begin{eqnarray}
\mathcal{S}_{gauge} &=& {1\over4} \int\!\!\! d^5x\!\! \int\!\!\! d^2\theta\,\, \left( W^{\alpha}W_{\alpha} + h.c. \right)
+ \int\!\!\! d^5x\!\! \int\!\!\! d^4\theta\,\,e^{-2\sigma(y)} \left( \partial_y V - {1\over\sqrt{2}}(\Omega + \overline{\Omega}) \right)^2 
\label{gaugeAction} \\
\nonumber\\
\mathcal{S}_{matter} &=& \!\!\!\! \int\!\!\! d^5x\!\! \int\!\!\! d^4\theta\,\, e^{-2\sigma(y)}  
\left(\overline{\Phi}_L e^{-2 g q_L V} \Phi_L + \Phi_L^{--} e^{2 g q_L V} \overline{\Phi}_L^{--} + \overline{\Phi}_L^c e^{- 2 g q_R V} \Phi_L^c 
+ \Phi_L^{c--} e^{2 g q_R V} \overline{\Phi}_L^{c--} \right) \nonumber\\
&+& \!\!\!\! \int\!\!\! d^5x\!\!\! \int\!\!\! d^2\theta\,\, e^{-3\sigma(y)} \left( \Phi_L^{--} [ D_5 - \sqrt{2} g q_L\Omega ] \Phi_L + \Phi_L^{c--} [ D_5 - \sqrt{2} g q_R\Omega] 
\Phi_L^c \right) + h.c. 
\label{matterAction} \\
\nonumber \\ 
\mathcal{S}_{Higgs}\Big|_{brane} &=& \!\!\!\! \int\!\!\! d^5x\!\!\! \int\!\!\! d^4\theta\,\, e^{-2\sigma(y)} \left( \overline{H^0_u} e^{-(2 g q_{H_u^0}) V} 
H^0_u + \overline{H^0_d} e^{-(2 g q_{H_d^0}) V} H^0_d \right)\delta(y-\pi R_c) \nonumber\\
&+& \!\!\!\! \int\!\!\! d^5x\!\!\! \int\!\!\! d^2\theta\,\, e^{-3\sigma(y)} \left( \mu\, H_u^0 H^0_d + {\cal Y} H^0_u\Phi_L\Phi_L^c \right)\delta(y-\pi R_c) + h.c. 
\label{muAction}
\end{eqnarray}
where $g$ (${\cal Y}$) is the 5D gauge (Yukawa) 
coupling constant, $D_5 = \partial_y - ({3\over2}-c_{L/R})\sigma'$ [with $\sigma'=\partial_y\sigma(y)=sign(y) \times k$, 
$sign(y)$ being a step function] and  
the ${\rm U(1)}$ charges of the superfields $H_{u,d}^0$, $\Phi_L$, $\Phi_L^{c}$ must obey $q_{H_u^0}+q_L+q_R= 0$, $q_{H_u^0}+q_{H_d^0}= 0$. 
The above $c_{L/R}$ terms represent 
5D mass terms in the superpotential of the superfields $\Phi_L/\Phi_L^{c}$ that will lead to 5D fermion and 5D scalar mass terms, as will appear soon.
In the above action, the fundamental parameter $\mu$ is of order $k$.

\subsection{Auxiliary field Lagrangians}

The 5D Lagrangian for the gauge auxiliary field $D$ is given by (from now on, $\sigma$ stands for $\sigma(y)$):
\begin{eqnarray}
\mathcal{L}_{D} &=& {1\over2} D^2 + e^{-2\sigma} D \left( \partial_y - 2\sigma' \right) \Sigma \nonumber\\
&-& g e^{-2\sigma} D \left( q_L(\overline{\phi}_L  \phi_L  - \phi_L^{c} \overline{\phi}_L^{c}) + q_R(\overline{\phi}_R \phi_R - \phi_R^c \overline{\phi}_R^c) \right) \nonumber\\
&-& g e^{-2\sigma} D \left( q_{H_u^0} \overline{\phi}_{H_u^0}  \phi_{H_u^0}  + q_{H_d^0} \overline{\phi}_{H_d^0} \phi_{H_d^0} \right) \delta(y-\pi R_c) 
\end{eqnarray}
$\phi_{H_u^0}$ and $\phi_{H_d^0}$ being the scalar field components of the superfields $H_{u}^0$ and $H_{d}^0$, respectively.

The 5D Lagrangian for the auxiliary field $F_{\Omega}$ is given by:
\begin{eqnarray}
\mathcal{L}_{\Omega} &=& e^{-2\sigma} \overline{F}_{\Omega} F_{\Omega}
- \sqrt{2} g e^{-3\sigma} \left( (q_L\,\phi_L^{c} \phi_L  + q_R\,\phi_R^c \phi_R) F_{\Omega} + h.c. \right).
\end{eqnarray}

The 5D Lagrangian for the matter auxiliary fields, namely $F_L, F_R, F_L^c, F_R^c$, is given by:
\begin{eqnarray}
\mathcal{L}_{F} &=& e^{-2\sigma} \left(\overline{F}_L F_L  + F_L^ {c}\overline{F}_L^ {c} + \overline{F}_RF_R + F_R^c\overline{F}_R^c \right) \nonumber\\
&-& e^{-3\sigma} \left( F_L^ {c} D_5 \phi_L  - F_L  D_5' \phi_L^ {c} + F_R^c D_5 \phi_R - F_R D_5' \phi_R^c \right) + h.c. \nonumber\\
&+& e^{-3\sigma} 2 g \left( q_L (\phi_L^ {c} F_L  + F_L^ {c} \phi_L  ) + q_R (\phi_R^c F_R + F_R^c \phi_R) \right) \Sigma + h.c. \nonumber\\
&-& e^{-3\sigma} {\cal Y} \left( \phi_{H_u^0}  \phi_R F_L  + \phi_{H_u^0}  \phi_L  F_R \right) \delta(y-\pi R_c) + h.c.
\end{eqnarray}
where $D_5' = \partial_y - ({3\over2}+c_{L/R})\sigma'$. This Lagrangian is obtained after integrating by part (for convenience).

The 5D Lagrangian for the Higgs auxiliary fields, $F_{H_u^0},F_{H_d^0}$, is given by:
\begin{eqnarray}
\mathcal{L}_{H} &=& e^{-2\sigma} \left(\overline{F}_{H_u^0} F_{H_u^0}  + \overline{F}_{H_d^0} F_{H_d^0}  \right)\delta(y-\pi R_c) \nonumber\\
&-& e^{-3\sigma} \left( \mu\,F_{H_u^0} \phi_{H_d^0}  + \mu\,\phi_{H_u^0} F_{H_d^0}  + {\cal Y}\,F_{H_u^0} \phi_L \phi_R \right)\delta(y-\pi R_c) + h.c.
\end{eqnarray}

\subsection{Auxiliary field solutions}

The solutions of the equations of motion for the auxiliary fields are the following ones, 
\begin{eqnarray}
&& D = - e^{-2\sigma} \left\{ (\partial_y - 2\sigma')\Sigma - g \left( q_L(\overline{\phi}_L  \phi_L  - \phi_L^{c} \overline{\phi}_L^{c}) + q_R(\overline{\phi}_R \phi_R - \phi_R^c \overline{\phi}_R^c) \right) \right. \nonumber\\
&& \left. \qquad\qquad\qquad\qquad\qquad\qquad\qquad - g 
\left( q_{H_u^0} \overline{\phi}_{H_u^0}  \phi_{H_u^0}  + q_{H_d^0} \phi_{H_d^0}  \overline{\phi}_{H_d^0}  \right) \delta(y-\pi R_c) \right\} \\
&& F_{\Omega} = - \sqrt{2} g e^{-\sigma} \left( q_L\, \overline{\phi}_L^{c} \overline{\phi}_L  + q_R\, \overline{\phi}_R^c \overline{\phi}_R \right) \\
&& F_L  = - e^{-\sigma} \left( (D_5' + g \Sigma) \overline{\phi}_L^ {c} - \bar {\cal Y}\delta(y-\pi R_c)\, \overline{\phi}_{H_u^0} \overline{\phi}_R \right) \quad,\quad
\overline{F}_L^ {c} = e^{-\sigma} (D_5 - g \Sigma) \phi_L  \\
&& F_R = - e^{-\sigma} \left( (D_5' + g \Sigma) \overline{\phi}_R^c - \bar {\cal Y}\delta(y-\pi R_c)\, \overline{\phi}_{H_u^0} \overline{\phi}_L  \right) \quad,\quad
\overline{F}_R^c = e^{-\sigma} (D_5 - g \Sigma) \phi_R \\
&& F_{H_u^0}  = e^{-\sigma} \left( \bar \mu\, \overline{\phi}_{H_d^0}  + \bar {\cal Y}\, \overline{\phi}_L \overline{\phi}_R \right)
\quad,\quad
F_{H_d^0}  = e^{-\sigma} \bar \mu\, \overline{\phi}_{H_u^0} .
\end{eqnarray}

Plugging those back into the above Lagrangians, we get:
\begin{eqnarray}
\mathcal{L}_{D} &=& - {1\over2} D^2 \qquad,\qquad
\mathcal{L}_{\Omega} = - e^{-2\sigma} \overline{F}_{\Omega} F_{\Omega} \\
\mathcal{L}_{F} &=& - e^{-2\sigma} \left(\overline{F}_L F_L  + F_L^ {c}\overline{F}_L^ {c} + \overline{F}_RF_R + F_R^c\overline{F}_R^c \right) \\
\mathcal{L}_{H} &=& - e^{-2\sigma} \left(\overline{F}_{H_u^0} F_{H_u^0}  + \overline{F}_{H_d^0} F_{H_d^0}  \right)\delta(y-\pi R_c).
\end{eqnarray}

\subsection{Scalar field Lagrangian}

The whole 5D Lagrangian -- all SUSY Lagrangians given in terms of 5D fields in this Appendix \ref{lagrangianpart} 
are provided before field redefinition through warp factors -- for scalar fields, $\mathcal{L}_{kin.} + \mathcal{L}_{scalar}$ with  
$\mathcal{L}_{scalar} =  \mathcal{L}_D + \mathcal{L}_{\Omega} + \mathcal{L}_F + \mathcal{L}_H$, can be rewritten in explicit forms as:
\begin{eqnarray}
\mathcal{L}_{kin.} &=& e^{-2\sigma} \partial_\mu \bar \phi_{H_u^0} \partial^\mu \phi_{H_u^0} \delta(y-\pi R_c)
+ e^{-2\sigma} \partial_\mu \bar \phi_{H_d^0} \partial^\mu \phi_{H_d^0} \delta(y-\pi R_c) \nonumber\\
&+& e^{-2\sigma} \partial_\mu \bar \phi_L \partial^\mu \phi_L + e^{-2\sigma} \partial_\mu \bar \phi_L^c \partial^\mu \phi_L^c 
+ e^{-2\sigma} \partial_\mu \bar \phi_R \partial^\mu \phi_R + e^{-2\sigma} \partial_\mu \bar \phi_R^c \partial^\mu \phi_R^c ,
\end{eqnarray}
\begin{eqnarray}
- \sqrt{G}^{\,-1} \mathcal{L}_{scalar} 
&=& D_5\overline{\phi}_L  D_5\phi_L  + D_5\overline{\phi}_R D_5\phi_R + \vert \mu \vert ^2\, \overline{\phi}_{H_u^0} \phi_{H_u^0} \, \delta(y-\pi R_c) \nonumber\\
&+& \left|{\cal Y}\delta(y-\pi R_c)\,\phi_{H_u^0} \phi_L  - D_5'\phi_R^c\right|^2 + \left|{\cal Y}\delta(y-\pi R_c)\,\phi_{H_u^0} \phi_R - D_5'\phi_L^ {c}\right|^2 \nonumber\\
&+& \left| \mu\,\phi_{H_d^0} + {\cal Y}\, \phi_L \phi_R \right|^2 \delta(y-\pi R_c) + 2 g^2 \left| q_L\,\phi_L^{c} \phi_L  + q_R\,\phi_R^c \phi_R \right|^2 \nonumber\\
&+& {1\over2} \left| (\partial_y - 2\sigma')\Sigma - g \Big( q_L \Delta_{\phi_L} + q_R \Delta_{\phi_R} 
+ [ q_{H_d^0} \overline{\phi}_{H_d^0} \phi_{H_d^0} 
+ q_{H_u^0} \overline{\phi}_{H_u^0} \phi_{H_u^0} ] \delta(y-\pi R_c) \Big) \right|^2 \nonumber\\
\label{ScaFieLagPrim}
\end{eqnarray}
where e.g. $\Delta_{\phi_R} \equiv \overline{\phi}_R \phi_R - \phi_R^c \overline{\phi}_R^c$.
By developing the $|D_5\phi|^2$, $|D_5'\phi|^2$ and $|(\partial_y - 2\sigma')\Sigma|^2$ terms from above, one finds the 5D scalar masses:
\begin{eqnarray}
m^2_{\phi_{L/R},\phi_{L/R}^c} \equiv (c_{L/R}^2 \pm c_{L/R} - {15\over4})k^2 + ({3\over 2}\mp c_{L/R}) \partial_y(\partial_y\sigma) 
\quad,\quad m^2_{\Sigma} \equiv -4k^2 +2\partial_y(\partial_y\sigma).
\label{BulScalMass}
\end{eqnarray}
If one now develops the last two lines proportional to the $g$ gauge coupling in the above Lagrangian, one finds (after few simplifications):
\begin{eqnarray}
- \sqrt{G}^{\,-1} \mathcal{L}_{scalar} &=& D_5\overline{\phi}_L D_5\phi_L  + D_5\overline{\phi}_R D_5\phi_R 
+ \vert \mu \vert ^2\, \overline{\phi}_{H_u^0} \phi_{H_u^0} \, \delta(y-\pi R_c) \nonumber\\
&+& \left|{\cal Y}\delta(y-\pi R_c)\,\phi_{H_u^0} \phi_L  - D_5'\phi_R^c\right|^2 + \left|{\cal Y}\delta(y-\pi R_c)\,\phi_{H_u^0} \phi_R - D_5'\phi_L^ {c}\right|^2 \nonumber\\
&+& \left| \mu\, \phi_{H_d^0} + {\cal Y}\, \phi_L \phi_R \right|^2 \delta(y-\pi R_c) + {1\over2} \left\vert D^k_5 \Sigma \right\vert^2     \nonumber\\
&-& g \Big( q_L (\overline{\phi}_L \phi_L - \phi_L^c \overline{\phi}_L^c) + q_R (\overline{\phi}_R \phi_R - \phi_R^c \overline{\phi}_R^c) 
+ \delta(y-\pi R_c) (q_{H_u^0} \overline{\phi}_{H_u^0} \phi_{H_u^0} + q_{H_d^0} \overline{\phi}_{H_d^0} \phi_{H_d^0}) \Big) D^k_5 \Sigma \nonumber\\
&+& {g^2\over2} \Big( q_L^2 (|\phi_L|^4 + |\phi_L^c|^4) + q_R^2 (|\phi_R|^4 + |\phi_R^c|^4) 
+ \delta^2(y-\pi R_c) (q_{H_u^0}^2|\phi_{H_u^0}|^4 + q_{H_d^0}^2|\phi_{H_d^0}|^4) \Big) \nonumber \\
&+& g^2 \Big( q_L^2 |\phi_L^c \phi_L|^2 + q_R^2 |\phi_R^c \phi_R|^2 + q_{H_u^0}q_{H_d^0}\delta^2(y-\pi R_c) |\phi_{H_u^0} \phi_{H_d^0}|^2 \Big) \nonumber \\
&+& g^2 q_L q_R \Big( |\phi_L\phi_R|^2 - |\phi_L\phi_R^c|^2 - |\phi_L^c\phi_R|^2 + |\phi_L^c\phi_R^c|^2 + 2 (\phi_L^c \phi_L\overline{\phi}_R^c \overline{\phi}_R + \phi_R^c \phi_R\overline{\phi}_L^c \overline{\phi}_L) \Big) \nonumber\\
&+& g^2 \Big( q_L (q_{H_u^0} |\phi_L\phi_{H_u^0}|^2 + q_{H_d^0} |\phi_L\phi_{H_d^0}|^2) + q_R ( q_{H_u^0} |\phi_R\phi_{H_u^0}|^2 
+ q_{H_d^0} |\phi_R\phi_{H_d^0}|^2) \Big)\delta(y-\pi R_c) \nonumber \\
\label{ScaFieLag}
\end{eqnarray}
with $D^k_5 = \partial_y - 2\sigma'$.

One can also simply add the soft bilinear terms for the Higgs fields on the TeV-brane and having, as usually in RS, soft squared Higgs masses and $B\mu$ (Higgs mixing) 
scales of order $k$ squared leads to 4D effective soft terms with a scale at the TeV. The soft scalar trilinear $A$ couplings are discussed in details in the main text.

\subsection{Fermion field Lagrangian}

We now derive the fermionic part of the 5D Lagrangian issued from the actions in Eq.(\ref{gaugeAction})-(\ref{muAction}):
\begin{eqnarray}
\mathcal{L}_{fermion} &=& 
- e^{-4\sigma} \left( \chi_L D_5''\zeta_L  + \overline{\chi}_L D_5''\overline{\zeta}_L  + \chi_R D_5''\zeta_R + \overline{\chi}_R D_5''\overline{\zeta}_R \right) \nonumber\\ 
& - & i e^{-3\sigma} \left( \zeta_L \sigma^\mu \partial_\mu \overline{\zeta}_L + \zeta_R \sigma^\mu \partial_\mu \overline{\zeta}_R \right) 
- i e^{-3\sigma} \left( \chi_L \sigma^\mu \partial_\mu \overline{\chi}_L + \chi_R \sigma^\mu \partial_\mu \overline{\chi}_R \right) \nonumber\\
&-& i e^{-3\sigma} \left( \zeta_{H_u^0} \sigma^\mu \partial_\mu \overline{\zeta}_{H_u^0} + \zeta_{H_d^0} \sigma^\mu \partial_\mu \overline{\zeta}_{H_d^0} \right) \delta(y-\pi R_c) \nonumber\\ 
&-& e^{-4\sigma} \left( \lambda_2 (\partial_y-{3\over2}k) \lambda_1 + \overline{\lambda}_2 (\partial_y-{3\over2}k) \overline{\lambda}_1 \right) 
- i e^{-3\sigma} \left( \lambda_1 \sigma^\mu \partial_\mu \bar \lambda_1 + \lambda_2 \sigma^\mu \partial_\mu \bar \lambda_2 \right) \nonumber\\
&-& e^{-4\sigma} \left( \mu\, \zeta_{H_u^0} \zeta_{H_d^0}  + {\cal Y} \phi_{H_u^0} \zeta_L \zeta_R + {\cal Y} \phi_L \zeta_{H_u^0} \zeta_R + {\cal Y} \phi_R\zeta_{H_u^0} \zeta_L  \right)\delta(y-\pi R_c) \nonumber\\
&-& i \sqrt{2} e^{-4\sigma} g \left( q_{H_u^0} (\overline{\phi}_{H_u^0} \lambda_1 \zeta_{H_u^0} - \phi_{H_u^0} \overline{\zeta}_{H_u^0} \overline{\lambda}_1) 
+ q_{H_d^0} (\overline{\phi}_{H_d^0} \lambda_1 \zeta_{H_d^0} - \phi_{H_d^0} \overline{\zeta}_{H_d^0} \overline{\lambda}_1) \right) \delta(y-\pi R_c) \nonumber\\
&-& i \sqrt{2} e^{-4\sigma} g \bigg ( q_L (\overline{\phi}_L \lambda_1 \zeta_L - \phi_L \overline{\zeta}_L \overline{\lambda}_1) + q_R (\overline{\phi}_R \lambda_1 \zeta_R - \phi_R \overline{\zeta}_R \overline{\lambda}_1) \bigg ) \nonumber\\
&+& i \sqrt{2} e^{-4\sigma} g \bigg ( q_L (\overline{\phi}^c_L \lambda_1 \chi_L - \phi^c_L \overline{\chi}_L \overline{\lambda}_1) + q_R (\overline{\phi}^c_R \lambda_1 \chi_R - \phi^c_R \overline{\chi}_R \overline{\lambda}_1) \bigg ) \nonumber\\
&-& i \sqrt{2} e^{-4\sigma} g \bigg ( q_L (\phi_L^c \lambda_2 \zeta_L + \phi_L \chi_L \lambda_2) + q_R (\phi_R^c \lambda_2 \zeta_R + \phi_R \chi_R \lambda_2) \bigg ) + h.c.
\label{FermLag}
\end{eqnarray}
where $D_5'' = \partial_y - (2-c_{L/R})\sigma'$ and $\zeta_{H_{u,d}^0}$ are the higgsino two-component spinors.

\subsection{Gauge field Lagrangian}

Finally, the gauge interactions encoded in the superfield actions of Eq.(\ref{gaugeAction})-(\ref{muAction}) read in terms of the 5D gauge fields as,
\begin{eqnarray}
\mathcal{L}_{gauge} &=& - {1\over4} F_{\mu\nu}F^{\mu\nu} + {1\over2} e^{-2\sigma} \partial_y A_{\mu} \partial_y A^{\mu}   
+ \frac{1}{2} e^{-2\sigma} \partial_\mu \Sigma \partial^\mu \Sigma 
\nonumber\\
&+& e^{-2\sigma} \left(\overline{D_{\mu}\phi}_L D^{\mu}\phi_L + \overline{D_{\mu}\phi}_L^c D^{\mu}\phi_L^c + \overline{D_{\mu}\phi}_R D^{\mu}\phi_R + \overline{D_{\mu}\phi}_R^c D^{\mu}\phi_R^c \right) \nonumber\\
&-& e^{-3\sigma} \left(i \zeta_L \sigma^{\mu} D_{\mu} \overline{\zeta}_L + i \chi_L \sigma^{\mu} D_{\mu} \overline{\chi}_L + i \zeta_R \sigma^{\mu} D_{\mu} \overline{\zeta}_R + i \chi_R \sigma^{\mu} D_{\mu} \overline{\chi}_R \right) \nonumber\\
&+& e^{-2\sigma} \left( \overline{D_{\mu}\phi}_{H_u^0} D^{\mu}\phi_{H_u^0} + \overline{D_{\mu}\phi}_{H_d^0} D^{\mu}\phi_{H_d^0} \right) \delta(y-\pi R_c) \nonumber\\
&-& e^{-3\sigma} \left( i \zeta_{H_u^0} \sigma^{\mu} D_{\mu} \overline{\zeta}_{H_u^0} + i \zeta_{H_d^0} \sigma^{\mu} D_{\mu} \overline{\zeta}_{H_d^0} \right) \delta(y-\pi R_c)
\label{AppZpar}
\end{eqnarray}
where
\begin{eqnarray}
D_{\mu} &\equiv& \partial_{\mu} + i g q_{L/R} A_{\mu} \ / \ \partial_{\mu} + i g q_{H_{u,d}^0} A_{\mu}  
\nonumber \\
\overline{D_{\mu} \phi} D^{\mu}\phi &\equiv& \partial_{\mu}\bar \phi \partial^{\mu}\phi 
+ i g q \left( \phi\partial_{\mu}\overline{\phi} - \overline{\phi}\partial_{\mu}\phi \right) A^{\mu} + g^2 q^2 \overline{\phi}\phi A_{\mu}A^{\mu} 
\nonumber \\
i \zeta \sigma^{\mu} D_{\mu} \overline{\zeta}  &\equiv& i \zeta \sigma^{\mu} \partial_{\mu} \overline{\zeta} - g q \zeta \sigma^{\mu} \overline{\zeta} A_{\mu}.
\nonumber \end{eqnarray}
To compare with another parametrization often found in literature, one can rewrite the second term of Eq.(\ref{AppZpar}) 
in terms of the new coordinate $z=\frac{e^{\sigma(y)}}{k}$ ($y=0 \Leftrightarrow z={1\over k}\equiv R$ and $y=\pi R_c \Leftrightarrow z=\frac{e^{k\pi R_c}}{k}\equiv R'$):
\begin{eqnarray}
{1\over2} e^{-2\sigma} \left( \partial_y A_{\mu} \right)^2 = - {1\over2} A_{\mu} \left(\partial_z^2 - {1\over z} \partial_z\right) A^{\mu}.
\end{eqnarray}


\renewcommand{\thesubsection}{B.\arabic{subsection}} 
\renewcommand{\theequation}{B.\arabic{equation}} 
\setcounter{subsection}{0} 
\setcounter{equation}{0} 

\section{Wave functions} 
\label{WF} 

\subsection{Generic relations} 
\label{Generic} 

Any wave function $f_n(y)$, for a $n$th KK state along the fifth dimension $y$, satisfies the orthonormalization condition
(after the usual RS field redefinition) 
\bea
\int_{-\pi R_c}^{\pi R_c} \bar f_n(y) f_m(y) dy = \delta_{nm} ,
\label{orthonormalization}
\eea
showing that $f_n(y)$ has dimension 1/2, and the completeness relation (see for instance Ref.~\cite{Falkowski})
\bea
\sum_{n=0}^{\infty} \bar f_n(y) f_n(y') =  \delta(y-y') .
\label{completeness}
\eea

\subsection{Solutions for free vectorial fields} 
\label{SolutionFreeVect} 

For the equations of motion in the warped SUSY background, the solutions for the wave functions along the fifth dimension of the free 5D vectorial field in Eq.(\ref{SuperVectorial}), with KK decomposition 
$$
A_{\mu}(x^{\mu},y) = \sum_{n=0}^{\infty} A_{\mu}^{(n)}(x^{\mu}) g^{++}_n(y), 
$$ 
are \cite{PomGher} for $n \geq 1$:
\bea
g^{++}_n(y) = \frac{1}{\sqrt{2\pi R_c}} \frac{e^{\sigma}}{N_n} 
\bigg [J_1\bigg (\frac{M^{(n)} e^\sigma}{k}\bigg ) + b_1^{++}(M^{(n)}) Y_1\bigg (\frac{M^{(n)} e^\sigma}{k}\bigg )\bigg ], \nonumber
\\ b_1^{++}(M^{(n)}) = - \frac{J_1(M^{(n)} /k)+(M^{(n)} /k)J'_1(M^{(n)} /k)}{Y_1(M^{(n)} /k)+(M^{(n)} /k)Y'_1(M^{(n)} /k)}
\eea
where $\sigma=\sigma(y)=k\vert y \vert$, $J_1,Y_1$ ($J'_1,Y'_1$) are the (differentiated) Bessel functions, 
$M^{(n)}=M^{(n)}_{KK}$ is the n-{\it th} KK gauge mass and $N_n$ is the normalization constant. Note that these wave functions are given here
before the field redefinition usually done in the RS model.
The solutions for the wave functions of the associated 5D scalar field in Eq.(\ref{SuperVectorialbis}), with KK decomposition
\bea
\Sigma(x^{\mu},y) = \sum_{n=1}^{\infty} \Sigma^{(n)}(x^{\mu}) g^{--}_n(y), 
\label{SigmaDecomp}
\eea 
are \cite{PomGher} for $n \geq 1$:
\begin{eqnarray}
g^{--}_n(y) = sign(y) \frac{1}{\sqrt{2\pi R_c}} \frac{e^{2\sigma}}{N_n} 
\bigg [J_0\bigg (\frac{M^{(n)} e^\sigma}{k}\bigg ) + b_0^{--}(M^{(n)}) Y_0\bigg (\frac{M^{(n)} e^\sigma}{k}\bigg )\bigg ], \nonumber\\
b_0^{--}(M^{(n)}) = - \frac{J_0(M^{(n)} /k)}{Y_0(M^{(n)} /k)} = b_1^{++}(M^{(n)}). 
\end{eqnarray}
For completeness, we present the wave functions of the would-be $A_5$ component, 
even if we work in the gauge where $A_5=0$ together with the constraint $\partial^\mu A_{\mu}=0$, 
$$
A_5(x^{\mu},y) = \sum_{n=1}^{\infty} A_5^{(n)}(x^{\mu}) a^{--}_n(y), 
$$ 
with, for $n \geq 1$,
\begin{eqnarray}
a^{--}_n(y) = e^{-\sigma} \ g^{--}_n(y).
\end{eqnarray}

Using the various relations on Bessel function derivatives, one obtains the useful following relation:
\begin{eqnarray}
(\partial_y - 2 \sigma') g^{--}_n(y) = - M^{(n)} e^{2 \sigma} g^{++}_n(y).
\label{UsefulRelat}
\end{eqnarray}
There is also a term $-(\sigma''/\sigma')g^{--}_n$ but this term gives rise to a vanishing contribution when replaced in couplings
and integrated over $y$ cause $\sigma''=2k[\delta(y)-\delta(y-\pi R_c)]$ and $g^{--}_n$ vanishes at $y=0,\pi R_c$.
One finds also the relation 
\begin{eqnarray}
\partial_y  g^{++}_n(y) = M^{(n)} g^{--}_n(y).
\label{UsefulRelatBis}
\end{eqnarray}

Eq.(\ref{UsefulRelat})-(\ref{UsefulRelatBis}) allow to recover the differential equations \cite{ChangRSbulk}
\begin{eqnarray}
\partial_y \bigg ( \frac{1}{e^{2 \sigma}} \partial_y g^{++}_n(y) \bigg ) = - (M^{(n)})^2 g^{++}_n(y)
\end{eqnarray}
and \cite{PomGher}
\begin{eqnarray}
- e^{-4 \sigma} \partial_y \bigg ( \frac{1}{e^{4 \sigma}} \partial_y g^{--}_n(y) \bigg ) - 4 k^2 g^{--}_n(y) = (M^{(n)})^2 e^{2 \sigma} g^{--}_n(y).
\end{eqnarray}

\subsection{Scalar/fermion fields and SUSY breaking} 
\label{SolutionFreeScal}

Similarly, the solutions for the wave functions along the fifth dimension of the 5D scalar 
fields introduced in Appendix \ref{lagrangianpart}, with KK decomposition
\begin{eqnarray}
\phi_{L/R}(x^{\mu},y) = \sum_{n=0}^{\infty} \phi_{L/R}^{(n)}(x^{\mu}) f^{++}_n(c_{L/R};y), 
\label{PhiDecomp}
\end{eqnarray}
\begin{eqnarray}
\phi^c_{L/R}(x^{\mu},y) = \sum_{n=1}^{\infty} \phi_{L/R}^{c(n)}(x^{\mu}) f^{--}_n(c_{L/R};y), 
\label{PhiDecompBis}
\end{eqnarray}
read as \cite{PomGher} 
\begin{eqnarray}
f^{++}_0(c_{L/R};y) = \sqrt\frac{k}{2} \, \sqrt{{1-2c_{L/R}}\over{1-e^{-(1-2c_{L/R})k\pi R_c}}} \, e^{k({1\over2}-c_{L/R})(y-\pi R_c)}
\label{ZeroWaveFerm}
\end{eqnarray}
for the unique zero-mode and as [defining $\alpha^\pm_{L/R} = \vert c_{L/R} \pm 1/2\vert$]
\bea
f^{++}_n(c_{L/R};y) = \frac{1}{\sqrt{2\pi R_c}} \frac{e^{2\sigma}}{{\cal N}_n} 
\bigg [J_{\alpha^+_{L/R}}\bigg (\frac{m_{L/R}^{(n)} e^\sigma}{k}\bigg ) 
+ b_{\alpha^+_{L/R}}^{++}(m_{L/R}^{(n)}) Y_{\alpha^+_{L/R}}\bigg (\frac{m_{L/R}^{(n)} e^\sigma}{k}\bigg )\bigg ], 
\nonumber
\\ b_{\alpha^+_{L/R}}^{++}(m_{L/R}^{(n)}) = 
- \frac{[2-(3/2-c_{L/R})]J_{\alpha^+_{L/R}}(m_{L/R}^{(n)} / k)+(m_{L/R}^{(n)} / k)J'_{\alpha^+_{L/R}}(m_{L/R}^{(n)} /k)}{[2-(3/2-c_{L/R})]Y_{\alpha^+_{L/R}}(m_{L/R}^{(n)} / k)+(m_{L/R}^{(n)} / k)Y'_{\alpha^+_{L/R}}(m_{L/R}^{(n)} /k)} \nonumber \\ 
\eea
\bea
f^{--}_n(c_{L/R};y) = sign(y) \frac{1}{\sqrt{2\pi R_c}} \frac{e^{2\sigma}}{{\cal N}_n} 
\bigg [J_{\alpha^-_{L/R}}\bigg (\frac{m_{L/R}^{(n)} e^\sigma}{k}\bigg ) 
+ b_{\alpha^-_{L/R}}^{--}(m_{L/R}^{(n)}) Y_{\alpha^-_{L/R}}\bigg (\frac{m_{L/R}^{(n)} e^\sigma}{k}\bigg )\bigg ], 
\nonumber
\\ b_{\alpha^-_{L/R}}^{--}(m_{L/R}^{(n)}) = 
- \frac{J_{\alpha^-_{L/R}}(m_{L/R}^{(n)} / k)}{Y_{\alpha^-_{L/R}}(m_{L/R}^{(n)} / k)}  \nonumber \\ 
\eea
for KK modes ($n \geq 1$), where $m_{L/R}^{(n)}=m^{(n)}_{KK}(c_{L/R})$ is the n-{\it th} KK scalar mass 
(the KK fermion spectrum is identical as we do not consider Sherk-Schwarz like mechanisms of SUSY breaking) 
and ${\cal N}_n$ is the n\-orma\-lization constant. At this level,
the field redefinition has not been performed for the n$th$ wave functions.

From the above wave function expressions, we deduce the useful relations
\begin{eqnarray}
(\partial_y - (c_{L/R}+\frac{3}{2})\sigma') f^{--}_n(c_{L/R};y) = - m^{(n)}_{KK}(c_{L/R}) \ e^{\sigma} f^{++}_n(c_{L/R};y),
\label{UsefulRelatII}
\end{eqnarray}
\begin{eqnarray}
(\partial_y - (-c_{L/R}+\frac{3}{2})\sigma') f^{++}_n(c_{L/R};y) = m^{(n)}_{KK}(c_{L/R}) \ e^{\sigma} f^{--}_n(c_{L/R};y).
\label{UsefulRelatIIBis}
\end{eqnarray}

Concerning fermions, an example of KK decomposition is (in the two-component notation),
\begin{eqnarray}
\zeta_L(x^{\mu},y) = \sum_{n=0}^{\infty} \zeta_L^{(n)}(x^{\mu}) \omega^{++}_n(c_L;y), 
\label{PsiDecompI}
\end{eqnarray}
\begin{eqnarray}
\chi_L(x^{\mu},y) = \sum_{n=1}^{\infty} \chi_L^{(n)}(x^{\mu}) \omega^{--}_n(c_L;y), 
\label{PsiDecompII}
\end{eqnarray}
where $c_L$ parametrizes the 5D mass of the $\Phi_L$ superfield in Eq.(\ref{matterAction}). The wave function $\omega^{++}_n(c_L;y)$
(respectively $\omega^{--}_n(c_L;y)$) is exactly equal [after the RS field redefinition] to the scalar superpartner wave function $f^{++}_n(c_L;y)$ 
(respectively $f^{--}_n(c_L;y)$) as imposed by SUSY. This remains true as long as the SUSY breaking does not occur through the Scherk-Schwarz  
mechanism. In fact, these scalar wave functions $f^{++/--}_n$ depend on the scalar bulk mass $a$ and boundary mass $b$ which are imposed by 
5D SUSY to be the following functions of the fermionic superpartner bulk mass $c_L \sigma'$ [{\it c.f.} Eq.(\ref{FermLag})], 
$a(c_L)=c_L^2 \pm c_L-15/4$, $b(c_L)=3/2 \mp c_L$ [{\it c.f.} Eq.(\ref{BulScalMass})], 
relations rendering the scalar and fermion wave functions identical in terms of the $c_L$ parameter.
Analog remarks hold for the $c_R$ case as well as for the KK masses $m^{(n)}_{KK}(c_{L/R})$.

A possible SUSY breaking framework -- that we consider in this paper without specifying the underlying breaking mechanism -- 
is that additional bulk/brane masses arise for the scalar fields, encoded e.g. in $a(c_L)+\delta a$
and $b(c_L)+\delta b$: these corrective masses spoil the SUSY relations between scalar and fermion bulk/brane masses and in turn break SUSY. 
It is more convenient for the numerical calculation parts to define e.g. a new $c_L^s$ parameter for scalars such that $a(c_L)+\delta a = a(c_L^s)$ and 
$b(c_L)+\delta b = b(c_L^s)$, $c_L^s$ being different from the fermion (or superfield) mass parameter $c_L$ in this framework. Then the
fermion wave functions can still be written $\omega^{++/--}_n(c_L;y)$ while the scalar wave functions $f^{++/--}_n(c_L^s;y)$ are now controlled by an
independent mass parameter related to the amount of SUSY breaking.
\\ Note that such a SUSY breaking framework resembles the situation where additional (w.r.t. the pure 5D SUSY Lagrangian) bulk masses e.g. $\delta c_L \sigma'$
would appear for fermions. Indeed, the SUSY bulk masses for fermions $c_L \sigma'$ would be shifted to $c_L^f \sigma'= (c_L + \delta c_L) \sigma'$, 
a new mass independent from the scalar ones depending on $a(c_L), b(c_L)$ [one would deal here with $\omega^{++/--}_n(c_L^f;y)$ and $f^{++/--}_n(c_L;y)$]. 
This framework would be quantitatively equivalent to the previous one for final physical masses but would lead to an interesting alternative
to usual SUSY breaking frameworks (4D or 5D): here typically both fermions and their scalar superpartners would be initially heavy (mainly due to their 
large and identical effective Yukawa couplings generated by equal wave functions), i.e. respecting the lower bounds on scalar masses, while the fermions would 
become lighter (reduced down to their measured masses) because of the effect of their arising SUSY breaking bulk masses [$\delta c_{L/R} >0$] on wave functions.


\renewcommand{\thesubsection}{C.\arabic{subsection}} 
\renewcommand{\theequation}{C.\arabic{equation}} 
\setcounter{subsection}{0} 
\setcounter{equation}{0} 

\section{4D versus 5D propagators} 
\label{propapator}

For a generic bulk field (expressed in terms of $z=\frac{e^{\sigma(y)}}{k}$)
\begin{eqnarray}
\Phi(x^{\mu},z) = \sum_{n} \phi^{(n)}(x^{\mu})  S^{(n)}_a(z) \nonumber
\end{eqnarray}
satisfying the orthonormalization condition
\begin{eqnarray}
\int_R^{R'}\!\!\!\!\!dz \left(\frac{R}{z}\right)^3 S^{(n)}_a(z) S^{(m)}_a(z) = \delta_{nm} ,
\nonumber
\end{eqnarray}
the 5D action and the corresponding KK decomposed 4D action are given by:
\begin{eqnarray}
{\cal S}_5 &=& - \frac{1}{2} \int d^4x \int_R^{R'}\!\!\!\!\!dz \left(\frac{R}{z}\right)^3 \bar \Phi(x^{\mu},z) \left[ \eta^{\mu\nu} \partial_{\mu} \partial_{\nu} - \partial_z^2 + {3\over z} \partial_z + \frac{a^2}{z^2} \right] \Phi(x^{\mu},z) \\
{\cal S}_4 &=& - \frac{1}{2} \int d^4x \sum_n \bar \phi^{(n)}(x^{\mu}) \Bigg( \eta^{\mu\nu} \partial_{\mu} \partial_{\nu} + m_n^2 \Bigg) \phi^{(n)}(x^{\mu}) .
\end{eqnarray}
The equation of motion reads as
\begin{eqnarray}
\left[ -\partial_z^2 + {3\over z} \partial_z + \frac{a^2}{z^2} \right] S^{(n)}_a(z) = m_n^2 S^{(n)}_a(z) \nonumber 
\nonumber
\end{eqnarray}
and the 2-point functions are defined as follows,
\begin{eqnarray}
\mbox{From ${\cal S}_5$ :} && \left[ p^2 + \partial_z^2 - {3\over z} \partial_z - \frac{a^2}{z^2} \right] G^{S_a}_5(p^2;z,z') = \delta(z-z') \\
\mbox{From ${\cal S}_4$ :} && \left[ p^2 - m_n^2 \right] G_4^{(n)}(p^2) = 1 .
\end{eqnarray}
From above, one can find the relation between the 4D and 5D propagators:
\begin{eqnarray}
G^{S_a}_5(p^2;z,z') &=& \sum_{n=0}^{\infty} G_4^{(n)}(p^2)  S^{(n)}_a(z) S^{(n)}_a(z') = \sum_{n=0}^{\infty} \frac{S^{(n)}_a(z) S^{(n)}_a(z')}{p^2 - m_n^2}  \\
 &=& \frac{S^{(0)}_a(z) S^{(0)}_a(z')}{p^2} + \sum_{n\geq1} \frac{S^{(n)}_a(z) S^{(n)}_a(z')}{p^2 - m_n^2}.
\label{5Dprop}
\end{eqnarray}
Indeed, one can check that:
\begin{eqnarray}
&& \left[ p^2 + \partial_z^2 - {3\over z} \partial_z - \frac{a^2}{z^2} \right] G^{S_a}_5(p^2;z,z') =
\left[ p^2 + \partial_z^2 - {3\over z} \partial_z - \frac{a^2}{z^2} \right] \sum_n G_4^{(n)}(p^2)  S^{(n)}_a(z) S^{(n)}_a(z') \nonumber\\
&=& \sum_n G_4^{(n)}(p^2) \left[ p^2 + \partial_z^2 - {3\over z} \partial_z - \frac{a^2}{z^2} \right] S^{(n)}_a(z) S^{(n)}_a(z') =
\sum_n G_4^{(n)}(p^2) \left[ p^2 - m_n^2 \right] S^{(n)}_a(z) S^{(n)}_a(z') \nonumber\\
&=& \sum_n S^{(n)}_a(z) S^{(n)}_a(z') = \delta(z-z').
\end{eqnarray}


\newpage


\begin{thebibliography}{999} 

\bibitem{RS} L.~Randall and R.~Sundrum, Phys. Rev. Lett. \textbf{83} (1999)
3370. See also, M.~Gogberashvili, Int. J. Mod. Phys. \textbf{D11} (2002)
1635.
 
\bibitem{PomGher} T.~Gherghetta and A.~Pomarol, Nucl. Phys.  \textbf{B586} (2000) 141.

\bibitem{GHunif} See e.g. M.~Carena, E.~Ponton, J.~Santiago and C.~E.~M.~Wagner, 
Nucl. Phys. \textbf{B759} (2006) 202; 
Phys. Rev. \textbf{D76} (2007) 035006.

\bibitem{Higgsless} C.~Cs\'{a}ki, C.~Grojean, H.~Murayama, L.~Pilo and J.~Terning, 
Phys. Rev. \textbf{D69} (2004) 055006; 
C.~Cs\'{a}ki, C.~Grojean, L.~Pilo and J.~Terning, Phys. Rev. Lett. \textbf{92} (2004) 101802; 
G.~Cacciapaglia, C.~Cs\'{a}ki, C.~Grojean and J.~Terning, Phys. Rev. \textbf{D70} (2004) 075014.

\bibitem{AdSCFT} J.~M.~Maldacena, Adv. Theor. Math. Phys. \textbf{2} (1998) 231; 
Int. J. Theor. Phys. \textbf{38} (1999) 1113; S.~S.~Gubser, I.~R.~Klebanov and A.~M.~Polyakov, 
Phys. Lett. \textbf{B428} (1998) 105; 
E.~Witten, Adv. Theor. Math. Phys. \textbf{2} (1998) 253; 
N.~Arkani-Hamed, M.~Porrati and L.~Randall, JHEP \textbf{0108} (2001) 017.

\bibitem{MHCM} R.~Contino, Y.~Nomura and A.~Pomarol, Nucl. Phys.  \textbf{B671} (2003) 148; 
K.~Agashe, R.~Contino and A.~Pomarol, Nucl. Phys. \textbf{B719} (2005) 165; 
K.~Agashe and R.~Contino, Nucl. Phys. \textbf{B742} (2006) 59;
R.~Contino, L.~Da Rold and A.~Pomarol, Phys. Rev. \textbf{D75} (2007) 055014. 

\bibitem{EWSBI} I. Antoniadis Phys. Lett.  \textbf{B246} (1990) 377; 
G. Dvali and M. Shifman Nucl. Phys. {\textbf B504} (1997) 127.

\bibitem{EWSBII} A. Delgado, M. Quir\'os, Nucl. Phys. \textbf{B607} (2001) 99.
  
\bibitem{hidden} E. A. Mirabelli, M. E. Peskin, Phys. Rev. \textbf{D58} (1998) 065002;
W. D. Goldberger, Y. Nomura, D. R. Smith, Phys. Rev. \textbf{D67} (2003) 075021;
H.P. Nilles, TASI97 lectures as published in Supersymmetry, Supergravity and Supercolliders, 
World Scientific, Ed. J. A. Bagger. 

\bibitem{Steve} S.~Abel and T.~Gherghetta, JHEP \textbf{1012} (2010) 091.

\bibitem{BCmech} T. Gherghetta, A. Pomarol, Nucl. Phys. \textbf{B602} (2001) 3.

\bibitem{SSmech} J. Scherk and J. H. Schwarz, Phys. Lett. \textbf{B82} (1979) 60; 
Nucl. Phys. \textbf{B153} (1979) 61.

\bibitem{SSmssm} I. Antoniadis, Phys. Lett. \textbf{B246} (1990) 377.

\bibitem{SShorava} E. Dudas and C. Grojean, Nucl. Phys. \textbf{B507} (1997) 553. 

\bibitem{newSUSYbRS} 
C. P. Burgess, P. G. Camara, S. P. de Alwis, S. B. Giddings, A. Maharana, F. Quevedo, K. Suruliz, 
JHEP \textbf{0804} (2008) 053;
H. Abe, Y. Sakamura, JHEP \textbf{0703} (2007) 106;
M. R. Douglas, J. Shelton, G. Torroba, {\tt arXiv: hep-th/0704.4001};
N. Uekusa, Mod. Phys. Lett. \textbf{A23} (2008) 603. 

\bibitem{tension} J. Casas, J. R. Espinosa, I. Navarro, Nucl. Phys. \textbf{B620} (2002) 195;
R. C. Myers, O. Tafjord, JHEP \textbf{0111} (2001) 009.

\bibitem{Schmidhuber} C. Schmidhuber, Nucl. Phys. \textbf{B585} (2000) 385; \textbf{B619} (2001) 603.

\bibitem{unification} K. R. Dienes, E. Dudas, T. Gherghetta, Phys. Lett. \textbf{B436} (1998) 55; 
Nucl. Phys. \textbf{B537} (1999) 47.

\bibitem{AdSUSY} P. K. Townsend, Phys. Rev. \textbf{D15} (1977) 2802; 
S. Deser and B. Zumino, Phys. Rev. Lett. \textbf{38} (1977) 1433.

\bibitem{RSUSYspec} E. Shuster, Nucl. Phys. \textbf{B554} (1999) 198.  
 
\bibitem{RSUSYgrav} R. Altendorfer, J. Bagger, D. Nemeschansky, Phys. Rev. \textbf{D63} (2001) 125025; 
N. Alonso-Alberca, P. Meessen, T. Ortin, Phys. Lett. \textbf{B482} (2000) 400.

\bibitem{RSUSYunif} 
A. Pomarol, Phys. Rev. Lett. \textbf{85} (2000) 4004; 
W. D. Goldberger, Y. Nomura, D.R. Smith, Phys. Rev. \textbf{D67} (2003) 075021; 
H.-Y. Guo, C.-G. Huang, Z. Xu, B. Zhou, Phys. Lett. \textbf{A331} (2004) 1; 
Y. Nomura, D.R. Smith, Phys. Rev. \textbf{D68} (2003) 075003;
O. Saremi, A. W. Peet, Phys. Rev. \textbf{D70} (2004) 026008.

\bibitem{RString}H. Verlinde, Nucl. Phys. \textbf{B580} (2000) 264; 
A. Bacchetta, P. J. Mulders, Phys. Rev. \textbf{D62} (2000) 114004;
J.Perez-Peraza, J. Velasco, A. Gallegos-Cruz, M. Alvarez-Madrigal, 
A. Faus-Golfe, A. Sanchez-Hertz, {\tt arXiv: hep-th/0011167};
A. Falkowski, Z. Lalak, S. Pokorski, Phys. Lett. \textbf{B491} (2000) 172; 
R. N.Faustov, A.P.Martynenko, Phys. Atom. Nucl. \textbf{65} (2002) 265, 
Yad. Fiz. \textbf{65} (2002) 291;
E. Bergshoeff, R. Kallosh, A. Van Proeyen, JHEP \textbf{0010} (2000) 033; 
J. R. Pelaez, A. Gomez Nicola, AIP Conf. Proc. \textbf{602} (2001) 34; 
H. P\"as, T.W. Kephart, Phys. Rev. \textbf{D70} (2004) 086009;
G. A. Diamandis, B. C. Georgalas, P. Kouroumalou, A. B. Lahanas, Phys. Lett. \textbf{B602} (2004) 112.
  
\bibitem{Strathdee} J.~A.~Strathdee, Int. J. Mod. Phys. \textbf{A2} (1987) 273.

\bibitem{4DSuperFC} D.~Mart\'{\i} and A.~Pomarol, Phys. Rev. \textbf{D64} (2001) 105025.

\bibitem{Sharpe} E. Sharpe, Nucl. Phys. \textbf{B523} (1998) 211.

\bibitem{Garrie} M.~McGarrie and R.~Russo, {\tt arXiv:1004.3305 [hep-ph]};
M.~McGarrie and D.~C.~Thompson, {\tt arXiv:1009.4696 [hep-th]}.

\bibitem{Gero} A. Delgado, G. v. Gersdorff, P. John, M. Quir\'os, 
Phys. Lett. \textbf{B517} (2001) 445;
G. Grignani, M. Orselli, G. W. Semenoff, JHEP \textbf{0107} (2001) 004.

\bibitem{HewSa} J. L. Hewett and D. Sadri, Phys. Rev. \textbf{D69} (2004) 015001.    

\bibitem{NillesReport} H.~P.~Nilles, Phys. Rept. \textbf{110} (1984) 1.

\bibitem{DjouadiReviewII} A. Djouadi, Phys. Rept. \textbf{459} (2008) 1.

\bibitem{PDG} Particle Data Group,  
Phys. Lett. \textbf{B667} (2008) 1; J. Phys. \textbf{G37} (2010) 075021.

\bibitem{Gautam} G. Bhattacharyya, S. K. Majee, A. Raychaudhuri, Nucl. Phys. \textbf{B793} (2008) 114;	
G. Bhattacharyya, S. K. Majee, T.S. Ray, Phys. Rev. \textbf{D78} (2008) 071701; 
G. Bhattacharyya, T. S. Ray, JHEP \textbf{1005} (2010) 040.

\bibitem{DelgQuir} A. Delgado, M. Quir\'os, Phys. Lett. \textbf{B484} (2000) 355.

\bibitem{Aldo} A. Deandrea, P. Hosteins, M. Oertel, J. Welzel, Phys. Rev. \textbf{D75} (2007) 113005. 

\bibitem{GautamBIS} 	
    K. R. Dienes, E. Dudas, T. Gherghetta, Phys. Lett. \textbf{B436} (1998) 55; Nucl. Phys. \textbf{B537} (1999) 47;
   G. Altarelli, F. Feruglio, Phys. Lett. \textbf{B511} (2001) 257;
   G. Bhattacharyya, K. Sridhar, J. Phys. \textbf{G29} (2003) 993;
   Y. Nomura, D. Poland, B. Tweedie, JHEP \textbf{0612} (2006) 002.

\bibitem{Fichet}  
F. Br\"ummer, S. Fichet, A. Hebecker, S. Kraml, JHEP \textbf{0908} (2009) 011;
F. Br\"ummer, S. Fichet, S. Kraml, R.K. Singh, {\tt arXiv:1007.0321 [hep-ph]}.

\bibitem{Ben} M. Redi and B. Gripaios, {\tt arXiv:1004.5114 [hep-ph]}.

\bibitem{AKnochel} A.~Knochel and T.~Ohl, Phys. Rev. \textbf{D78} (2008) 045016; {\tt arXiv:0909.5330 [hep-ph]}; 
A.~Knochel, to appear in conference proceedings of Rencontres de Moriond 2010 (Electroweak), {\tt arXiv:1005.2103 [hep-ph]}.

\bibitem{pMSSM} A. Djouadi and S. Rosier-Lees (conv.), 
The Minimal Supersymmetric Standard Model: Group Summary Report, {\tt arXiv: hep-ph/9901246}.

\bibitem{RSmass} S.~J.~Huber and Q.~Shafi, Phys. Lett. \textbf{B498} (2001) 256; 
\textbf{B512} (2001) 365; \textbf{B544} (2002) 295; \textbf{B583} (2004) 293;
S.~Chang \textit{et al.}, Phys. Rev. \textbf{D73} (2006) 033002; 
G.~Moreau and J.~I.~Silva-Marcos, JHEP \textbf{0601} (2006) 048; \textbf{0603} (2006) 090;
K.~Agashe \textit{et al.}, Phys. Rev. \textbf{D71} (2005) 016002; 
Phys. Rev. Lett. \textbf{93} (2004) 201804; 
Phys. Rev.  \textbf{D75} (2007) 015002; Phys. Rev. \textbf{D74} (2006) 053011.

\bibitem{RSmassBIS} T.~Lari \textit{et al.}, Eur. Phys. J. \textbf{C57} (2008) 183; 
M.~Raidal \textit{et al.}, Eur. Phys. J. \textbf{C57} (2008) 13.

\bibitem{DelgQuiSS} A. Delgado, A. Pomarol, M. Quir\'os, Phys. Rev. \textbf{D60} (1999) 095008.

\bibitem{PartSusyRS} T. Gherghetta and A. Pomarol, Phys. Rev. \textbf{D67} (2003) 085018. 

\bibitem{DreesHag} M. Drees, K. Hagiwara, Phys. Rev. \textbf{D42} (1990) 1709.

\bibitem{RSewpt} C. Cs\'aki \textit{et al.}, Phys. Rev. \textbf{D66} (2002) 064021; 
G. Burdman, Phys. Rev. \textbf{D66} (2002) 076003; 
J. Hewett, F. Petriello and T. Rizzo, JHEP \textbf{0209} (2002) 030.

\bibitem{ADMS} K.~Agashe, A.~Delgado, M.~J.~May and R.~Sundrum, JHEP \textbf{0308} (2003) 050;
      A.~Delgado and A.~Falkowski, JHEP \textbf{0705} (2007) 097. 

\bibitem{CBGMI} C.~Bouchart and G.~Moreau, Nucl. Phys. \textbf{B810} (2009) 66; Phys. Rev. {\bf D80} (2009) 095022;
A.~Djouadi, G.~Moreau and F.~Richard, Nucl. Phys. {\bf B773} (2007) 43.

\bibitem{Hambye} A. Abada, C.Biggio, F. Bonnet, M.B. Gavela, T. Hambye, JHEP \textbf{0712} (2007) 061.

\bibitem{MultiBrane} G.~Moreau, Eur. Phys. J. \textbf{C40} (2005) 539. 

\bibitem{Ghilencea} D. Ghilencea, H.P. Nilles, Phys. Lett. \textbf{B507} (2001) 327;
E.M. Prodanov, Phys. Lett. \textbf{B530} (2002) 210.

\bibitem{Kim} H. Do Kim, {\tt arXiv: hep-ph/0106072};
B. Morariu, A.P. Polychronakos, JHEP \textbf{0107} (2001) 006.

\bibitem{KKreg} I. Antoniadis, S. Dimopoulos, A. Pomarol, 
M. Quir\'os, Nucl. Phys. \textbf{B544} (1999) 503;
A. Delgado, A. Pomarol, M. Quir\'os, Phys. Rev. \textbf{D60} (1999) 095008;
N. Arkani-Hamed, L. J. Hall, Y. Nomura, D. Smith, N. Weiner, Nucl. Phys. \textbf{B605} (2001) 81;
R. Barbieri, L. J. Hall, Y. Nomura, Phys. Rev. \textbf{D63} (2001) 105007.

\bibitem{Koba} T. Kobayashi, H. Terao, Prog. Theor. Phys. \textbf{107} (2002) 785;
Sunil Mukhi, Pramana \textbf{58} (2002) 21.

\bibitem{Gero2loops} A. Delgado, G. V. Gersdorff, M. Quir\'os, Nucl. Phys. \textbf{B613} (2001) 49;
M. Giovannini, Phys. Rev. \textbf{D64} (2001) 124004.

\bibitem{PV} R. Contino, L. Pilo, Phys. Lett. \textbf{B523} (2001) 347;
T. Fujita, K. Ohashi, Prog. Theor. Phys. \textbf{106} (2001) 221; 
A. Masiero, C.A. Scrucca, M. Serone, L. Silvestrini, Phys. Rev. Lett. \textbf{87} (2001) 251601;
A. Nayeri and A. Reynolds, {\tt arXiv: hep-ph/0107201}; 
V. Di Clemente, S. F. King, D. A. J. Rayner, Nucl. Phys. \textbf{B617} (2001) 71.

\bibitem{5D} 
K. Goeke, M. V. Polyakov, M. Vanderhaeghen, Prog. Part. Nucl. Phys. \textbf{47} (2001) 401; 
R. Emparan, D. Mateos, P. Townsend, JHEP \textbf{0107} (2001) 011;
T. Ueda, T. Uematsu, K. Sasaki, Phys. Lett. \textbf{B640} (2006) 188;
E. \'Alvarez, A. F. Faedo, Phys. Rev. \textbf{D74} (2006) 124029;
G. F. Giudice, R. Rattazzi, J. D. Wells, Nucl. Phys. \textbf{B630} (2002) 293;
R. Contino, A. Gambassi, J. Math. Phys. \textbf{44} (2003) 570.

\bibitem{anomaly} S. L. Adler and W. A. Bardeen, Phys. Rev. \textbf{182} (1969) 1517; 
R. Jackiw, Lectures on ÒCurrent Algebra and its ApplicationsÓ, Princeton University Press, 1972.
  
\bibitem{Ritesh} A. Djouadi, G. Moreau, R.K. Singh, Nucl. Phys. \textbf{B797} (2008) 1.

\bibitem{warpCS} T. Hirayama and K. Yoshioka, JHEP \textbf{0401} (2004) 032.

\bibitem{Stuck} C. T. Hill, Phys. Rev. \textbf{D73} (2006) 085001;
P. Anastasopoulos, M. Bianchi, E. Dudas and E. Kiritsis, JHEP \textbf{0611} (2006) 057.

\bibitem{BenBis} B.~Gripaios, Phys. Lett. \textbf{B663} (2008) 419.

\bibitem{Split} See e.g. L.~Senatore, Phys. Rev. \textbf{D71} (2005) 103510; D.~A.~Demir, {\tt arXiv:0410056 [hep-ph]}. 

\bibitem{AbdPLB} A.~Djouadi, Phys. Lett. \textbf{B435} (1998) 101.

\bibitem{RSfusion} A.~Djouadi and G.~Moreau, Phys. Lett. {\bf B660} (2008) 67.

\bibitem{SusyHiggsTev} L.~Scodellaro, the CDF, D0 Collaborations, {\tt arXiv:0905.2554 [hep-ex]}.

\bibitem{SmHiggsTev} Tevatron New Phenomena, Higgs working group, for the CDF collaboration, DZero collaboration, 
{\tt arXiv:0903.4001 [hep-ex]}.

\bibitem{SmHiggsErr} J.~Baglio and A.~Djouadi, {\tt arXiv:1012.2748 [hep-ph]}.

\bibitem{CCB} J.-M.~Fr\`ere, D.~R.~T.~Jones and S.~Raby, Nucl. Phys. \textbf{B222} (1983) 11; 
L.~Alvarez-Gaum\'e, J.~Polchinski and M.~Wise, Nucl. Phys. \textbf{B221} (1983) 495; 
M.~Claudson, L.~Hall and I.~Hinchliffe, Nucl. Phys. \textbf{B228} (1983) 501.

\bibitem{NMSSMccb} U.~Ellwanger and C.~Hugonie, Phys. Lett. \textbf{B457} (1999) 299.

\bibitem{UFB} H.~Komatsu, Phys. Lett. \textbf{B215} (1988) 323.

\bibitem{COSMOccb} A.~Strumia, Nucl. Phys. \textbf{B482} (1996) 24; 
T.~Falk, K.~A.~Olive, L.~Koszkowski, A.~Singh and M.~Srednicki, Phys. Lett. \textbf{B396} (1997) 50.

\bibitem{EWWGwebPage} See last results/references on {\tt http://lepewwg.web.cern.ch/LEPEWWG/} .

\bibitem{TDRatlas} ATLAS detector and physics performance, Technical Design Report, Volume II, 25 May 1999;
CMS Physics TDR: Volume II (PTDR2), Physics Performance, 25 June 2006.

\bibitem{firstLHCexclusion} CMS Collaboration, {\tt arXiv:1101.1628 [hep-ex]};
ATLAS Collaboration, {\tt arXiv:1102.2357 [hep-ex]}; {\tt arXiv:1102.5290 [hep-ex]}; {\tt arXiv:1103.6214 [hep-ex]}; 
ATLAS-CONF-2011-039; ATLAS-CONF-2011-064; ATLAS-CONF-2011-086.

\bibitem{TevWebPage} CDF and D$0$ experiment webpages: {\tt http://www-cdf.fnal.gov/} and {\tt http://www-d0.fnal.gov/} .

\bibitem{spec} ATLAS Collaboration, {\tt arXiv:1103.6208 [hep-ex]}.  

\bibitem{gravLSP} CMS Collaboration, {\tt arXiv:1103.0953 [hep-ex]}. 

\bibitem{SQsupGL} ATLAS Collaboration, {\tt arXiv:1103.4344 [hep-ex]}. 

\bibitem{SQeqGL} CMS Collaboration, {\tt arXiv:1105.3152 [hep-ex]}. 

\bibitem{Bartl} A.~Bartl \textit{et al.}, Eur. Phys. J. direct \textbf{C2} (2000) 6. 

\bibitem{FRICHARD} F.~Richard, Private communication. 

\bibitem{Feng} J.~L.~Feng and D.~E.~Finnell, Phys. Rev. \textbf{49} (1994) 2369.  

\bibitem{LHCgluonKK} 
K.~Agashe \textit{et al.}, Phys. Rev. \textbf{D77} (2008) 015003;  
B.~Lillie, J.~Shu and T.~M.~P.~Tait, Phys. Rev. \textbf{D76} (2007) 115016;
B.~Lillie, L.~Randall and L.-T.~Wang, JHEP \textbf{0709} (2007) 074;
M.~Guchait, F.~Mahmoudi and K.~Sridhar, Phys. Lett. \textbf{B666} (2008) 347;
A.~Djouadi, G.~Moreau and R.~K.~Singh, Nucl. Phys. \textbf{B797} (2008) 1.

\bibitem{LHCbosonKK} 
K.~Agashe \textit{et al.}, Phys. Rev. \textbf{D81} (2010) 096002; \textbf{D80} (2009) 075007; \textbf{D76} (2007) 115015; 
F.~Ledroit, G.~Moreau and J.~Morel, JHEP \textbf{0709} (2007) 071.

\bibitem{Sher} E.~De~Pree and M.~Sher, Phys. Rev. \textbf{D73} (2006) 095006.

\bibitem{4DSuperFA} N.~Marcus, A.~Sagnotti and W.~Siegel, Nucl. Phys. \textbf{B224} (1983) 159.

\bibitem{4DSuperFB} N.~Arkani-Hamed, T.~Gregoire and J.~Wacker, JHEP \textbf{0203} (2002) 055.

\bibitem{Falkowski} A.~Falkowski, Phys. Rev. \textbf{D77} (2008) 055018.

\bibitem{ChangRSbulk} S.~Chang \textit{et al.}, Phys. Rev. \textbf{D62} (2000) 084025.


\end{thebibliography}
\end{document}